\documentclass[aps,pre,floatfix,superscriptaddress,showpacs,nobibnotes,twocolumn]{revtex4-1}
\usepackage{CJK}
\usepackage{amsmath,amssymb,graphicx,psfrag}
\usepackage{bm}
\usepackage[CJKbookmarks=true]{hyperref}

\usepackage{color}
\usepackage[squaren,Gray, cdot]{SIunits}        %
\usepackage{soul}
\newcommand{\nedit}[1] {{\color{blue} #1}}
\definecolor{coloreditll}{RGB}{218,97,0}
\newcommand{\editll}[1] {{\color{blue} #1}}
\usepackage{ulem}
\usepackage{xcolor}
\newcommand\editsout{\bgroup\markoverwith{\textcolor{blue}{\rule[0.5ex]{10pt}{0.4pt}}}\ULon}
\renewcommand{\st}[1]{\unskip}
\usepackage{multirow}
\usepackage{array}
\newcolumntype{L}[1]{>{\raggedright\let\newline\\\arraybackslash\hspace{0pt}}m{#1}}
\newcolumntype{C}[1]{>{\centering\let\newline\\\arraybackslash\hspace{0pt}}m{#1}}
\newcolumntype{R}[1]{>{\raggedleft\let\newline\\\arraybackslash\hspace{0pt}}m{#1}}

\addtocounter{secnumdepth}{2}

\begin{document}

\title{Microstructural pattern formation during liquid metal dealloying: Phase-field simulations and theoretical analyses}

\author{Longhai~Lai }
\affiliation{Department of Physics and Center for Interdisciplinary Research on Complex Systems, Northeastern University, Boston, MA 02115 USA}
\author{Pierre-Antoine~Geslin}
\affiliation{Univ. Lyon, CNRS, INSA Lyon, UCBL, MATEIS, UMR5510, 69621 Villeurbanne, France}
\author{Alain~Karma}
\email[]{a.karma@northeastern.edu} 
\affiliation{Department of Physics and Center for Interdisciplinary Research on Complex Systems, Northeastern University, Boston, MA 02115 USA}
\date{\today}

\begin{abstract}
In recent years, liquid metal dealloying has emerged as a promising material processing method to generate micro and nano-scale bicontinuous or porous structures. 
Most previous studies focused on the experimental characterization of the dealloying process and on the properties of the dealloyed materials, leaving the theoretical study incomplete to fully understand the fundamental mechanisms of the liquid metal dealloying process. 
In this paper, we use theoretical models and phase-field simulations to clarify the kinetics and pattern formation during liquid metal dealloying. 
Our investigation starts from a theoretical analysis of the 1D dissolution of a binary precursor alloy, which reveals that the 1D dissolution process involves two regimes. In the first regime, due to the low solubility of one of the elements in the melt, it accumulates at the solid-liquid interface, which reduces the dissolution kinetics. In the second regime, the interface kinetics reaches a stationary regime where both elements of the precursor alloy dissolve into the melt. %
Previous works revealed that in the early dealloying stage, the dealloying front is destabilized by an interfacial spinodal decomposition, which triggers the formation of interconnected ligaments.
We extend this line of work by proposing a linear stability analysis able to predict the initial length-scale of the ligaments formed in the initial stage of the dealloying. Combining this analysis with the 1D dissolution model proposed here enables us to better understand the initial conditions (composition of the precursor alloy and the melt) leading to a planar dissolution without interface destabilization. 
Finally, we report a strong influence of solid-state diffusion on dealloying that was overlooked in previous studies. Although the solid-state diffusivity is four to five orders of magnitude smaller than in the liquid phase, it is found to affect both dissolution kinetics and ligament morphologies.
\end{abstract}

\maketitle
\section{Introduction}
Dealloying is a well-known process used to manufacture nanoporous materials. The fundamental mechanism of this process is selective dissolution, where one element is dissolved from an alloy with two or more components, leaving the rest of the component(s) to form a nanoporous structure.
Due to the high porosity and interfacial area, those nanoporous metals have been shown to display outstanding properties and to find potential applications in various fields. It was used to fabricate a wide range of functional materials such as actuators \cite{biener2009surface}, catalytic materials \cite{wittstock2010nanoporous, fujita2012atomic, zugic2017dynamic}, sensors \cite{hu2008electrochemical}, fuel cells \cite{zeis2007platinum, snyder2010oxygen}, electrolytic capacitors \cite{lang2011nanoporous, kim2015optimizing, chen2013toward}, radiation-damage resistant materials \cite{bringa2011nanoporous}, composites with superior mechanical properties \cite{mccue2016size, gaskey2019self}, and high-capacity battery materials with improved mechanical stability \cite{wada2014bulk, wada2017preparation}.

Dealloying was first employed as an electrochemical dealloying technique \cite{erlebacher2001evolution, erlebacher2004atomistic, erlebacher2009hard, chen2013spontaneous, wang2014magnesium, zhang2009generalized, li2010achieving} in which a less noble element is selectively dissolved from an alloy by an acid bath. This selective dissolution leads to the reorganization of the noble component into islands, eventually leading to a nano-porous structure. 
One limitation of this technique remains the requirement of a high chemical potential difference between the components of the precursor alloy, limiting its application to less noble metals \cite{mccue2018pattern, weissmuller2018dealloyed}. 
To overcome this limitation, liquid metal dealloying (LMD) was rediscovered \cite{harrison1959attack, wada2011dealloying} and allows to expand the range of dealloyable materials to less noble compounds such as Ti, Ta and Fe. 
Instead of using an acid solution to leach out the less noble element away, LMD relies on a liquid metal (e.g. Cu \cite{mccue2016size}, Bi \cite{wada2014bulk}, Mg \cite{wada2011dealloying, wada2013three, kim2015optimizing, joo2020beating}, Ge \cite{greenidge2020porous}, etc.) to selectively dissolve miscible elements, leaving immiscible elements to form a topologically connected structure \cite{geslin2015topology,  zhao2017three, mccue2018pattern, gaskey2019self}. 
As a fast developing technique, LMD experiments have been applied to fabricate various porous materials such as Ti \cite{wada2011dealloying}, Si \cite{wada2014bulk}, Nb \cite{kim2015optimizing}, FeCr \cite{wada2013three, mokhtari2018microstructure}, Ta \cite{mccue2016size}, graphite \cite{greenidge2020porous}, or even high-entropy alloys \cite{joo2020beating}. 
Extending the basic idea of selective dissolution, other dealloying methods have been developed: solid-state dealloying \cite{wada2016evolution, mccue2017alloy, zhao2019bi}, which uses a solid instead of a liquid melt to selectively dissolve the precursor alloy; and vapor phase dealloying where one of the components of the precursor alloy selectively evaporates \cite{lu2018three, Han2019vapor}.

Along with experimental studies, theoretical approaches have also been developed to understand the fundamental mechanisms of dealloying.
Electrochemical dealloying has been studied theoretically with kinetic Monte Carlo (KMC) simulations \cite{erlebacher2001evolution, zinchenko2013nanoporous, erlebacher2004atomistic, erlebacher2011mechanism}, bringing valuable insights to understand the fundamental mechanisms of the dealloying process and the following coarsening mechanisms. 
However, the KMC method is not suitable to study LMD because of its inadequacy to model a liquid phase and its length and time-scale limitations. 
Moreover, simulations of LMD with atomistic techniques often require the development of quantitative inter-atomic potentials parameterized to reproduce the thermodynamic properties of the ternary system, which remains a long and challenging task.
Continuous approaches such as phase-field models appear more suited to investigate the LMD process because they can simulate the free boundary problems associated with solid/liquid interfaces on diffusive time scales while easily incorporating thermodynamic properties of multicomponent alloys \cite{Chen1994, Bhattacharyya2003}. 
In particular, this class of phase-field model has been successfully applied to predict the microstructure development during solidification of binary and multi-components alloys  \cite{chen2002phase, boettinger2002phase, kobayashi2003phase, Chen2004}.

Recently, the phase-field method has also been applied to model the pattern formation of LMD \cite{geslin2015topology, mccue2018pattern} and the subsequent coarsening process \cite{geslin2019phase}. These phase-field studies along with experimental results revealed the fundamental processes of the pattern formation of LMD.
First, the planar dissolution is destabilized by interfacial spinodal decomposition, where the solid-liquid interface becomes corrugated due to the redistribution of the immiscible element along the interface.
Also, the dealloying kinetics is limited by diffusion of the miscible element away from the dealloying front, leaving the immiscible element to reorganize and form interconnected ligaments. 
Lastly, after its formation, the ligament coarsens by bulk and surface diffusion, leading to an increase of the microstructure length-scale.

Despite these seminal studies, a deeper understanding of some aspects of the LMD process requires further investigation. 
This paper focuses on three distinct but interconnected parts. 
We first focus on the 1D dissolution kinetics of the ternary alloy system, which serves as a theoretical framework for the initial stage of LMD. 
The previous studies of dealloying demonstrated a diffusion-limited kinetics $x_i(t) \sim \sqrt{D_l t}$, where $x_i$ is the position of solid-liquid interface and $D_l$ the diffusion constant in the liquid \cite{geslin2015topology, mccue2016kinetics}. 
However, this calculation only considered the diffusion of the miscible element as in a binary system, thereby discarding the ternary nature of the problem at hand.
Here, we propose a \nedit{1D} ternary dissolution model that incorporates the diffusion of both the immiscible and the miscible elements in the liquid and the evolution of the equilibrium compositions at the solid-liquid interface.

Second, we use the linear stability analysis of the interfacial spinodal decomposition\nedit{\cite{morral1971spinodal, de1972analysis}} to investigate the early stage of the morphological evolution in 2D simulations. Here we extend the work of Ref.~\cite{geslin2015topology} to track the development of the interfacial instability and predict its initial wavelength. 
Also, connecting the 1D ternary diffusion model with the linear stability analysis provides a criterion for the initial interface destabilization and the development of interconnected morphologies as a function of the precursor and melt compositions. 
\nedit{The predicted boundary between the planar dissolution regime and the development of connected morphologies is found to be in good agreement with 2D phase-field simulations. We also analyze how this boundary is modified when chemical equilibrium is established at the solid-liquid interface, as expected on long experimental time scales when solid-state diffusion is taken into account.}

The third part is dedicated to the role of solid-state diffusivity on the dealloying process, which was ignored in the previous phase-field studies \cite{geslin2015topology, mccue2016kinetics}, based on the fact that it is four to five orders of magnitudes smaller than liquid-state diffusivity \cite{dantzig2016solidification}.
Using 1D and 2D phase-field simulations with varying solid-state diffusivity, we show that this parameter affects significantly the composition profiles at the dealloying front. In particular, solid-state diffusivity allows for the development of wider concentration profiles in the solid, thereby reducing the influence of the gradient terms and promoting the convergence of the interfacial concentrations towards the prediction of the phase diagram.
Furthermore, we discuss the discrepancy between numerical and experimental results on the equilibrium concentrations of the dealloying front. We show that a more quantitative thermodynamic model enables to improve the comparison and that this equilibrium concentration evolves slowly with time, enabling us to connect numerical and experimental results.

This paper is structured as follows. In Section \ref{secpfmodel}, we first provide the phase-field model and the corresponding parameters used in the following simulations. We then present the 1D dissolution model for the ternary system in Section \ref{sec1Dana}. 
In Section \ref{stability_analysis}, we present a linear stability analysis of the initial spinodal decomposition and compare the theoretical results with 2D phase-field simulations.  We also use the criterion of spinodal decomposition to better understand the planar dissolution regime obtain for some melt compositions. 
Moreover, in Section \ref{solid_diff}, we propose a discussion of the effect of the finite solid-state diffusivity on dealloying kinetics and morphologies. 
Finally, conclusions are presented in the last section.

\section{Phase-field model for ternary alloy}
\label{secpfmodel}

We use a phase-field model for ternary alloys to simulate the dealloying process. 
Even though the dissolution process of a solid in a liquid is generally endothermic, its dynamics is controlled by solute diffusion, which is orders of magnitude slower than thermal diffusion in both phases. Hence, we consider the adiabatic limit and assume the temperature constant in the system.
This phase-field model relies on the coupling between concentration fields and an order parameter describing the order of the phase (liquid or solid). It naturally incorporates interactions between the different species as well as inter-diffusion mechanisms. 
We first introduce the order parameter $\phi(\mathbf{x})$ describing the crystalline order of the phase: $\phi(\mathbf{x})=0$ (respectively $\phi(\mathbf{x})=1$) if $\mathbf{x}$ is in the liquid (solid). The solid/liquid interface is described through a smooth variation of the field $\phi$. In order to describe the variation of composition between the different points of the system, we introduce the atomic concentration fields $c_1(\mathbf{x})$, $c_2(\mathbf{x})$ and $c_3(\mathbf{x})$ with the constraint $c_1(\mathbf{x})+c_2(\mathbf{x})+c_3(\mathbf{x})=1$ at any position $\mathbf{x}$.
The total free energy functional describing the state of the system is $\mathcal{F}=\int_V f (\phi,c_i) \mathrm{d}V $, where the energy density $f (\phi,c_i)$ is defined as
\begin{equation}
	f  (\phi,c_i)= \frac{\sigma_{\phi}}{2} |\nabla \phi|^2 +  f_{do}(\phi) + \sum_{i=1}^{3} \frac{\sigma_i}{2}|\nabla c_i|^2 + f_{ch}(\phi,c_i).
	\label{eq:free_energy}
\end{equation}
The first term is the gradient contribution of the phase field, which preserves a finite interface thickness.
The second term is a double-obstacle potential characterized by two minima located at $0$ and $1$:
\begin{align}
	f_{do}(\phi) &= +\infty 		     & \text{for } \phi<0	\nonumber	\\	
	f_{do}(\phi) &= \lambda \phi(1-\phi) & \text{for } 0 \leq \phi \leq 1	\\
	f_{do}(\phi) &= +\infty 			 & \text{for } \phi>1	\nonumber
	\label{eq:do_potential}
\end{align}
The parameters $\lambda$ and $\sigma_{\phi}$ can be chosen to obtain a solid-liquid interface of a specific width and energy. 
The double-obstacle potential presents the advantage of fixing the value of $\phi$ in the bulk phases to be exactly $1$ for the solid phase and $0$ for the liquid phase. In contrast, with a double-well potential, $\phi$ reaches $0$ in the liquid and $1$ in the solid only asymptotically. If we consider the diffusivity as a function of $\phi$ (as done below), the double obstacle potential then allows to control exactly the diffusivity in both solid and liquid phases.
The third term represents the gradient energy associated with spatial composition variations. For simplicity reasons, we assume $\sigma_1=\sigma_2=\sigma_3$. 
Finally, the last term represents the chemical contribution, i.e. the thermodynamic model of the free-energy density, which is defined as
\begin{eqnarray}
	f_{ch}  (\phi,c_i)= && \sum_{i=1}^{3} \left[ \phi c_i L_i \left(\frac{T-T_i}{T_i}\right)  + \frac{kT}{V_a} c_i\log(c_i)  \right]   \nonumber \\
	&&  + \sum_{i<j}^{i,j\leq3} \Omega_{ij} c_i c_j %
	\label{eq:free_energy_ch}
\end{eqnarray}

The first term couples the concentration field to the phase-field through the temperature $T$, the melting point of the compounds $T_i$ and the latent heat of pure material $L_i$. The coupling is assumed to be linear in temperature (which is true close to the melting point) and in concentration (which is true close to pure metals). 
The second term is the entropy term of each element. Let us notice that the atomic volume $V_a$ is assumed to be the same for all the elements and does not change between the solid and liquid phases. In other words, we neglect the dilatation or contraction due to the changes in phase and concentration. 
The last term represents the mixing enthalpy between the different species. The magnitude of the parameters $\Omega_{ij}$ controls the strength of partitioning for the different binary systems. For simplicity reasons, Eq.~\ref{eq:free_energy_ch} is chosen to rely on a small set of parameters while enabling to reproduce the main features of the dealloying process. However, we note that it is straightforward to parameterize the phase-field model with more complex free energies taken from thermodynamics databases \cite{lukas2007computational} as presented in the last part of this paper.

With the constraint $c_1+c_2+c_3=1$, we note that Eq.~(\ref{eq:free_energy}) can be written as a function of three degrees of freedom, which are the concentration fields $c_1$ and $c_2$ and the phase field $\phi$. 
The evolution equations of the order parameters can be derived from the variations of the total free energy. In particular, the concentration fields $c_1$ and $c_2$ are assumed to follow Cahn-Hilliard equations \cite{cahn1961spinodal}:
\begin{equation}
	\dot{c}_i = \nabla \cdot M_{ij} \nabla \mu_j
	\label{eq:diffusion}
\end{equation}
where $\mu_j=\delta \mathcal{F}/\delta c_j$ denotes the chemical potential of element $c_j$ and $M_{ij}$ the elements of the mobility matrix, symmetric because of the Onsager reciprocal relations. These components are expressed as $M_{ij}=M_0(\phi) c_i (\delta_{ij}-c_j)$ \cite{andersson1992models, nestler2005multicomponent}. This choice enables to reproduce a Fickian diffusion equation in the diluted limit. The parameter $M_0(\phi)$ depends on the order parameter to account for phase-dependent mobilities. It is considered as a linear function of $\phi$ chosen such that the mobility reaches $M_s$ in the solid and $M_l$ in the liquid:
\begin{equation}
	M_0(\phi) = \phi(x) (M_s-M_l) + M_l
\end{equation}
The values of $M_l$ and $M_s$ are chosen according to the diffusion coefficients in both phases: $M_l=D_l V_a/kT$ and $M_s=D_s V_a/kT$.
The kinetics equations are completed with a simple dissipative dynamics on the field $\phi$ \cite{cahn1977microscopic} :
\begin{equation}
	\dot{\phi} = -L_{\phi} \frac{\delta \mathcal{F}}{\delta \phi}
	\label{eqdtphi}
\end{equation}	

We consider the model system of a Ti-Ta precursor alloy immersed in liquid Cu, where Ti dissolves selectively in the Cu melt, resulting in an interconnected Ta structure \cite{geslin2015topology,mccue2016kinetics}. 
In the following, the indices $i=1,2,3$ represent respectively Cu, Ti, and Ta.
The parameters used in our simulations are listed in Table \ref{c3topar}. The parameters $\lambda$ and $\sigma_{\phi}$ are chosen to obtain an equilibrium profile of the phase-field with realistic width $w=\unit{2}{nm}$ and surface energy $\gamma=\unit{200}{mJ/m^2}$. \nedit{We note that the value of $\gamma$ is characteristic of the excess free-energy of the solid/liquid interface for pure metals. For interfaces between phases of different compositions as encountered in this work, the composition gradient terms of Eq.~(\ref{eq:free_energy}) also contribute to the total excess free-energy of the interface.}

\nedit{We would like to highlight that the interface width is considered here as a physical parameter. Indeed, the morphologies of the dealloyed microstructures depend on the diffusivity of the immiscible element within the interfacial layer between the solid and liquid phases. Enlarging the interface width would increase diffusive transport along the interface and therefore alter the resulting morphologies. Hence, the results presented in this paper are expected to depend on the specific value of $w$.}

\begin{figure}[htbp] 
	\centering
	\includegraphics[scale=0.45]{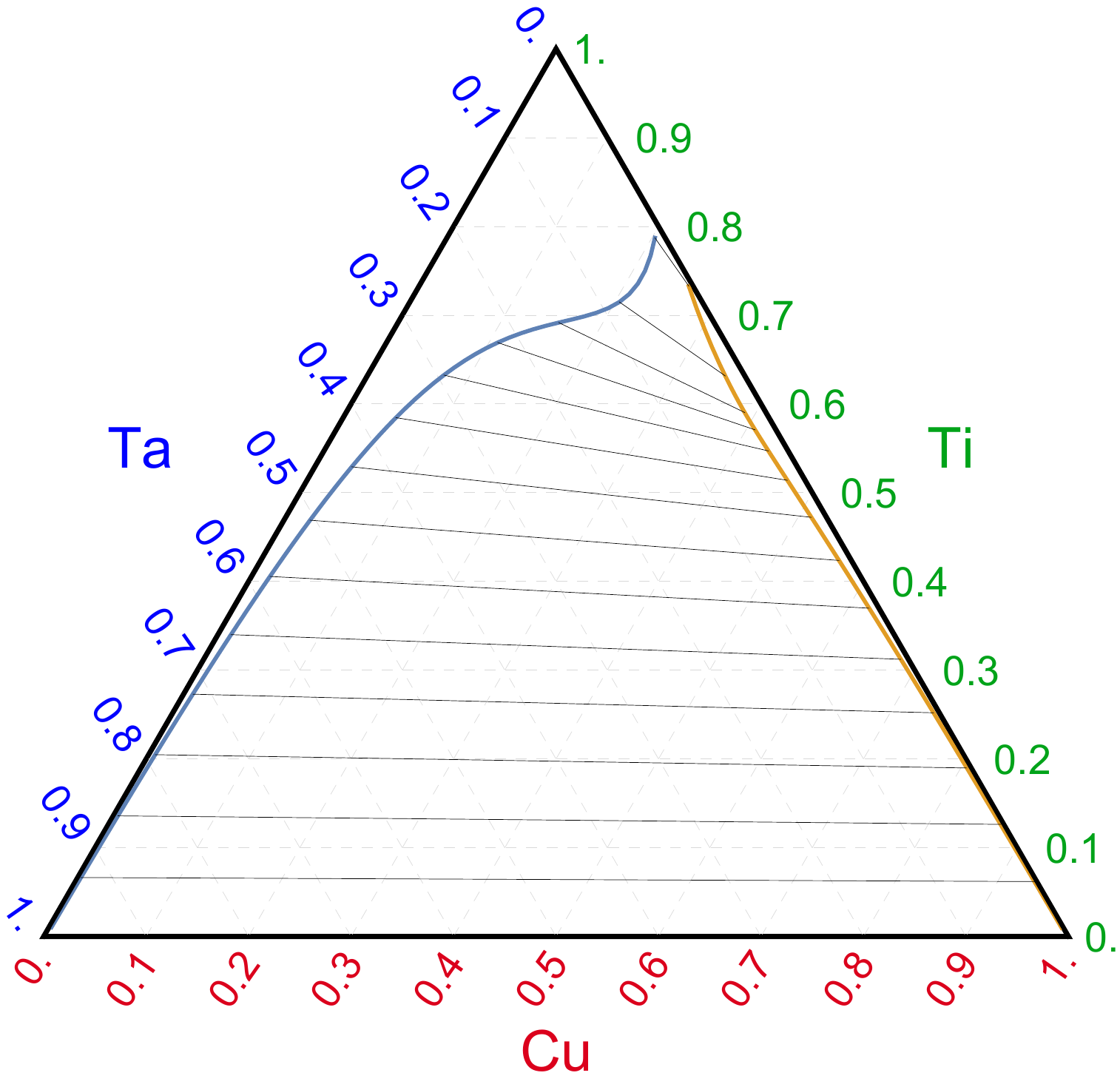}
	\caption[Cu-Ti-Ta ternary phase diagram.]{Cu-Ti-Ta ternary phase diagram obtained with the thermodynamic parameters listed in Table~\ref{c3topar}. It displays a large region of two-phase coexistence between the solidus (blue line) and the liquidus (orange line). Tie-lines are shown with black straight lines.}
	\label{figtCuTiTa}
\end{figure}

In the following, we set the temperature at $T=\unit{1775}{K}$, for which the equilibrium concentration of Ti in the Cu-Ti phase diagram is close to the experimental value (Fig.~\ref{figtCuTiTa}). 

For numerical purposes, the phase-field equations are normalized with the characteristic length-scale $l_c=w$ (the liquid-solid interface width) and the characteristic time scale $t_c=w^2/(M_l \lambda)$, where $\lambda$ defines the characteristic energy density.
We assume that the time-scale associated with the phase-change at the interface is much faster than the diffusive time scale \cite{dantzig2016solidification}, which translates into the dimension-less coefficient $\tilde{L}_{\phi} \gg 1$. In practice we take $\tilde{L}_{\phi}=10$.

\nedit{The phase-field equations are discretized in space and time and numerically integrated using an explicit Euler scheme. In this paper, the dimensionless space discretization is taken as $\mathrm{d}x=0.25$ for 2D and 3D simulations and $\mathrm{d}x=0.125$ for 1D simulations. The dimensionless time discretization is $\mathrm{d}t=10^{-5}$  in 1D, $\mathrm{d}t=1.5 \times 10^{-4}$ for 2D simulations, and $\mathrm{d}t=7\times 10^{-5}$ for 3D simulations.}

\begin{table}[]
\centering
\begin{tabular}{c|ccc}
\hline
$T$ (K)              & \multicolumn{3}{c}{1775}                           \\
$V_a$ (nm$^3$)             & \multicolumn{3}{c}{0.01}                           \\
$\sigma_\phi$ (eV/nm)      & \multicolumn{3}{c}{3.18}                           \\
$\lambda_\phi$ (eV/nm$^3$) & \multicolumn{3}{c}{1.59}                           \\
$L_\phi$ (nm$^3$/(eVs))      & \multicolumn{3}{c}{$1.14\times 10^9$}              \\ \hline \hline
                         	  & Cu              & Ti              & Ta              \\ \hline
$L_i$ (eV/nm$^3$)            & 11.5            & 11.8            & 17.6            \\
$T_i$ (K)                    & 1358            & 1941            & 3290            \\
$\sigma_i$ (eV/nm)           & 9.0             & 9.0             & 9.0             \\
$D_l$ (nm$^2$/s)           & $7 \times 10^9$ & $7 \times 10^9$ & $7 \times 10^9$ \\
$D_s$ (nm$^2$/s)           & 0               & 0               & 0               \\ \hline \hline
		            & Cu-Ti           & Ti-Ta           & Cu-Ta           \\ \hline
$\Omega_{ij}$ (eV/nm$^3$)       & 0               & 0               & 90              \\ \hline
\end{tabular}
\caption[Original parameters used in our simulations.]{Parameters used in the phase-field simulations in the following sections unless indicated otherwise.}
\label{c3topar}
\end{table}
The thermodynamics parameters $L_i$, $T_i$ and $\Omega_{ij}$ listed in Table \ref{c3topar} control the shape of the ternary phase diagram for our model Cu-Ti-Ta system.
The conditions for equilibrium between solid and liquid phases can be found by equating the chemical potentials of two species (the third component necessarily satisfies $c_1+c_2+c_3=1$) and equating the grand potential \cite{dantzig2016solidification} :

\begin{eqnarray}
&&\mu^s_1 (c^s_i)=\mu^l_1(c^l_i), \label{eqc3PhDEq1} \\
&&\mu^s_2  (c^s_i)=\mu^l_2(c^l_i) , \label{eqc3PhDEq2}\\
&&f^s (c^s_i)-c_1^s\mu^s_1  (c^s_i)-c_2^s\mu^s_2  (c^s_i)\nonumber \\
&&\quad \quad =f^l(c^l_i)-c_1^l\mu^l_1(c^l_i)-c_2^l\mu^l_2(c^l_i), \label{eqc3PhDEq3}
\end{eqnarray}

where $\mu_i^{s,l}(c^{s,l}_i)=\frac{\delta f^{s,l}(c_i)}{\delta c_i}|_{c_i=c_i^{s,l}}$ are the chemical potentials with $i=1,2$;
$f^s(c^s_i)=f_{ch}(1,c_i^s)$ and $f^l(c^l_i)=f_{ch}(0,c_i^l)$ are solid and liquid free energies taken from Eq.~(\ref{eq:free_energy_ch}).
There are four free parameters ($c^{s,l}_i$, $i=1,2$) and three equilibrium equations, such that multiple equilibrium conditions are possible.
Using the thermodynamic parameters considered here, these equilibrium conditions can be solved to express the equilibrium composition as function of a single degree of freedom (e.g. the Ta composition in the solid). Fig.~\ref{figtCuTiTa} represents the resulting phase-diagram obtained at the temperature of interest $T=1775$ K.
The tie-lines between the solidus and the liquidus represent possible equilibrium concentrations at the solid-liquid interface. 
Due to the large mixing enthalpy between Cu and Ta, the Ta solubility in the Cu melt is very small. However, there is no mixing enthalpy between Ta and Ti, so that the Ta solubility in the Cu-Ti melt increases with Ti concentration. 

\section{Theoretical analysis of 1D dissolution}
\label{1D}
\label{sec1Dana}
We start our investigation of the LMD process with a 1D implementation of the phase-field model. In 1D, the solid/liquid interface cannot destabilize through spinodal decomposition and remains necessarily planar. Its dynamics is then controlled only by the composition profiles in both liquid and solid phases. 
\nedit{The dissolution of ternary systems is a classic and challenging topic in materials science. Indeed, in contrast with binary systems, the interfacial equilibrium compositions are not uniquely defined for ternary systems. A dissolution model was proposed by Maugis et al. \cite{maugis1997multiple}: the authors assume a simple linear phase diagram and neglect the off-diagonal terms of the mobility matrix such that the diffusion is described by independent Fick equations. However, this latter assumption does not apply to the Cu-Ti-Ta system where the strong mixing enthalpy between Cu and Ta leads to non-negligible cross-interactions terms in the diffusion equations. It was therefore necessary to develop a more elaborate diffusion-limited dissolution model applicable to the Cu-Ti-Ta system, which considers both diagonal and off-diagonal terms of the mobility matrix.} 
This section is dedicated to the development of such model.

Phase-field simulations reveal that the dealloying kinetics follows two stages.
In the early stages, Ti dissolves in large quantities into the melt because it has a much larger solubility than Ta in Cu. As a consequence, Ta accumulates at the solid-liquid interface, slowing down the dissolution of Ti. During this stage, the dissolution rate decreases progressively, until Ta dissolves at the same rate as Ti in the melt resulting in a steady-state dissolution. These two stages are discussed separately in the following.

\subsection{Passivation}
\label{passivation}
For a ternary system, the movement of the solid/liquid \nedit{interface} is controlled by the diffusion of both Ta and Ti away from the interface. In contrast with binary systems, the interface concentrations on the solid and liquid side are not unique and can follow an infinite number of equilibrium conditions represented by tie-lines in Fig.~\ref{figtCuTiTa}.

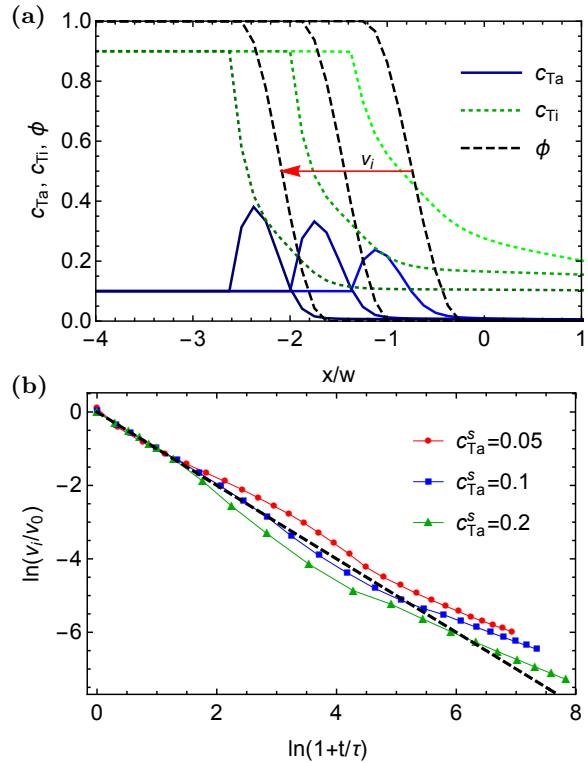
\begin{figure}[h]
	\begin{center}
		\includegraphics[width=0.87\linewidth]{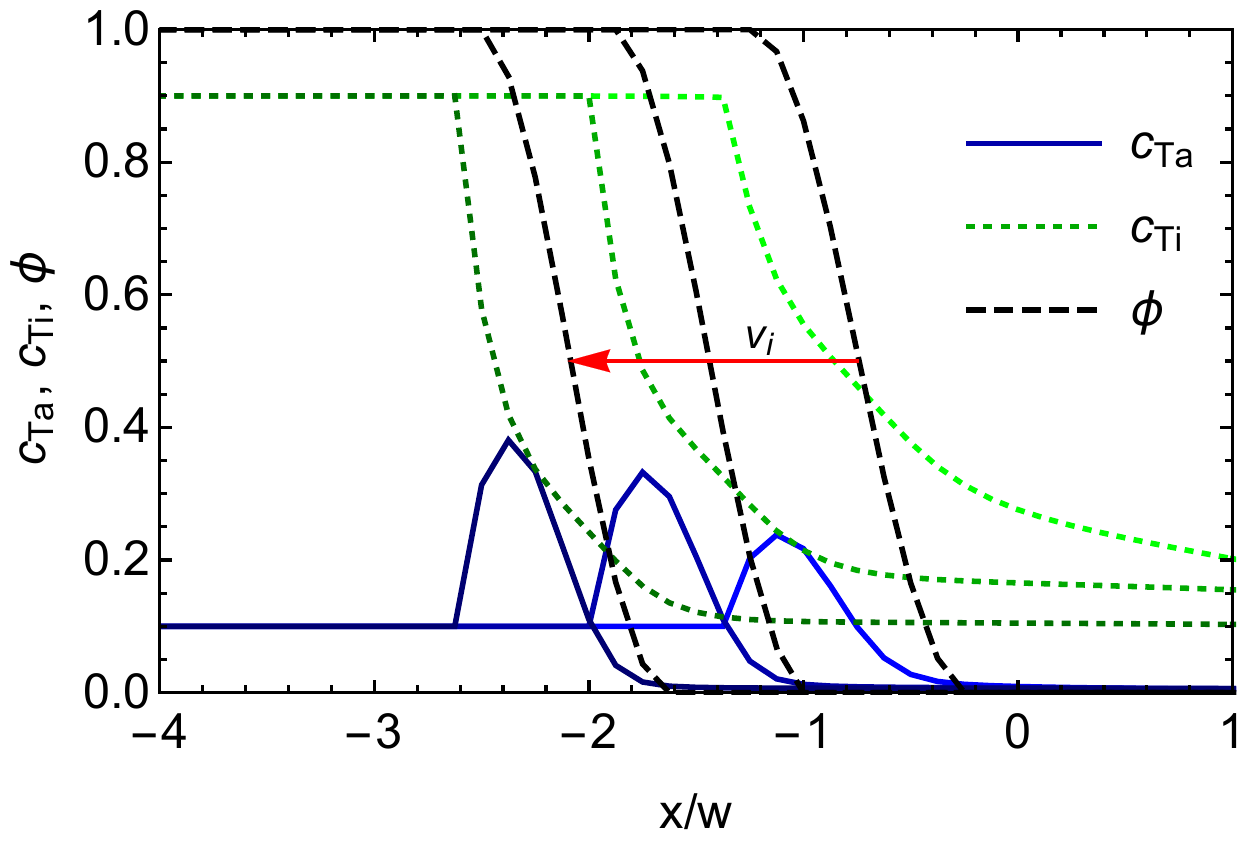}
		\hspace{0.06\linewidth}
		\includegraphics[width=0.87\linewidth]{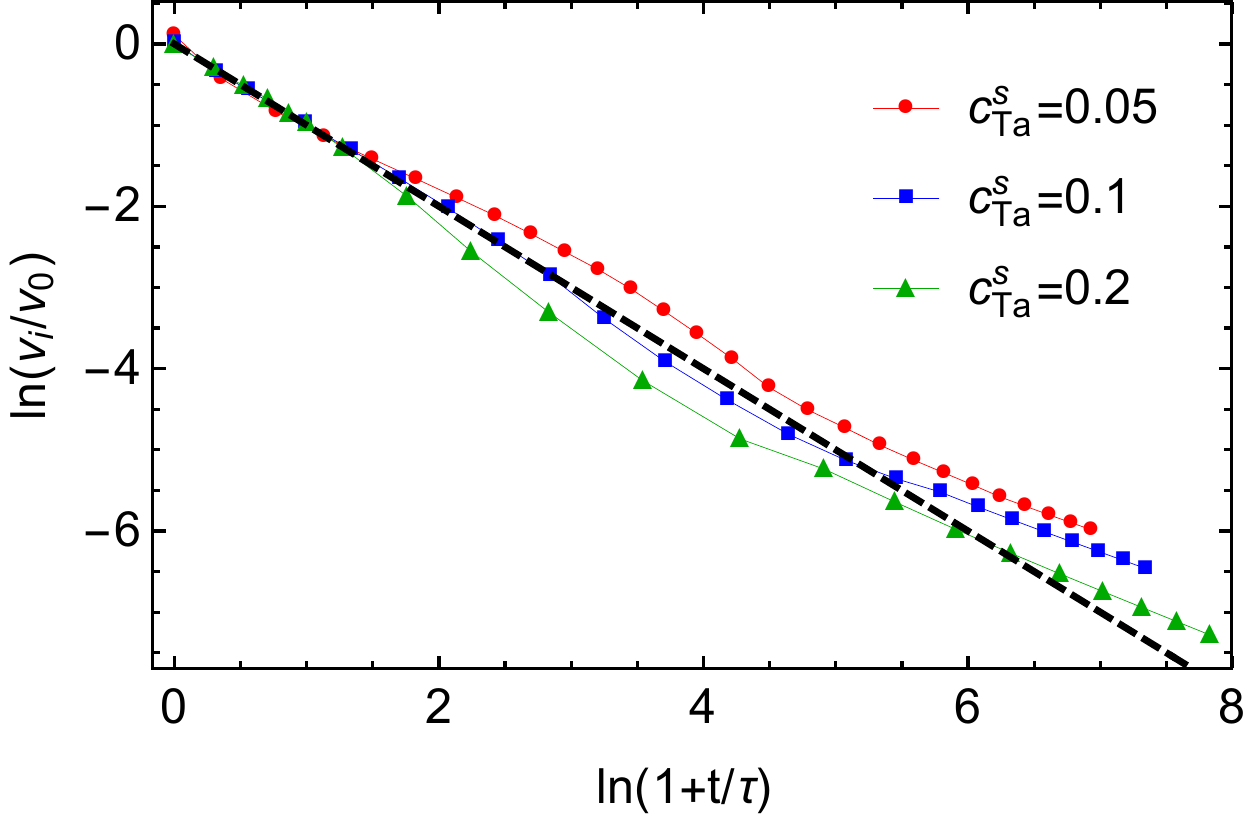}
		\begin{picture}(1,0)(0,0)
			\put(-220.,280.) {\text{{\textbf{(a)}}}}
			\put(-220.,140.) {\text{{\textbf{(b)}}}}
		\end{picture}
		\caption{(a) Evolution of the composition profiles of Ti and Ta during a passivation simulation of Ta$_{10}$Ti$_{90}$ dealloyed in the pure Cu melt. (b) Log-log plot of the interface velocity against time for three compositions of the precursor. The black dashed line is a guide to the eye of slope $-1$. }
		\label{fig:passivation_1D}
	\end{center}
\end{figure}

In the first stage, Ti dissolves much faster than Ta due to its larger solubility in the Cu melt and Ta therefore accumulates at the solid-liquid interface. This is demonstrated by the phase-field profiles reported in Fig.~\ref{fig:passivation_1D}.a representing the Ta (blue) and Ti (green) profiles at three simulation times together with the phase-field shown with a dash line. According to the phase diagram of the ternary system (Fig.~\ref{figtCuTiTa}), the accumulation of Ta on the solid side of the interface leads necessarily to the reduction of the interfacial concentration of Ti on the liquid side. As exemplified in Fig.~\ref{fig:passivation_1D}.a, this will reduce the flux of Ti leaving the interface, and slow down the dissolution. If we assume that Ta is completely immiscible with the Cu melt, the interface will become saturated in Ta and the dealloying process will eventually stop.

Let us note $c_p=c_3^s$ the concentration of Ta in the interface. As we assume that no Ta escapes in both liquid and solid phases, the height of the Ta peak in the interface can be easily related to the position of the interface:
\begin{equation}
	c_p(t) = \frac{c_s x_i(t)}{\xi}
	\label{eq:passivation_c_t}
\end{equation}
where $\xi$ is a length-scale related to the interface width and $x_i(t)$ denotes the position of the interface ($x_i(0)=0$) and $c_s=c_{30}^s$ is the Ta composition of the precursor. Following previous work \cite{geslin2015topology}, we assumed that the interface velocity is only controlled by the interfacial Ta content and decreases exponentially with $c_p$:
\begin{equation}
	v_i(c_p)=v_0 \exp \left( -c_p/c^* \right)
	\label{eq:passivation_v_c}
\end{equation}
where $v_0$ is the dissolution velocity for $c_i=0$ and $c^*$ is a characteristic concentration with $c^{*} \simeq 0.045$ \cite{geslin2015topology}. 
Injecting Eq.~\ref{eq:passivation_c_t} in Eq.~\ref{eq:passivation_v_c} and integrating the time, we find that the velocity should follow a $1/t$ behavior:
\begin{equation}
	v_i(t)=\frac{v_0}{1+t/\tau}
	\label{eq:passivation_v_t}
\end{equation}
where $\tau= \xi c^* /c_s v_0 $ is a characteristic time for passivation. 

Fig.~\ref{fig:passivation_1D}b presents the log-log relation of the interface velocity against time obtained from 1D phase-field simulations with different initial Ta solid concentrations. 
On short time scales (from $t=0$ to $t\simeq 20\tau$), the interface velocity follows a straight line of slope $-1$, demonstrating the $\sim 1/t$ dynamics. 
For three different values of Ta precursor composition, we have performed a fit of our data against Eq.~(\ref{eq:passivation_v_t}) to deduce $v_0$ and $\tau$. These fitting procedures are consistent and we find $v_0 \simeq 9.0 (w/t_c)$ and that the product $\xi c^*\simeq 0.05w$ varies marginally with the initial solid concentration of Ta. Despite the simplicity of the model, it reproduces accurately the early stage of the dissolution kinetics.

At longer times, the velocity keeps dropping but the slope changes \nedit{(see Fig.~\ref{fig:passivation_1D})}. This is attributed to the flux of Ta in the liquid. Indeed, the enthalpy of mixing between Cu and Ta is finite and allows \nedit{a small amount of} Ta to dissolve slowly into the liquid. 
This leakage of Ta is negligible on short time scales but affects the interface kinetics on longer time-scales.

\subsection{Self-similarity solution of dissolution}
\label{dealloying_kinetics_ternary}
\begin{figure}[htbp] 
	\centering
	\includegraphics[scale=0.73]{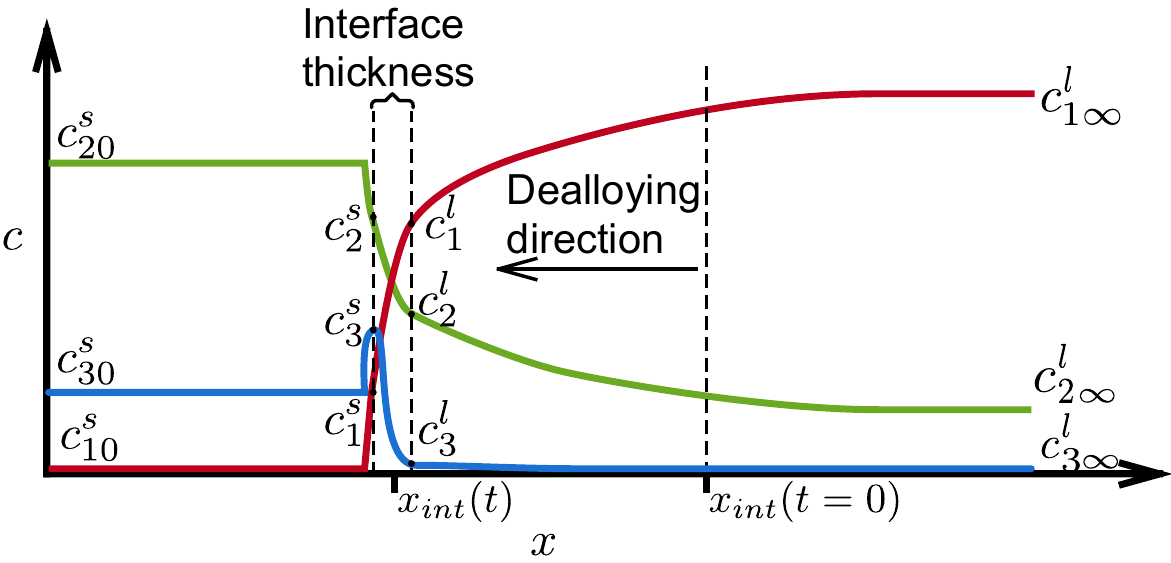}
	\caption[Schematic representation of the 1D concentration profiles during LMD.]{Schematic representation of the 1D concentration profiles during LMD. The initial composition of the precursor alloy is labeled $c^s_{i0}$ and the initial composition of the melt is labeled $c^l_{i\infty}$;  $c_i^s$ and $c_i^l$ denote the interfacial concentrations.}
	\label{figt1d}
\end{figure}

After the initial stage described above, the interface kinetics becomes function of the dissolution of both Ta and Ti and depends on the chemical equilibrium at the solid/liquid interface that is not uniquely defined.
The initial conditions of the problem include the concentrations of the base alloy ($c_{i0}^s$) and the melt ($c^l_{i\infty}$), but the concentrations near the solid-liquid interface ($c^s_i$ and $c^l_i$) remain unknown (see Fig.~\ref{figt1d}). The position of the solid-liquid interface is denoted $x_{int}(t)$, and we consider $x_{int}(t=0)=0$. \nedit{With the present definition, $x_{int}<0$ and equal in magnitude to the dealloying depth $x_i(t)=|x_{int}(t)|$.} As shown in Fig.~\ref{figt1d}, the dealloying front moves towards $x<0$ with a negative velocity $v_{int}<0$.  
\nedit{Analytical solutions for $x_{int}(t)$ and the concentration fields, which depend on both space and time, can be obtained by exploiting the self-similar nature of the of the dissolution kinetics. Namely, the concentration fields only depend on the scaled variable $x/x_{int}(t)$ where $x_{int}(t)\sim -\sqrt{D_lt}$, thereby enabling to map the time-dependent free-boundary problem of dissolution to a stationary problem as described in what follows.}

The phase-field model provides the evolution equations of the concentration fields (see Eq.~(\ref{eq:diffusion})). To simplify the calculation, we neglect the gradient terms on the composition profiles and assume that the mobilities of all components are the same ($M_l=M_{li}$ for $i=1,2,3$ and $M_s=0$). Therefore, the mobility matrix $M_{ij}$ is,

\begin{align}
    M_{ij} &= M_l c_i (\delta_{ij} - c_j) \quad \quad \mbox{in the liquid}\\
    M_{ij} &= 0 \quad \quad \quad \quad \quad \quad \quad \; \; \mbox{in the solid}
\end{align}

The chemical potential are derived from Eq.~(\ref{eq:free_energy}) with $\mu_j=\delta \mathcal{F}/\delta c_j$.\nedit{The concentration fields in the bulk solid are constant because of the zero solid state diffusivity, and the evolution equations in the liquid phase} become in 1D:
\begin{eqnarray}%
&& \partial_t c_1=\partial_x \left[ M_l c_1(1-c_1) \partial_x \mu_1 -M_l c_1 c_2 \partial_x \mu_2   \right], \label{eqevc1} \\
&& \partial_t c_2=\partial_x \left[ M_l c_2(1-c_2) \partial_x \mu_2 -M_l c_1 c_2 \partial_x \mu_1   \right],\label{eqevc2}
\end{eqnarray}
where the derivatives of the chemical potentials are expressed as function of composition gradients:
 \begin{align}
 \partial_x \mu_1  = &\frac{k_B T}{V_a} \left( \frac{\partial_x c_1}{c_1}+\frac{\partial_x c_1+\partial_x c_2}{1-c_1-c_2}   \right) \nonumber \\
&- \Omega_{13}\left( 2\partial_x c_1+\partial_x c_2 \right), \label{eqdmu1} \\
&\color{blue}{+ \partial_x \phi \left( L_1 \left( \frac{T-T_1}{T_1} \right) - L_3 \left( \frac{T-T_3}{T_3} \right) \right)} \nonumber \\
 \partial_x \mu_2 = &\frac{k_B T}{V_a} \left( \frac{\partial_x c_2}{c_2}+\frac{\partial_x c_1+\partial_x c_2}{1-c_1-c_2}   \right) - \Omega_{13} \partial_x c_1 \label{eqdmu2} \\
                    &\color{blue}{+\partial_x \phi \left( L_2 \left( \frac{T-T_2}{T_2} \right) - L_3 \left( \frac{T-T_3}{T_3} \right) \right)}, \nonumber
\end{align}
as we consider that $\Omega_{12}=\Omega_{23}=0$, and $\Omega_{13}$ is the only non-zero mixing enthalpy.

\nedit{We introduce a moving reference frame $x'(t)=x-x_{int}(t)$, such that $x'=0$ at the solid-liquid interface. The time dependent concentration fields can be written as
\begin{equation}
\frac{\partial c_i(x,t)}{\partial t}=\frac{\partial c_i(x'(t),t)}{\partial t}+\frac{\partial c_i(x'(t),t)}{\partial x'}\frac{\partial x'(t)}{\partial t}
\end{equation}
At the interface, we assume $c_i(x'(t),t)$ constant, so we have the relation $\partial_t c_1=-v_{int}\frac{\partial c_1}{\partial x}$ using the fact that $\partial_t x'(t)=-\frac{\mathrm{d}x_{int}}{\mathrm{d} t} = -v_{int}$. To obtain the mass conservation, we integrate Eq.~(\ref{eq:diffusion}) for $c_1$ over the solid-liquid interface, which yields
\begin{eqnarray}
	\int^\delta_{-\delta} && -v_{int} \frac{\partial c_1}{\partial x} \mathrm{d}x = v_{int} (c^s_{10}-c^l_1)  \nonumber \\
	&& = M_l c^l_1(1-c^l_1) \partial_x \mu^l_1-M_l c^l_1 c^l_2 \partial_x \mu^l_2 .
\label{eqddevc1}
\end{eqnarray}
}
Taking $\partial_x \mu^l_1$ and $\partial_x \mu_2^l$ from Eq.~(\ref{eqdmu1}) and Eq.~(\ref{eqdmu2}), we obtain:
\begin{eqnarray}
v_{int} (c^s_{10}-c^l_1) =&& \frac{M_l k_B T}{V_a}\partial_{x} c^l_1 - M_l \Omega_{13} c_1^l (2-2c_1^l-c_2^l)\partial_{x} c^l_1 \nonumber \\
&& - M_l \Omega_{13} c_1^l (1-c^l_1)\partial_{x} c^l_2,
\end{eqnarray}
where $\partial_{x} c^l_i$ denotes the concentration gradient on the liquid side taken at the solid-liquid interface. \nedit{We note that $\partial_x \phi=0$ on the solid and liquid sides of the interface, such that the terms of the chemical potential involving the latent heat of fusion cancel out.}
During the dealloying process, the concentration of the immiscible element is very small in the liquid ($c_3^l\ll 1$), and we can consider that $c_1^l+c_2^l \approx 1$. Therefore we can simplify the equation to obtain the boundary condition of $c_1^l$,
\begin{equation}
v_{int} (c^s_{10}-c^l_1) =  \frac{M_l k_B T}{V_a}\partial_{x} c^l_1 -  M_l \Omega_{13} c^l_1 c^l_2 (\partial_{x} c^l_1+ \partial_{x} c^l_2).\label{eqdbv1}
\end{equation}
The boundary conditions of $c_2^l$ and $c_3^l$ can be derived with the same procedures, yielding
\begin{flalign}
v_{int} (c^s_{20}-c^l_2)& =  \frac{M_l k_B T}{V_a}\partial_{x} c^l_2 +  M_l \Omega_{13} c^l_1 c^l_2 (\partial_{x} c^l_1+ \partial_{x} c^l_2),&\label{eqdbv2} \\
v_{int} (c^s_{30}-c^l_3)& =  \frac{M_l k_B T}{V_a}\partial_{x} c^l_3 .&\label{eqdbv3}
\end{flalign}
Eqs.~(\ref{eqdbv1}-\ref{eqdbv3}) provide boundary conditions for the PDEs Eq.~(\ref{eqevc1}-\ref{eqevc2}). To solve the PDEs, we first rewrite the equations by substituting the chemical potentials with Eq.~(\ref{eqdmu1}) and Eq.~(\ref{eqdmu2}):
 \begin{eqnarray}
 \partial_t c_1=D_l \partial_{xx} c_1 - M_l&& \Omega_{13} \big[ (1-3 c_1) \partial_x c_1 (\partial_x c_1+ \partial_x c_2)\nonumber \\
&&+c_1(1-c_1) (\partial_{xx} c_1 +\partial_{xx} c_2)  \big], \label{eq:kins2}\\
 \partial_t c_2= D_l \partial_{xx} c_2 + M_l&& \Omega_{13} \big[ (2c_2 \partial_x c_1 + c_1 \partial_x c_2 ) (\partial_x c_1+ \partial_x c_2)  \nonumber \\
 &&+ c_1 c_2 (\partial_{xx} c_1 +\partial_{xx} c_2) \big] \label{eq:kins3}.
\end{eqnarray}
These PDEs offer a compact expression of the ternary diffusion problem but are too complex to solve analytically. 

\begin{figure}[htbp] 
	\begin{center}
	\includegraphics[scale=0.6]{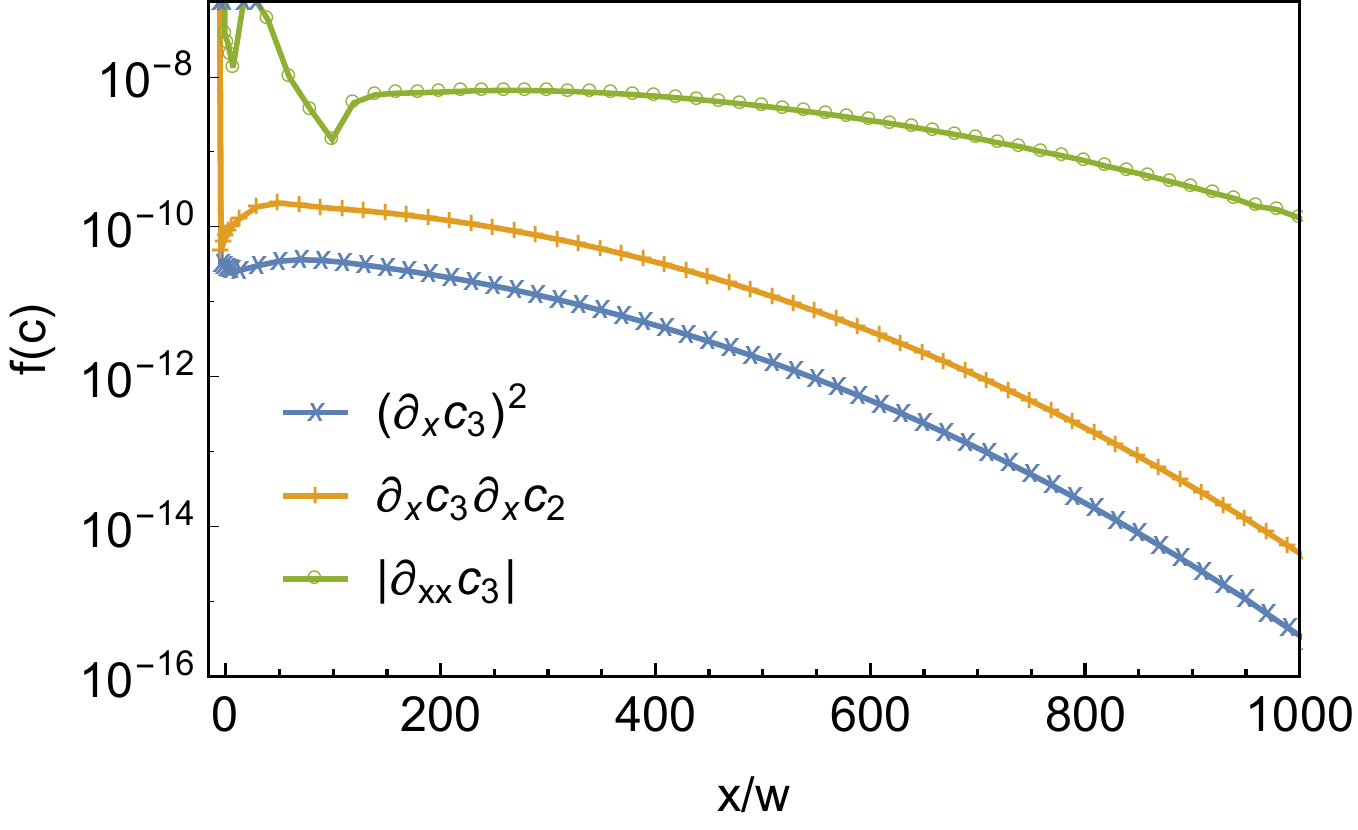}
	\caption[Comparison between the partial derivatives of the concentration profiles from a phase field simulation.]{Comparison between the partial derivatives of the concentration profiles obtained from a phase-field simulation of Ta$_{15}$Ti$_{85}$ dealloyed in the pure Cu melt.} 
	\label{figc3KinctP}
	\end{center}
\end{figure}

To further simplify the PDEs, it is necessary to neglect some nonlinear terms. To choose which terms can be discarded, we examine the concentration profiles obtained from phase-field simulations and compute the various gradient terms involved in Eqs.~(\ref{eq:kins2}-\ref{eq:kins3}) (Fig.~\ref{figc3KinctP} shows an example). We realize that $|\partial_{xx}c_{i}|\gg \partial_x c_{j}\partial_x c_{k}$ far away from the interface. 
Therefore, we choose to ignore all terms involving products of first derivatives (i.e. of the form $\partial_x c_{i}\partial_x c_{j}$). From the constraint $c_1+c_2+c_3=1$ and the fact that $c_3 \ll 1$ in the liquid phase, we have the approximation $c_1+c_2 \approx 1$ and the relation $\mathrm{d} c_1+\mathrm{d} c_2 =0$.  Based on these approximations, the evolution equations become
\begin{eqnarray}
&& \partial_t c_1=D_l \partial_{xx} c_1 + M_l \Omega_{13} c_1(1-c_1) \partial_{xx} c_3  ,\\
&& \partial_t c_2= D_l \partial_{xx} c_2 - M_l \Omega_{13}  c_2(1- c_2)\partial_{xx} c_3 , \label{eqdfc2}\\
&& \partial_t c_3= D_l \partial_{xx} c_3. \label{eqdfc3}
\end{eqnarray}
If there is no mixing enthalpy ($\Omega_{13}=0$), all components follow independent Fickian diffusion equations. If $\Omega_{13} \neq 0$, the evolution of $c_1$ and $c_2$ are coupled, while the evolution of $c_3$ remains Fickian. In the following calculations, we will use $c_2$ and $c_3$ as independent variables and assume the relation $c_1+c_2+c_3=1$ to hold.

To solve the time-dependent diffusion problem, we \nedit{seek a self-similar solution by introducing} a new coordinate $z(x,t)=x/x_{int}(t)$, such that the concentration fields $c_i(x,t)$ only depend on $z$. The transformations from $c_i(x,t)$ to $c(z)$ are given by:
\begin{eqnarray}
&& \frac{\partial c_i}{\partial t}=-\frac{zv_{int}}{x_{int}}  \frac{\partial c_i}{\partial z} ,\nonumber  \\
&& \frac{\partial c_i}{\partial x}= \frac{1}{x_{int}} \frac{\partial c_i}{\partial z},\label{eqdtrans} \\
&& \frac{\partial^2 c_i}{\partial x^2}= \frac{1}{x^2_{int}} \frac{\partial^2 c_i}{\partial z^2}.  \nonumber
\end{eqnarray}
The boundary conditions at $x=x_{int}$ can be rewritten as
\begin{eqnarray}
\partial_z c_2 \big|_{z=1} =&& \frac{x_{int}v_{int}}{D_l} \bigg[ (c^s_{20}-c^l_2)\nonumber \\
&& + \frac{1}{D_l} M_l \Omega_{13} c^l_2 (1-c^l_2) (c^s_{30}-c^l_3) \bigg],   \label{eqdfb2}\\
\partial_z c_3 \big|_{z=1} = &&\frac{x_{int}v_{int}}{D_l} (c^s_{30}-c^l_3). \label{eqdfb3}
\end{eqnarray}
If we define the constants 
\begin{flalign}
	B_2&=(c^s_{20}-c^l_2) + \frac{ M_l \Omega_{13}}{D_l} c^l_2 (1-c^l_2) (c^s_{30}-c^l_3) & \\
	B_3&= c^s_{30}-c^l_3, &&
\end{flalign}
and introduce a dimensionless Peclet number $p=x_{int}v_{int}/2D_l$, the boundary condition can be written as $\partial_z c_2 \big|_{z=1} =2pB_2$ and $\partial_z c_3 \big|_{z=1} =2pB_3$. The evolution equations (\ref{eqdfc2}) and (\ref{eqdfc3}) become:
\begin{eqnarray}
	&& 2pz\partial_z c_2 + \partial_{zz} c_2 -\frac{M_l \Omega_{13} c_2 (1-c_2) }{D_l}  \partial_{zz} c_3 =0 , 	\label{eqdffc2}\\
&& 2pz \partial_z c_3 + \partial_{zz} c_3 =0 . \label{eqdffc3}
\end{eqnarray}
We focus first on Eq.~(\ref{eqdffc3}) that only contains first and second order derivatives and can be solved with the boundary condition (Eq.~\ref{eqdfb3}):
\begin{equation}
	\partial_z c_3 = 2 p B_3 \mathrm{e}^p\mathrm{e}^{-pz^2}, \label{eqdsolc3z}
\end{equation}
which can then be substituted into Eq.~(\ref{eqdffc2}). To solve Eq.~(\ref{eqdffc2}), we consider that the coefficient
\begin{equation} 
	K=\frac{M_l \Omega_{13}}{D_l} c^l_2 (1-c^l_2)
\end{equation}
 is constant, leading to:
\begin{equation}
	\partial_z c_2 = \big[2pB_2 + 2 p^2 K B_3 (1- z^2) \big] \mathrm{e}^p \mathrm{e}^{-pz^2}. \label{eqdsolc2z}
\end{equation}
Finally, we integrate Eq.~(\ref{eqdsolc3z}) with the boundary conditions $c_3( z=1 )= c^l_3$ and $c_3(z=-\infty) = c^l_{3\infty}$ to obtain the concentration profile of $c_3$,
\begin{equation}
	c_3(z)= B_3 \mathrm{e}^p \sqrt{\pi p} (\mathrm{erf} (\sqrt{p} z) +1 ) +c^l_{3\infty} ,
\end{equation}
and integrate Eq.~(\ref{eqdsolc2z}) with boundary conditions $c_2 (z=1)=c^l_2$ and $c_{2}( z= -\infty) =c^l_{2\infty}$ to obtain the concentration profile of $c_2$,
\begin{eqnarray}
	c_2(z)= &&\big[ B_2+KB_3(p-\frac{1}{2}) \big] \mathrm{e}^p \sqrt{\pi p} (\mathrm{erf} (\sqrt{p} z) +1 ) \nonumber \\
	&&+ K B_3 pz \mathrm{e}^p  \mathrm{e}^{-pz^2}+ c^l_{2\infty}.
\end{eqnarray}
From the definition of the Peclet number, we can calculate the interface position $x_{int}(t)= -\sqrt{4pD_lt}$. Therefore the coordinate $z(x,t)$ in the concentration profiles can be substituted by $z(x,t)=-x/\sqrt{4pD_lt}$ and the Peclet number and the concentration $c_2$ and $c_3$ at the interface must satisfy the constraints:
\begin{flalign}
c^l_2 -c^l_{2\infty}&= K B_3 p&\nonumber \\
&+\big[ B_2 +K B_3 (p- \frac{1}{2}) \big] \mathrm{e}^p \sqrt{\pi p} (\mathrm{erf} (\sqrt{p} ) +1 ),  &\label{c3eqkinc1}\\
c^l_3 - c^l_{3\infty}&=B_3 \mathrm{e}^p \sqrt{\pi p} (\mathrm{erf} (\sqrt{p} ) +1 )  .&\label{c3eqkinc2}
\end{flalign}

The analysis detailed above provides a theoretical prediction of the concentration profiles in 1D. Eqs.~(\ref{c3eqkinc1}-\ref{c3eqkinc2}) represent two constraints of the diffusion problem but involve three unknown ($c_2^l$, $c_3^l$, and $p$). \nedit{A way to overcome this limitation is to consider the Peclet number obtained from the phase-field simulation starting from the same initial conditions}; then, the interfacial compositions $c_2^l$, $c_3^l$ can be obtained from Eqs.~(\ref{c3eqkinc1}-\ref{c3eqkinc2}), fully determining the dissolution kinetics and the diffusion profiles.

\begin{figure}[htbp] 
	\centering
	\includegraphics[scale=0.6]{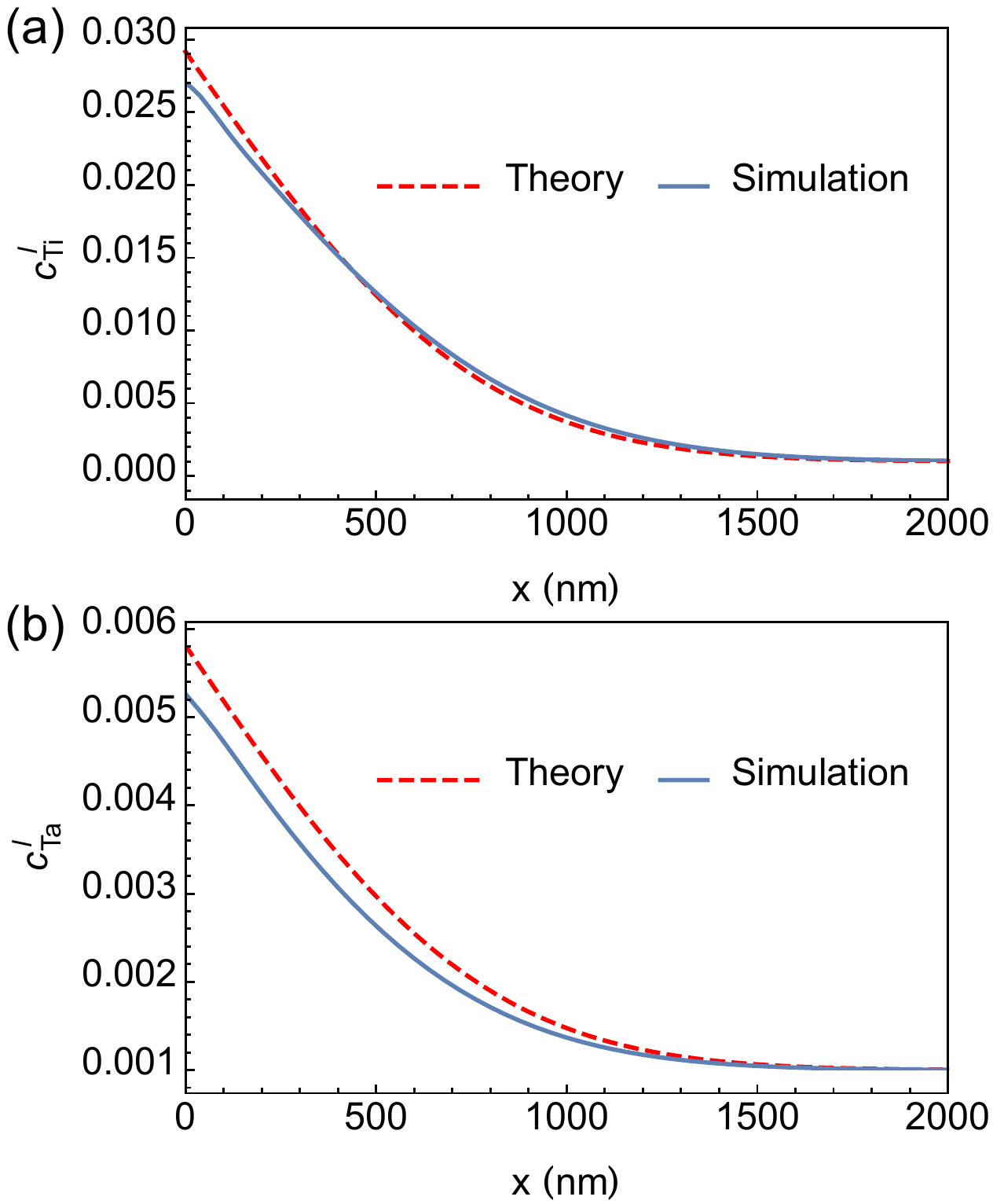}
	\caption[Concentration profiles of the 1D liquid metal dealloying results from phase field simulations and analytical calculations.]{Concentration profiles obtained from the phase field simulation (blue line) and analytical calculations (red dashed line) with the initial conditions $c^s_{20}=0.849$, $c^s_{30}=0.149$, $c^l_{2\infty}=0.001$, and $c^l_{3\infty}=0.001$. The snapshots are taken at $t=27.46 \; \mu s$ and the composition profiles follow a Fickian diffusion.}
	\label{figc3KinctC1dlTi}
\end{figure}

\begin{figure}[htbp] 
	\centering
	\includegraphics[scale=0.6]{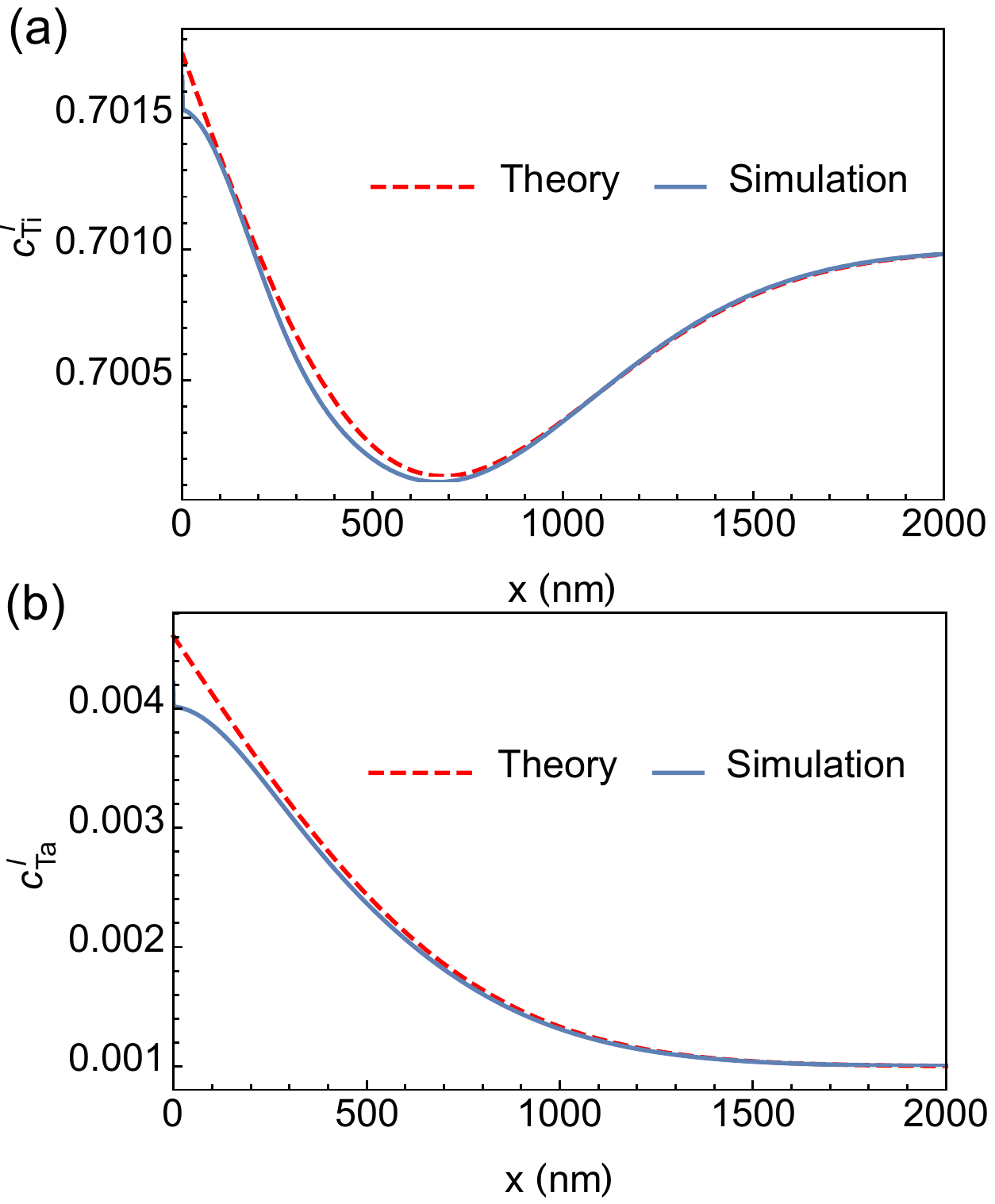}
	\caption[Concentration profiles of the 1D liquid metal dealloying results from phase field simulations and analytical calculations.]{Concentration profiles obtained from the phase field simulation (blue line) and analytical calculations (red dashed line) in the case of a non-Fickian diffusion. The initial conditions are $c^s_{20}=0.499$, $c^s_{30}=0.499$, $c^l_{2\infty}=0.701$, and $c^l_{3\infty}=0.001$. The snapshots are taken at $t=24.03 \; \mu s$.}
	\label{figc3KinctC1dhTi}
\end{figure}

Fig.~\ref{figc3KinctC1dlTi} and Fig.~\ref{figc3KinctC1dhTi} compare composition profiles thus obtained from the theoretical analysis with phase-field results. Fig.~\ref{figc3KinctC1dlTi} displays results obtained for a Ta$_{15}$Ti$_{85}$ precursor dealloyed in pure Cu melt; the corresponding Peclet number obtained from the phase-field simulation is $p=3.6\times 10^{-4}$. Because both the concentrations of Ti and Ta in the solid ($c^s_{20}$ and $c^s_{30}$) are larger than in the liquid ($c^l_{2\infty}$ and $c^l_{3\infty}$), Ti and Ta diffuse away from the solid-liquid interface. This diffusion is Fickian because it occurs from high to low concentrations regions. \nedit{Table~\ref{tc1ddis} compares the interface compositions obtained from the phase-field simulations and from the diffusion model (see ``Dissolution model 1" row ).}

In Fig.~\ref{figc3KinctC1dhTi}, we report an other example, which represents a case of non-Fickian diffusion. In this simulation, a Ta$_{50}$Ti$_{50}$ precursor dissolves into a Cu$_{30}$Ti$_{70}$ melt. The Peclet number of the dissolution kinetics obtained from the phase-field simulation is $p=1.69 \times 10^{-5}$. Since the concentration of Ti in the solid is smaller than in the liquid, Ti should diffuse from the liquid pool to the interface. However, on the liquid side of the interface, the Ti concentration reaches a value larger than $c^l_{2\infty}$. This promotes the formation of a concave Ti profile as shown in Fig.~\ref{figc3KinctC1dhTi}.a, characterizing a non-Fickian diffusion profile. In both cases, the theoretical approach successfully predicts the concentration profiles obtained during dissolutions in both Fickian and non-Fickian cases, which validates the ternary diffusion model developed above.

\subsection{\nedit{Self-similarity solution of dissolution with phase equilibrium conditions}}
\label{dis_phaseEq}

\begin{table}[]
\centering
\nedit{
\begin{tabular}{l |c|C{1.1cm}|C{1.1cm}}
\hline
             	 & $p$       			& $c^l_2$  	    & $c^l_3$ 	\\   \hline
{\small Phase-field simulation}  &  $3.6\times10^{-4}$   	& 0.0271		& 0.00533	\\  
{\small Dissolution model 1}  	&  $3.6\times10^{-4}$   	& 0.0298		& 0.00592 	\\   
{\small Dissolution model 2} 	&  $4.2\times10^{-5}$   	& 0.0104		& 0.00216 	\\   \hline
\end{tabular}
}
\caption[Comparison of the predictions.]{\nedit{Comparison of Peclet number and interfacial compositions obtained for the dissolution of a Ta$_{15}$Ti$_{85}$ precursor alloy in a pure Cu melt. Dissolution model 1 indicates the data obtained from the dissolution model when the Peclet number is taken from the phase-field simulation. Dissolution model 2 indicates the data obtained from the dissolution model coupled with phase equilibrium conditions (see section \ref{dis_phaseEq}).}}
\label{tc1ddis}
\end{table}

Instead of using phase-field simulations to identify the Peclet number of the dissolution, another strategy consists in combining the conditions~(\ref{c3eqkinc1}-\ref{c3eqkinc2}) with the phase equilibrium conditions~(\ref{eqc3PhDEq1}-\ref{eqc3PhDEq3}) to obtain analytical predictions without resorting to phase-field simulations. We then have five equations with five unknown variables (Peclet number $p$ and interfacial concentrations $c_2^l$, $c_3^l$, $c_2^s$, and $c_3^s$), which can be solved numerically. In other words, the dealloying kinetics and the interfacial concentrations can be uniquely determined from the initial compositions of the base alloy and the melt ($c_{20}^s$, $c_{30}^s$, $c_{2\infty}^l$, and $c_{3\infty}^l$). 

\begin{figure}[htbp] 
	\centering
	\includegraphics[scale=0.6]{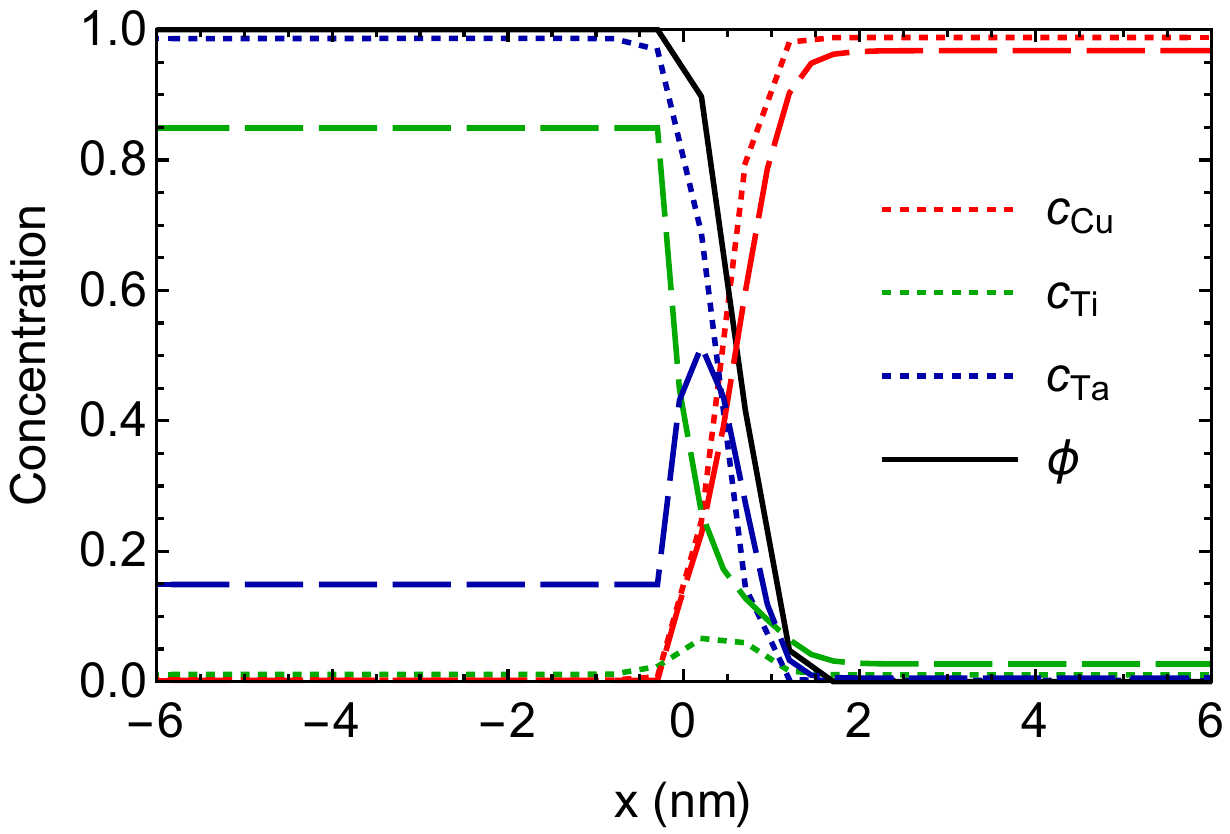}
	\caption[Interfacial concentration profiles]{\nedit{Interfacial concentration profiles from the phase-field simulation of the Ta$_{15}$Ti$_{85}$ precursor dealloyed in pure Cu melt (dashed line) and phase equilibrium conditions from theoretical prediction 2 in Table \ref{tc1ddis} (dotted line).}}
	\label{figintDis}
\end{figure}

\nedit{
As shown in Table~\ref{tc1ddis} (see row ``Dissolution model 2''), this calculation provides a very different prediction than the phase-field model for the dealloying kinetics of a Ta$_{15}$Ti$_{85}$ precursor dealloyed in pure Cu melt. 

To look more closely at this difference, we compare the composition profiles across the interface resulting from both approaches. To deduce composition profiles from the equilibrium compositions obtained by the dissolution model, we proceed as follows: we initialize a 1D phase-field simulation with the expected compositions on the solid and liquid sides of the interface. Then, the phase-field model is numerically integrated with a constant diffusivity in both phases ($D_s = D_l$) to allow the system to relax quickly towards an equilibrium configuration (given by $\partial_t{\phi}=0$ and $\partial_t{c}_i=0$). This method consists in using the phase-field model as a free energy minimizer to obtain equilibrium interfacial profiles.

The resulting profiles are shown with dotted lines in Fig.~\ref{figintDis} while phase-field results obtained from a dealloying simulation are shown with dash lines. This comparison demonstrates that the phase-field simulation does not follow the interface equilibrium expected from the phase diagram. In phase-field simulations, there is no diffusion in the solid, such that the Ta peak is constrained inside the solid-liquid interface (Fig.~\ref{figintDis}). The shape of this peak is controlled by the interplay between the chemical free energy and the gradient terms acting on the composition (see Eq.~\ref{eq:free_energy}). Because of these gradient terms, the height of the Ta peak does not relax to the equilibrium composition expected from the phase diagram.}

\nedit{
In practice, the solid-state diffusivity is about four to five orders of magnitude smaller than the liquid-state diffusivity but remains finite. Therefore, we expect that the equilibrium interfacial concentrations will first be close to the phase-field simulation results, but 
 will eventually approach the phase equilibrium on typical experimental time scales ranging from seconds to minutes. As it will be detailed later in Section~\ref{solid_diff}, we can estimate that, on such time scales, a finite solid diffusivity allows the Ta peak to spread in the solid and the interfacial composition to reach chemical equilibrium. Our theoretical estimate developed in that section predicts that chemical equilibrium will be achieved for dealloying depths satisfying $x_i/w\gg 2pD_l /D_s$, for which the role of the solid diffusivity becomes dominant.
 
Our theoretical calculation combining the dissolution model and the phase equilibrium conditions therefore provides a direct prediction of the concentration profiles obtained on experimental time scales. 
}

\section{Spinodal decomposition}
\label{stability_analysis}
\subsection{Initial destabilization}
\label{initial_destabilization}

The 1D analysis presented in section \ref{passivation} shows that the first stage of dissolution leads to the build-up of a peak of Ta within the solid-liquid interface. Because of the composition gradient terms of the free energy, the concentration profiles spread over the interface width and an overlap region appears naturally between the Ta peak and the liquid Cu. If the system is not confined to 1D, the interface composition can spinodally decompose within the interface to create alternating Ta-rich and Cu-rich domains. If this spinodal decomposition occurs, the dealloying process continues in the Ta-poor regions while it is stopped in the Ta-rich regions because of the strong dependence of the dealloying velocity on the Ta content (see Eq.~(\ref{eq:passivation_v_c}) and Fig.~\ref{fig:passivation_1D}). This concept of interfacial spinodal decomposition establishes a framework to explain the destabilization of the planar dealloying front and the initial stage of the dealloying process \nedit{\cite{morral1971spinodal, de1972analysis}}.

In this section, \nedit{following the classical analysis of spinodal decomposition in multicomponent systems \cite{morral1971spinodal},} we present a linear stability analysis able to predict analytically \nedit{the occurrence of the spinodal decomposition}, the wave-length of the initial destabilization, and therefore the size of the initial microstructure. We consider a system with initially uniform concentrations noted $\bar{c}_1$, $\bar{c}_2$, and $\bar{c}_3$ (with $\bar{c}_3=1-\bar{c}_1-\bar{c}_2$) and investigate the stability of these homogeneous concentrations upon small perturbations. The diffusion equations (see Eq.~(\ref{eq:diffusion})) are linearized around $\bar{c}_1$, $\bar{c}_2$, and $\bar{c}_3$ and are written as:

\begin{align}
	\dot{c}_1 &= M_i \bar{c}_1 (1-\bar{c}_1) \nabla^2 \mu_1 - M_i \bar{c}_1 \bar{c}_2 \nabla^2 \mu_2 	\label{eq:dynamics_c1} \\
	\dot{c}_2 &= M_i \bar{c}_2 (1-\bar{c}_2) \nabla^2 \mu_2 - M_i \bar{c}_1 \bar{c}_2 \nabla^2 \mu_1	\label{eq:dynamics_c2}
\end{align}
where $M_i$ is the mobility of solute within the interface and we consider $M_i=M_0(\phi=\frac{1}{2})$ if we assume {that} the diffusivities of all the components are the same. In the following, we note $M_{lm}=M_i \bar{c}_l (\delta_{lm}-\bar{c}_m)$. The chemical potential $\mu_1$ and $\mu_2$ are also linearized  around ($\bar{c}_1$, $\bar{c}_2$):

\begin{align}
	\mu_1(c_1,c_2) &= \left.\frac{\partial f_{ch}}{\partial c_1}\right|_{\bar{c}_1,\bar{c}_2} 
					+ (c_1-\bar{c}_1) f_{11}
					+ (c_2-\bar{c}_2) f_{12} \nonumber \\
					&- (\sigma_1+\sigma_3) \nabla^2 c_1 - \sigma_3 \nabla^2 c_2  \\
	\mu_2(c_1,c_2) &= \left.\frac{\partial f_{ch}}{\partial c_2}\right|_{\bar{c}_1,\bar{c}_2} 
					+ (c_2-\bar{c}_2) f_{22}
					+ (c_1-\bar{c}_1) f_{12}  \nonumber \\
					&- (\sigma_2+\sigma_3) \nabla^2 c_2 - \sigma_3 \nabla^2 c_1
\end{align}
where $f_{ij}=\left.\frac{\partial^2 f_{ch}}{\partial c_i \partial c_j}\right|_{\bar{c}_1,\bar{c}_2}$ ($f_{ch}$ is defined in Eq.~(\ref{eq:free_energy_ch})). We then consider a small periodic variation of $c_1$ and $c_2$ around their equilibrium values:

\begin{align}
	u_1(\bm{r})&=c_1(\bm{r})-\bar{c}_1=u_1^{0} e^{\omega t + i \bm{k}.\bm{r}} \label{eq:Defu}\\
	u_2(\bm{r})&=c_2(\bm{r})-\bar{c}_2=u_2^{0} e^{\omega t + i \bm{k}.\bm{r}} \nonumber
\end{align}
where $\textbf{r}$ is a position in the $(y,z)$ plane perpendicular to the dealloying direction, $\textbf{k}$ is a wave vector, $\omega_k$ is the corresponding growth rate, $u_1^{0}$ and $u_2^{0}$ are the initial amplitudes of the perturbations. Injecting Eq.~(\ref{eq:Defu}) into Eqs.~(\ref{eq:dynamics_c1}-\ref{eq:dynamics_c2}), we obtain the relations:
\begin{eqnarray}
	&& u_1^0(\omega_k +A k^2)+Bk^2u_2^0=0 , \label{eq:SA_u1} \\
	&& Ck^2u_1^0+u_2^0(\omega_k+Dk^2)=0. \label{eq:SA_u2}
\end{eqnarray}
where $k$ is the norm of the wave vector $\textbf{k}$, and
 \begin{eqnarray}
	 A=&&M_{11}(f_{11}(\bar{c_1},\bar{c_2}) + k^2(\sigma_1+\sigma_3)) \nonumber \\
	 &&+ M_{12} \left(  f_{12}(\bar{c_1},\bar{c_2}) + k^2 \sigma_3\right)  \\
	 B=&& M_{11} \left(  f_{12}(\bar{c_1},\bar{c_2}) + k^2 \sigma_3\right) \nonumber \\
	 &&+ M_{12} \left(  f_{22}(\bar{c_1},\bar{c_2}) + k^2(\sigma_2+\sigma_3)\right) \\
	 C=&& M_{22} \left(  f_{12}(\bar{c_1},\bar{c_2}) + k^2 \sigma_3\right) \nonumber \\
	 &&+ M_{12} \left( f_{11}(\bar{c_1},\bar{c_2}) + k^2(\sigma_1+\sigma_3)\right) \\
	 D= &&M_{22} \left(  f_{22}(\bar{c_1},\bar{c_2}) + k^2(\sigma_2+\sigma_3)\right) \nonumber \\
	 &&+ M_{12} \left(  f_{12}(\bar{c_1},\bar{c_2}) + k^2 \sigma_3\right) .
\end{eqnarray}
Eqs.~(\ref{eq:SA_u1}-\ref{eq:SA_u2}) establish a linear system that can be solved for the fields $u_1(\bm{r})$ and $u_2(\bm{r})$. The system admits a non trivial solution (i.e. different than $(0,0)$) only if its determinant is nil, which leads to a second degree equation on $\omega_k$:
\begin{equation}
	\omega_k^2+(A+D)k^2\omega_k+(AD-BC)k^4=0.
\end{equation}

\begin{figure}[htbp]
	\begin{center}
		\includegraphics[scale=0.56]{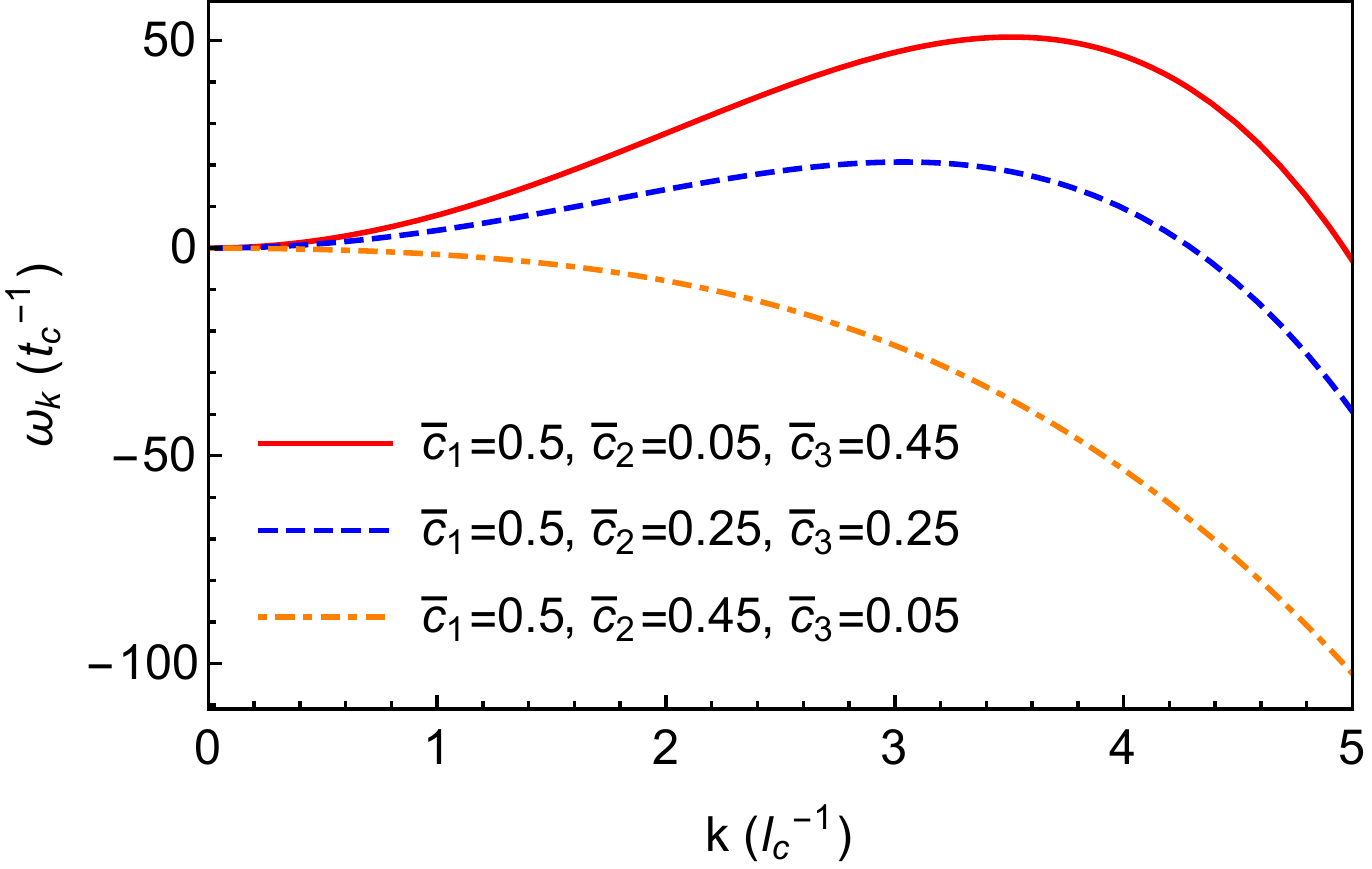}
		\caption[Growth rate $\omega_k$ versus the wave vector $k$ for different compositions.]{Growth rate $\omega_k$ versus the wave vector $k$ for three different compositions. The growth rate and wave vector are scaled by the characteristic time and length, respectively. }%
		\label{fig:omegak}
	\end{center}
\end{figure}

For any concentration $\bar{c}_1$, $\bar{c}_2$, this second degree equation can be solved analytically for $\omega_k$ to obtain the dispersion relation of the instability (the smaller solution of the quadratic equation can be discarded):
 \begin{equation}
	\omega_k=\frac{k^2}{2}\big[ -(A+D) +\sqrt{(A+D)^2-4(AD-BC)} \big],
	\label{eqc3spk}
\end{equation}
\begin{figure}[htbp]
	\begin{center}
		\includegraphics[scale=0.67]{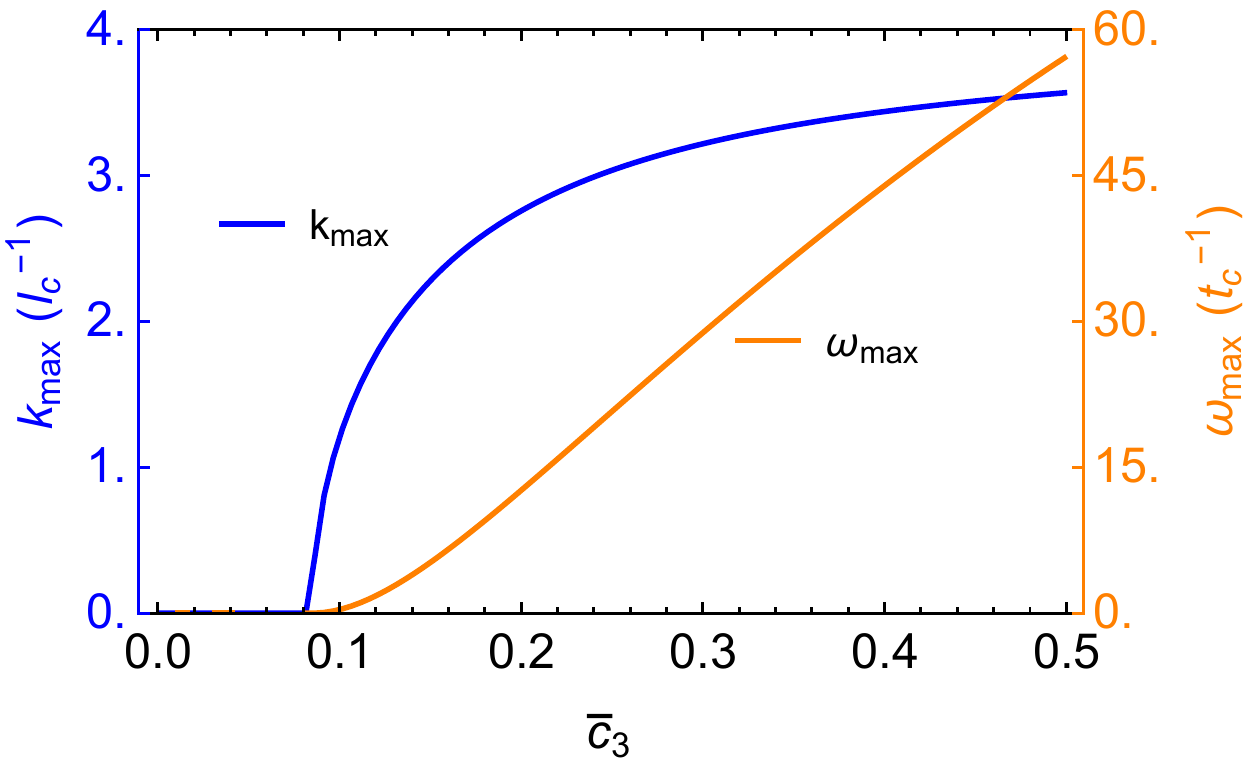}
		\caption[$\omega_{\max}$ and $k_{\max}$ versus the composition of Ta with a fixed Cu concentration of $50\%$.]{Maximum growth rate $\omega_{\max}$ and the corresponding wave-vector $k_{\max}$ as function of the composition of Ta with a fixed average Cu concentration $c_3 = 0.5$.}
		\label{fig:omegaTernary}
	\end{center}
\end{figure}
Fig.~\ref{fig:omegak} represents the dispersion relation of Eq.~(\ref{eqc3spk}) for three different compositions where the amount of Ta is gradually increased while the content of Cu is kept constant. Let us notice that in the case $\bar{c}_1=0.5$, $\bar{c}_2=0.45$ and $\bar{c}_3=0.05$, $\omega_k<0$ for any $k>0$. In other words, no perturbation can develop for this concentration combination. However, for higher Ta compositions, the system presents an unstable domain where $\omega_k>0$. The growth rate presents a maximum, thus selecting the corresponding wave-length of the microstructure (given by the fastest growing wave-vector $k_{max}$).

\begin{figure}[htbp]
	\begin{center}
		\includegraphics[scale=0.55]{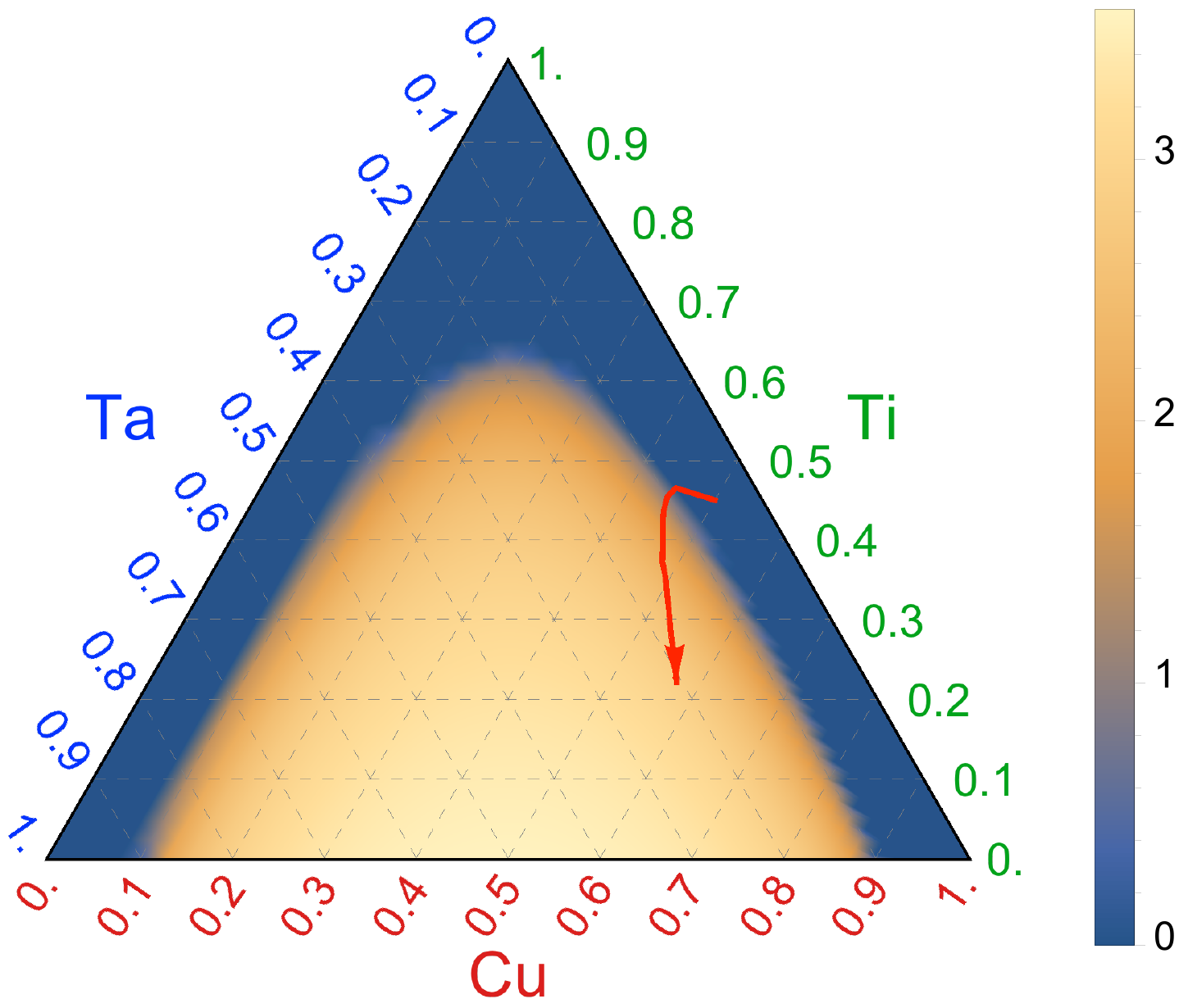}
		\caption[Maximum wave vector $k$ for varying interface concentrations in the ternary system.]{Maximum wave vector $k_{max}$ shown as a colormap for all possible interface concentrations in the ternary system. The red arrow shows the trajectory in the composition space obtained at the solid-liquid interface from the 1D phase-field simulation of Ta$_{10}$Ti$_{90}$ dealloyed in the pure Cu melt (Fig.~\ref{fig:passivation_1D}).  }%
		\label{figc3SpkTernary}
	\end{center}
\end{figure}

During the first stage of dissolution, the composition at the interface changes progressively with time and the Ta content increases in the interface (see e.g. Fig.~\ref{fig:passivation_1D}). 
Fig.~\ref{fig:omegaTernary} displays the evolution of the wave-vector and growth-rate with the Ta content, while the Cu content remains fixed at $50\%$. Fig.~\ref{fig:omegaTernary} shows that the maximum growth-rate $\omega_k$ quickly raises when $\bar c_3 $ exceeds a threshold.
While the growth rate $\omega_k$ raises, the corresponding wave vector $k_{max}$ first increases fast from $0$, and \nedit{quickly plateaus in the range $3l_c^{-1} - 4l_c^{-1}$}, where $l_c$ is the characteristic length-scale. This result is validated by extending the composition range to any possible concentration combinations of the ternary system. We display the results in the form of a ternary plot shown in Fig.~\ref{figc3SpkTernary}. The magnitude of the fastest growing wave-vector varies from $0$ for low Ta and Cu contents where no destabilization can occur to \nedit{$k_{max} \sim 3.5 l_c^{-1}$} along the binary Ta-Cu line.

\begin{figure}[htbp] 
	\centering
	\includegraphics[scale=0.45]{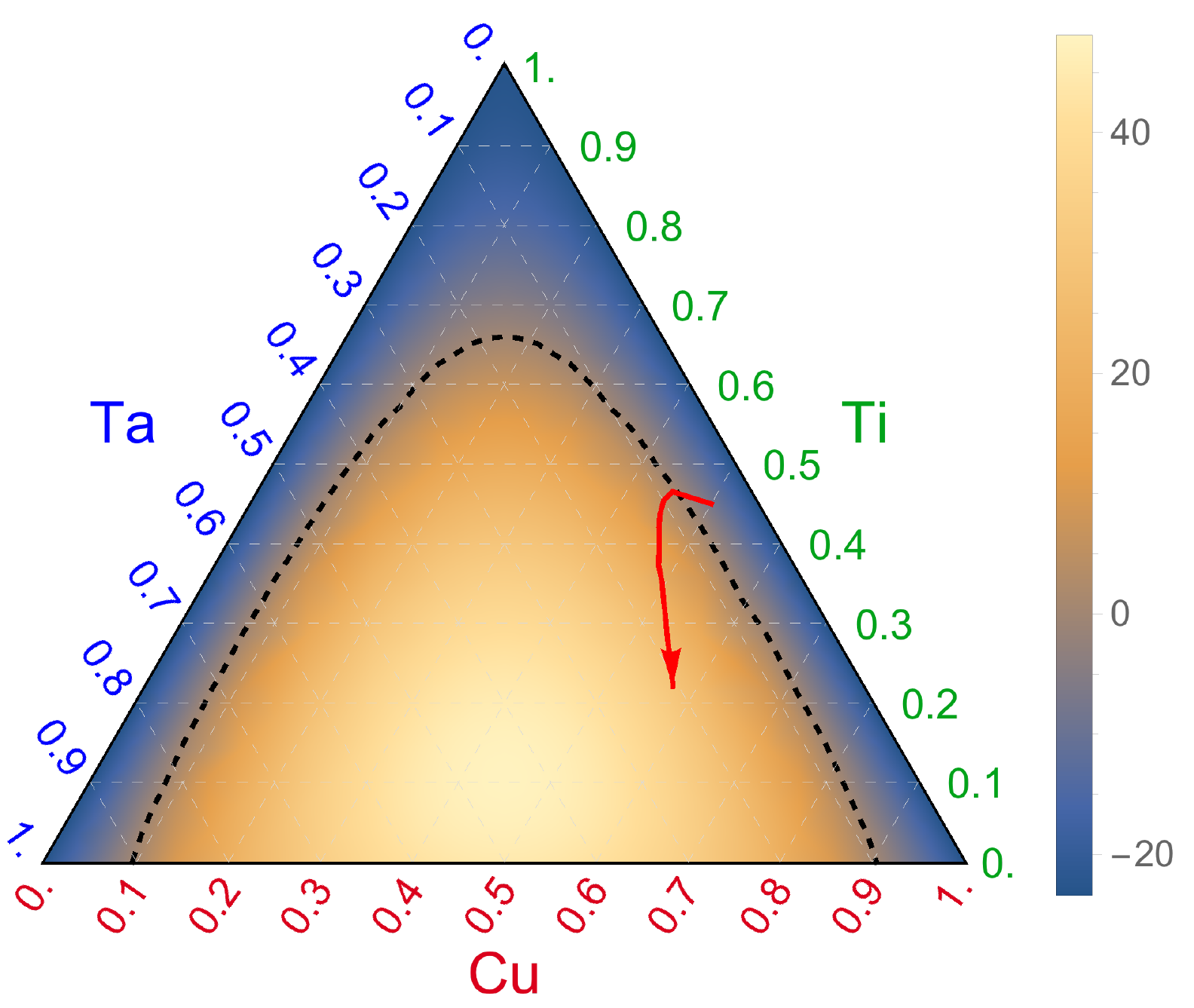}
	\caption[Driving force of spinodal decomposition.]{The driving force for spinodal decomposition. The black dashed line separates positive and negative regions. The red arrow shows the trajectory of the driving force extracted from the solid-liquid interface of the corresponding phase-field simulation of Ta$_{10}$Ti$_{90}$ dealloyed in the pure Cu melt (Fig.~\ref{fig:passivation_1D}). The color map represents the magnitude of the driving force. } %
	\label{figspinodal}
\end{figure}
To find the critical point where the instability develops, we consider the limit of vanishing wave-vector $k \rightarrow 0$. In this limit, the condition to have a positive growth-rate at a finite $k$ is 
\begin{equation}
    \frac{\mathrm{d} \omega}{\mathrm{d} k}\bigg |_{k \rightarrow 0}>0,
\end{equation}
which translates into
\begin{equation}
	\frac{1}{2}\big[ -(A+D) +\sqrt{(A+D)^2-4(AD-BC)} \big] >0,
\end{equation}
where higher order terms in $k$ are neglected.
Therefore, we can obtain the driving force for spinodal decomposition \cite{geslin2015topology} from this simplified criterion:
 \begin{equation}
	f_s=(M_{11}M_{22}-M_{12}^2)(f^2_{12}-f_{11}f_{22}).
	\label{eqspdf}
\end{equation}
\nedit{This criterion also matches the stability criterion from the analysis of the lattice model of multicomponent solid solutions \cite{de1972analysis}.}%

By combining this criteria with the results of the dealloying simulations, it is possible to predict the initial conditions leading to the destabilization of the dealloying front during the first stage of dissolution.
In Figure \ref{figspinodal}, we show the driving force defined by Eq.~(\ref{eqspdf}) in the ternary diagram. During the dissolution of a precursor dealloyed in Cu melt, the composition at the interface changes with time as discussed in section \ref{1D} and follows a trajectory in the compositional space. The red arrow shown in Fig.~\ref{figspinodal} shows the change of interfacial composition (obtained at the level-set $\phi=0.5$) during a 1D phase-field simulation of the dissolution of a Ta$_{10}$Ti$_{90}$ precursor dealloyed in the pure Cu melt. At the beginning of the dissolution, the interface composition contains a small amount of Ta and the driving force for the spinodal decomposition is negative (the blue region on Fig.~\ref{figspinodal}). Along the dissolution, the content of Ta at the interface builds up, and the driving force becomes positive, leading to the spinodal decomposition of the system.

\begin{figure}[htbp] 
	\begin{center}
	\includegraphics[scale=0.44]{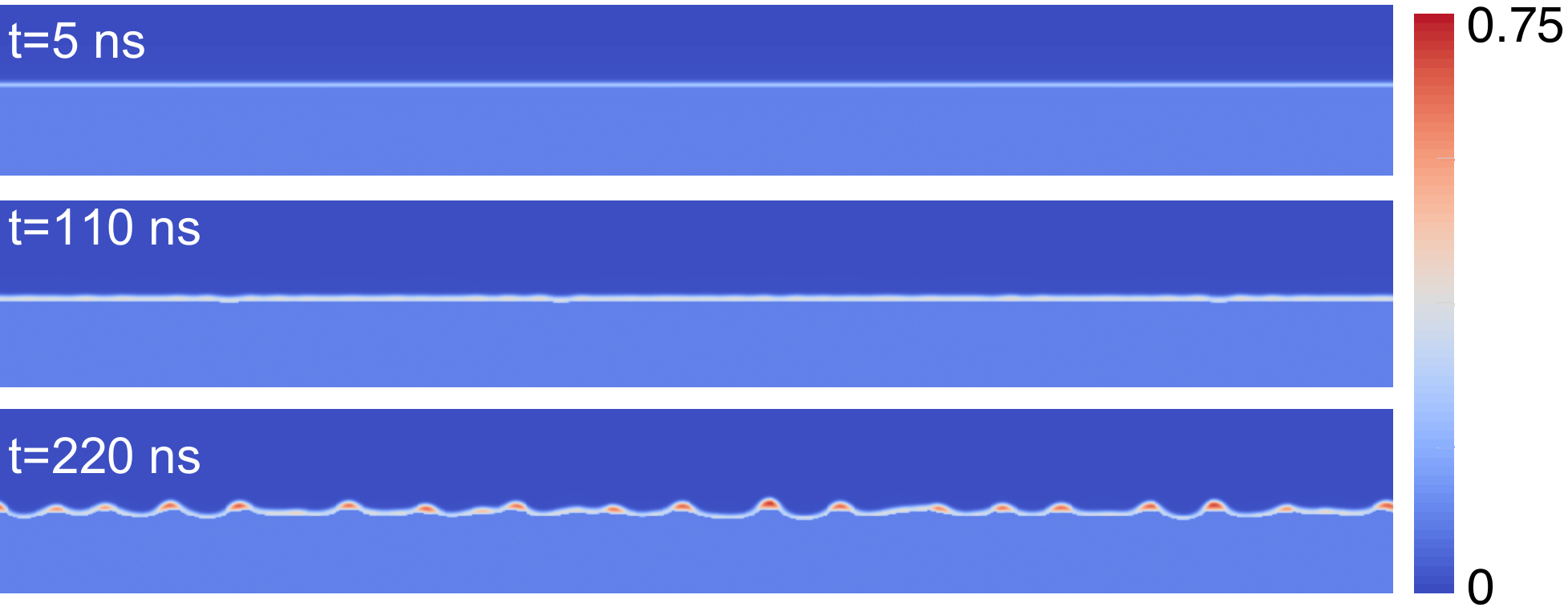}
	\caption[Morphological evolution during spinodal decomposition. ]{2D phase-field simulation showing the transition from a planar dealloying front to a corrugated interface promoted by spinodal decomposition. The color map represents the concentration of Ta in the system. The domain size is 256 nm $\times$ 32 nm.}
	\label{figspd2D}
	\end{center}
\end{figure}

\begin{figure}[htbp] 
	\begin{center}
	\includegraphics[scale=0.33]{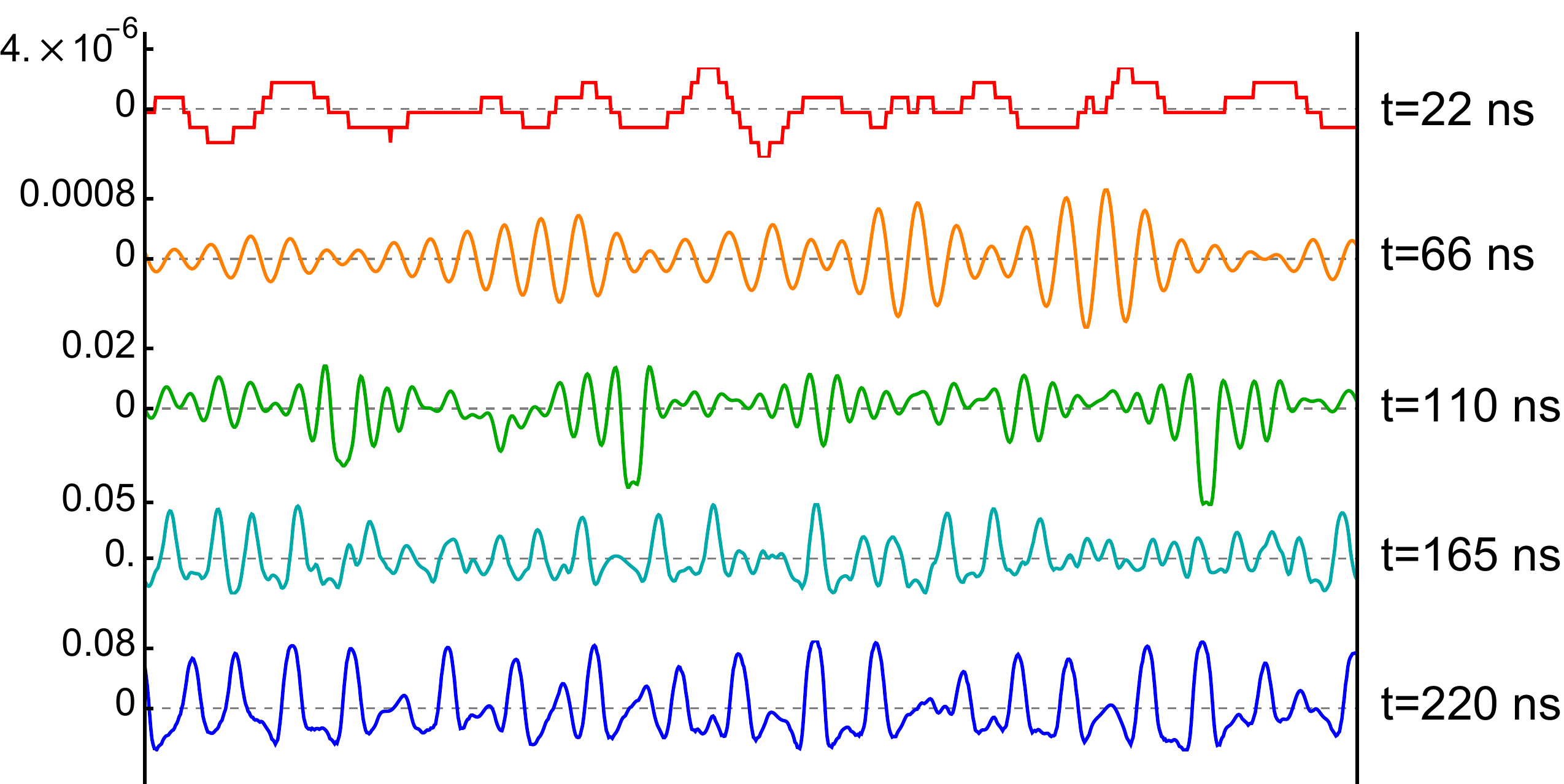}
	\includegraphics[scale=0.51]{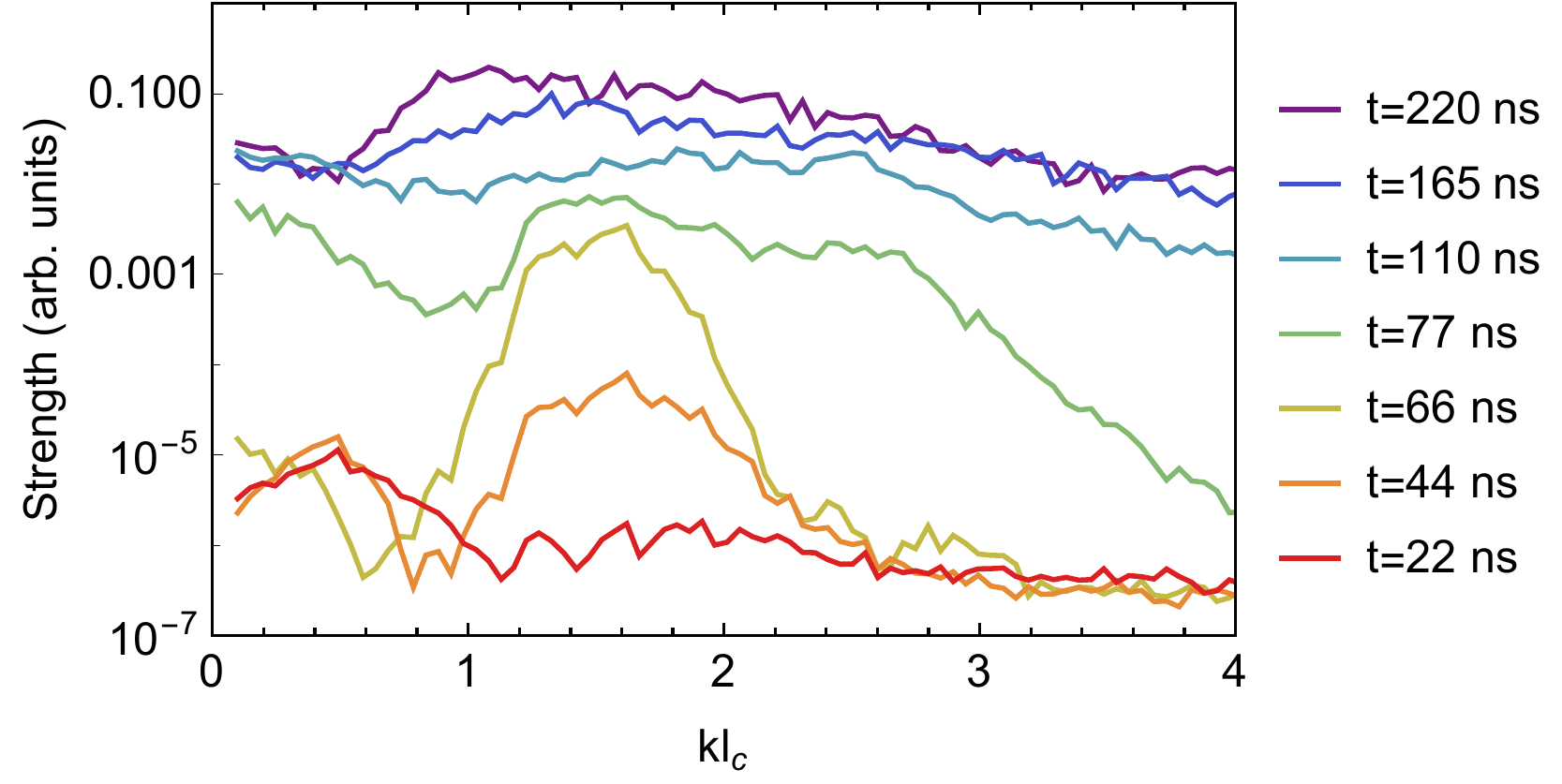}
	\begin{picture}(1,0)(0,0)
   			\put(-130.,255.) {\text{{\textbf{(a)}}}}
			\put(-130.,125.) {\text{{\textbf{(b)}}}}
		\end{picture}
	\caption[The periodic oscillation of Ta concentration at the solid-liquid interface and the evolution of the spectrum of the interface in the log plot. ]{(a) The periodic oscillation of Ta concentration at the solid-liquid interface ($\phi=0.5$) for five different stages (the concentration amplitudes are rescaled) obtained from the simulation shown on Fig.~\ref{figspd2D}. (b) Power-spectra of the interfacial concentration profile obtained at different times.}
	\label{figspdosc}
	\end{center}
\end{figure}
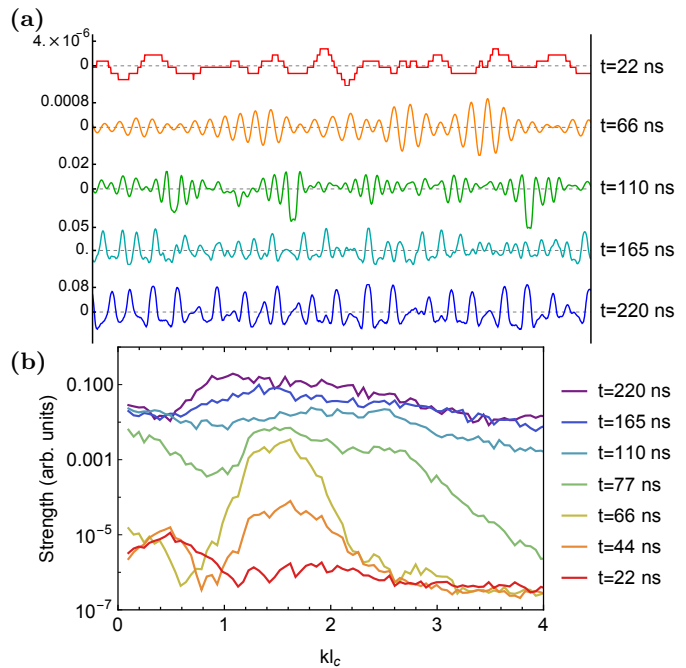

The scenario described above is verified by 2D phase-field simulations, where the spinodal decomposition can develop along the interface. Fig.~\ref{figspd2D} shows the evolution of the interface morphology obtained from the dissolution of a Ta$_{10}$Ti$_{90}$ precursor in a pure Cu melt. 
We also show the corresponding evolution of the interfacial Ta composition and its power spectrum in Fig.~\ref{figspdosc}. 
We observe that, at the first stage ($t=0$ ns to $t=110$ ns), the solid-liquid interface remains planar while the interfacial Ta content increases. The driving force for spinodal decomposition increases with Ta content and small composition fluctuations are amplified. During the second stage ($t=110$ ns to $t=220$ ns), the increasing composition fluctuations eventually leads to the interfacial spinodal decomposition and to the formation of Ta-rich blobs along the interface.

\begin{figure}[htb] 
	\begin{center}
	\includegraphics[scale=0.57]{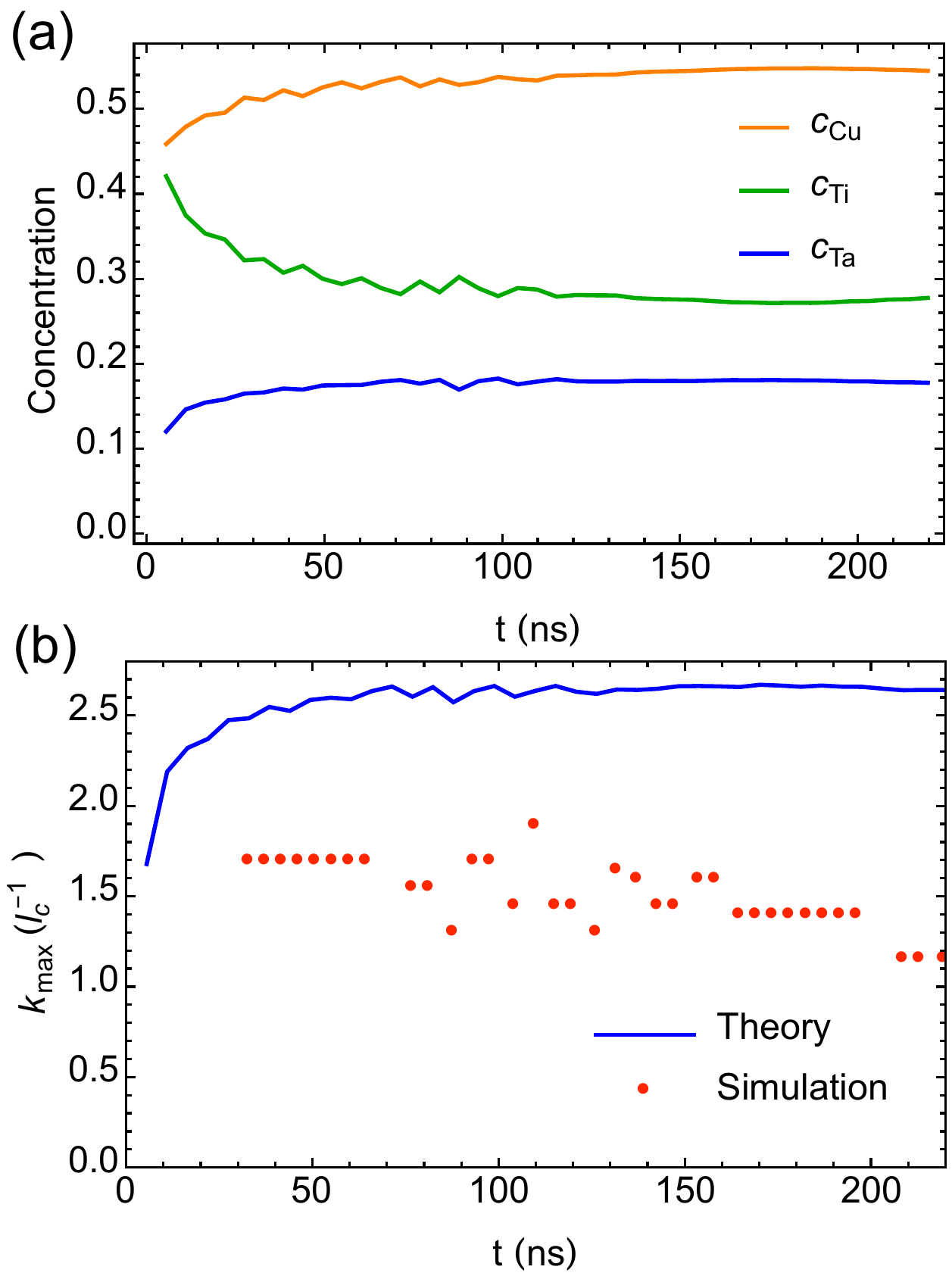}
	\caption[The evolution of the interfacial concentrations and the maximum wave vector. ]{(a) Time evolution of the averaged interfacial concentrations obtained from the phase-field simulation. (b) Comparison of the maximum wave vector $k$ calculated using the linear stability analysis with the interfacial concentrations obtained from (a) and measured from the phase-field simulation.}
	\label{figspdklc}
	\end{center}
\end{figure}

To quantitatively analyze this initial destabilization, we extract the fastest growing wave vector obtained from phase-field simulations by computing the power spectrum of the interfacial Ta composition profiles (see Fig.~\ref{figspdosc}) to extract the dominant wave-length. As shown in Fig.~\ref{figspdosc}b, the power-spectra are rather irregular but clearly present a peak around $k \sim 1 - 2 l_c^{-1}$. The position of the peak $k_{max}$ is extracted and its time-evolution is shown with red dots on Fig.~\ref{figspdklc}.b and compared to the fastest growing wave-vector obtained from the linear stability analysis (Eq.~\ref{eqc3spk}). Fig.~\ref{figspdklc}a display the evolution of the interfacial composition used as an input of the linear stability analysis.

As shown with a blue line on Fig.~\ref{figspdklc}b, the linear stability analysis predicts a sharp increase of the fastest-growing wave vector that stabilizes around $k=2.56l^{-1}_c$. The selected wave vector obtained from the phase-field simulation is of the order of $1.5 l^{-1}_c$ at the beginning of the simulation and decreases slightly with time to reach $1.2 l^{-1}_c$. 
The linear stability analysis is therefore able to predict the order of magnitude of the characteristic wave-length for the microstructure developing at the first stage of dealloying. 
The discrepancy between both results is attributed to the simplicity of the linear stability analysis that does not incorporate any non-linearities, nor the complexity of the composition fields captured with phase-field modeling.

This spinodal decomposition constitutes a framework to understand the initial stages of the dealloying process, which is difficult to observe and study experimentally. 
\nedit{Different from the classical spinodal decomposition where the reference state is a single phase of spatially uniform composition, in the present LMD application, the reference state is bi-phasic and hence has a spatially varying composition in the direction normal to the solid-liquid interface. Therefore, it is not obvious that the occurrence of spinodal decomposition can be quantitatively predicted by an analysis that treats the solid-liquid interfacial layer as a uniform phase with compositions corresponding to a constant value of the phase-field. Phase-field simulations presented in this section demonstrate that this approximation is reasonably quantitative, thereby providing a theoretical framework to predict the occurrence of spinodal decomposition and the initial length-scale of the microstructure.}
After spinodal decomposition, the interface is made of Ta-rich and Ta-poor regions (see last panel of Fig.~\ref{figspd2D}).  As the interface velocity decreases exponentially when the Ta concentration of the interface increases, the dealloying of Ta-rich areas is interrupted while it is facilitated in Ta-poor areas. This dependence of the interface velocity on the Ta-content leads to the corrugation of the interface and to the development of a dealloyed microstructure.

\subsection{\editll{Spinodal decomposition versus planar dissolution in phase-field simulations with vanishing solid-state diffusivity}}
\label{planar_diss}

As seen in the previous section, the dealloying process is triggered by an interfacial spinodal decomposition that can develop only for high Ta and Cu interfacial content for which the driving force for spinodal decomposition becomes positive (see Fig.\ref{figspinodal}). On the other hand, we have seen in section \ref{dealloying_kinetics_ternary} that the interfacial concentrations evolve in time and depend strongly on the initial compositions of the TaTi alloy and the CuTi melt. 
It seems therefore possible to investigate whether or not the system will dealloy (and develop a connected morphology) as function of the initial compositions of the alloy and the melt. \nedit{In this section, we use 2D phase-field simulations to investigate the dealloying process as a function of the base alloy and melt compositions. The simulations show that spinodal decomposition leading to dealloying only occurs below a critical concentration of Ti in the melt that depends weakly on base alloy composition, and planar dissolution occurs above this critical concentration. We then compare the results of 2D phase-field simulations to the theoretical predictions obtained by the analysis of spinodal decomposition with interfacial concentrations (i.e. concentrations at a position corresponding to $\phi=1/2$) extracted from 1D phase-field simulations. Both 2D and 1D phase-field simulations are performed with zero solid-state diffusivity.}

\begin{figure}[htbp] 
	\centering
	\includegraphics[scale=0.092]{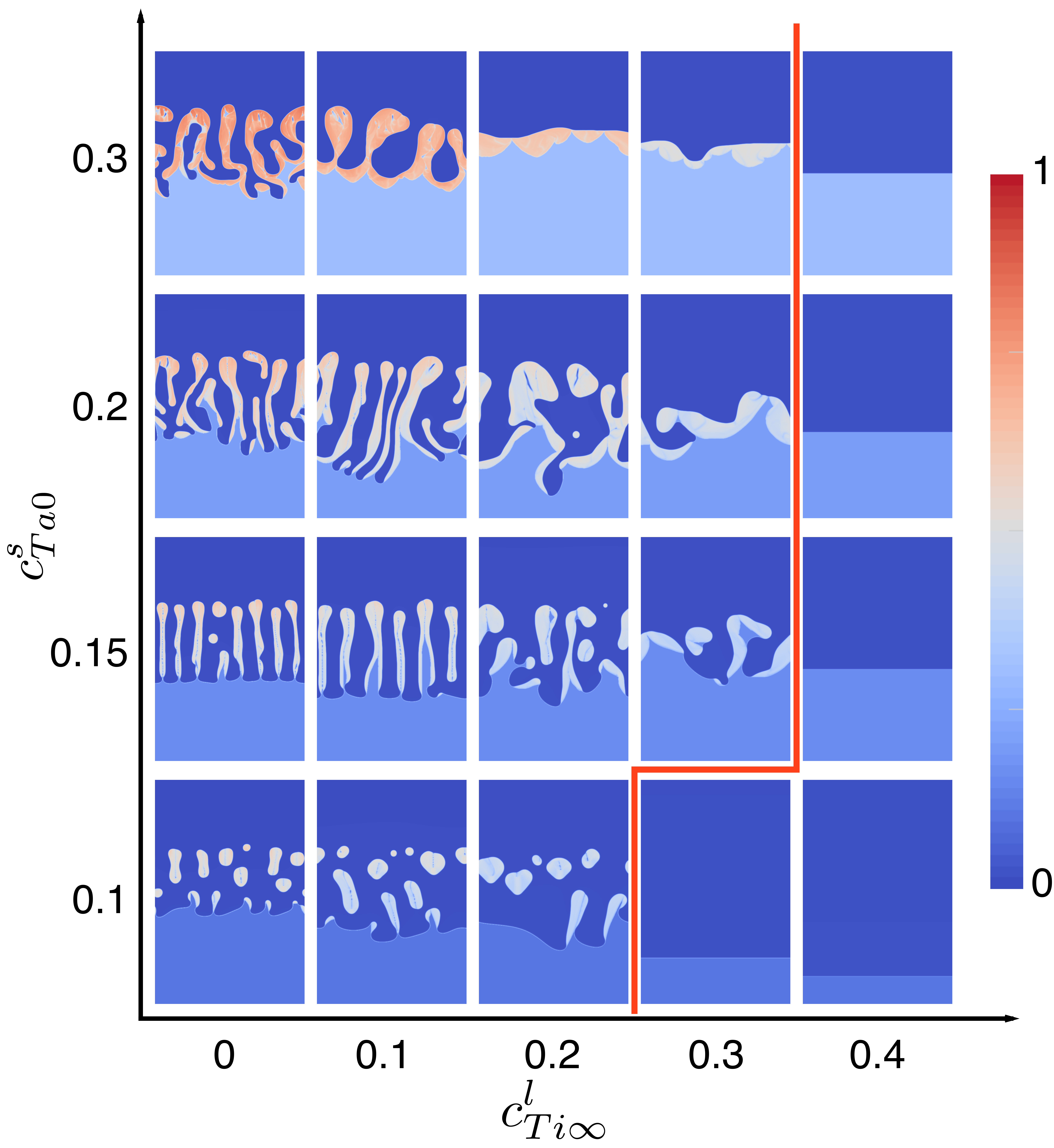}
	\caption[2D phase-field simulations of liquid metal dealloying showing stable planar dissolution.]{Results of 2D phase-field simulations of liquid metal dealloying as a function of Ta concentration in the precursor alloy (\nedit{$c_{Ta0}^s$ along $y$-axis}) and Ti concentration in the liquid melt (\nedit{$c_{Ti\infty}^l$ along the $x$-axis}). The red line shows the boundary between unstable spinodal decomposition and stable planar dissolution. The colormap on the snapshots represents the Ta composition field varying from $0$ to $1$. The simulation domain size is 256 nm $\times$ 384 nm for all simulations.}
	\label{figplandis}
\end{figure}

The results of the phase-field simulations, performed for various Ta contents in the precursor and Ti contents in the melt, are summarized in Fig.~\ref{figplandis}. When $c_{Ta0}^s$ is increased, the dealloyed morphology evolves from disconnected islands to filaments as discussed in an earlier publication \cite{geslin2015topology}. For $c_{Ta0}^s$ above a critical value, phase separation does not occur at the interface and the dissolution remains planar. Simulation results also reveal that the scale of the morphology (ligament size) increases with Ti concentration in the melt up to an upper limit beyond which spinodal decomposition does not occur. This limit corresponds to the thick red line shown in Fig.~\ref{figplandis}. To the left of this boundary, spinodal decomposition drives the formation of Ta-rich and Ta-poor regions inside the solid-liquid interfacial layer, and to the right of this boundary, planar dissolution occurs without spinodal decomposition at the solid-liquid interfacial layer. 

\begin{figure}[htbp] 
	\centering
	\includegraphics[scale=0.65]{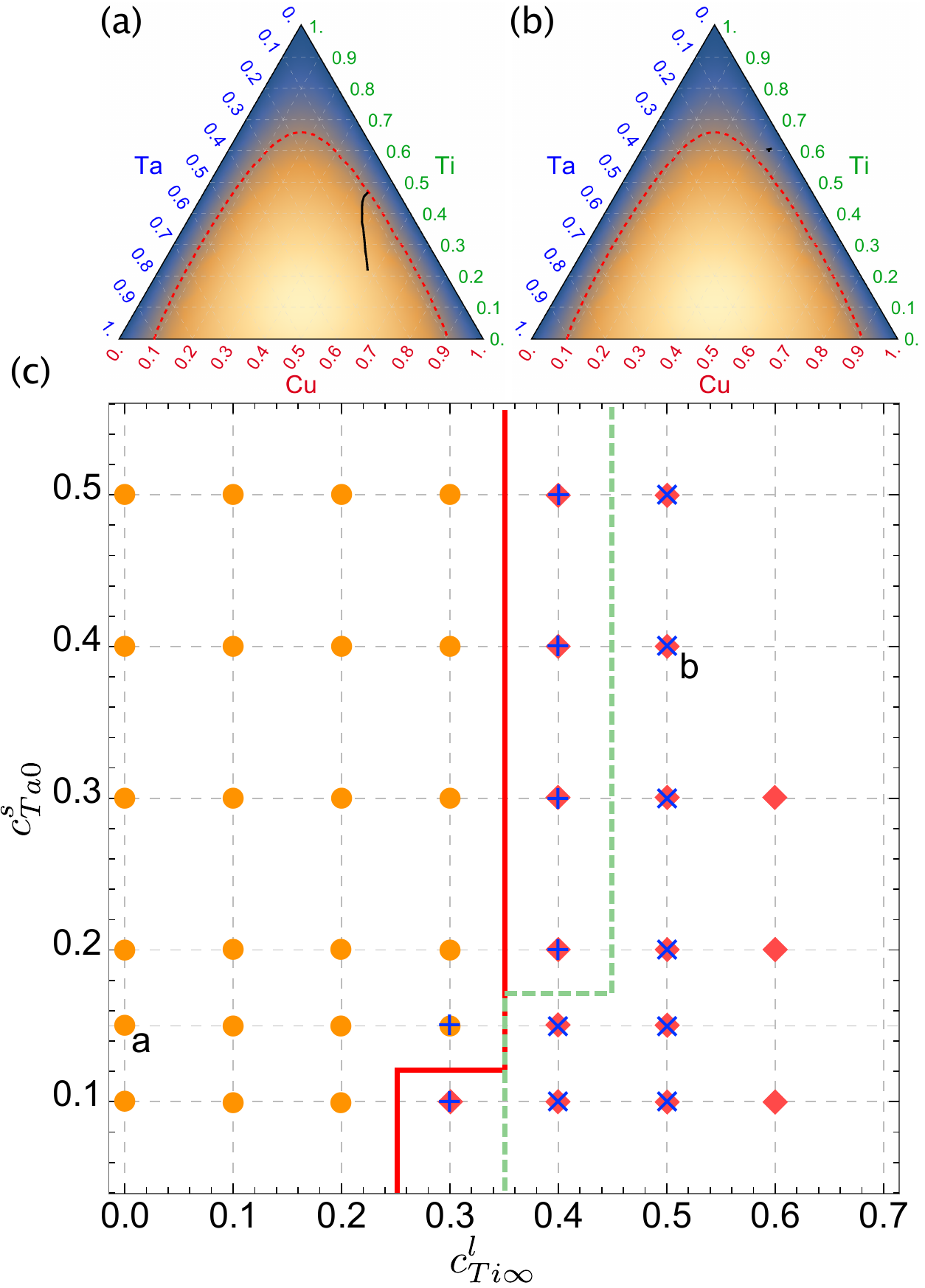}
	\caption[1D phase-field simulations of liquid metal dealloying showing the driving force of spinodal decomposition.]{(a) Trajectory (black line) in the composition space obtained from a 1D simulation of a Ta$_{15}$Ti$_{85}$ precursor dealloyed in a pure Cu melt. In this case, the system reaches the region for positive driving force for spinodal decomposition. (b) Trajectory (black line) obtained for a Ta$_{40}$Ti$_{60}$ precursor in a Cu$_{50}$Ti$_{50}$ melt that remains in the region of negative driving force. Panels (a) and (b) are plotted in the same Gibbs triangles as \nedit{Fig.~\ref{figspinodal}}. (c) Results of the evaluation of spinodal decomposition from 1D phase-field simulations and theoretical calculations as function of \nedit{initial} Ta concentration in the base alloy (\nedit{$c_{Ta0}^s$}) and Ti concentration in the liquid melt (\nedit{$c_{Ti\infty}^l$}). The orange dots represent phase-field simulations for which the trajectory in the ternary composition space reach positive driving forces (such as shown in (a)). The \nedit{orange diamonds} represent the simulations where the trajectories remain in the negative region \nedit{(such as shown in (b))}. The red line is the \nedit{same boundary as in Fig.~\ref{figplandis} and is obtained from 2D simulations}. \nedit{The plus ``$+$" (cross ``$\times$") symbols represent the positive (negative) driving force for spinodal decomposition obtained by combining the 1D dissolution model and the spinodal decomposition analysis. The green dashed lines represent the boundary between both regimes.}}
	\label{figplandis1D}
\end{figure}

To rationalize the occurrence of planar dissolution for high values of $c_{Ti0}^l$, we use the linear stability analysis of spinodal decomposition detailed in the previous section and developed a time-dependent analysis of compositional stability within the interfacial layer.
The driving force for interfacial spinodal decomposition is
 \begin{equation}
 f _ { s }(\bar c_1,\bar c_2) = \left( M _ { 11 } M _ { 22 } - M _ { 12 } ^ { 2 } \right) \left( f _ { 12 } ^ { 2 } - f _ { 11 } f _ { 22 } \right),
 \label{eqfsc1c2}
 \end{equation}
where $M_{ij}=M_0(\phi=\frac{1}{2}) c_i (\delta_{ij}-c_j)$ are the  components of the mobility matrix defined in Section \ref{secpfmodel} and $f_{ij}$ is defined as $f_{i j}=\left.\frac{\partial^{2} f_{c h}}{\partial c_{i} \partial c_{j}}\right|_{\bar{c}_{1}, \bar{c}_{2}}$, and $\bar c_1$ and $\bar c_2$ are the concentration at the of solid-liquid interface ($\phi=0.5$). If the driving force remains negative ($f_s<0$), the dealloying front remains planar, otherwise, an instability will develop through spinodal decomposition, possibly leading to the formation of a connected microstructure.

The results of the analysis are shown on Fig.~\ref{figplandis1D} that distinguishes regions of the \nedit{initial} composition plane (\nedit{$c_{Ta0}^s$, $c_{Ti\infty}^l$}) where spinodal decomposition occurs (filled orange circles) or does not occur (blue crosses) during planar front dissolution. 
To predict the boundary between those two regions, we first compute the range of solid-liquid interfacial compositions that are stable or unstable against compositional fluctuations, corresponding to regions above and below the spinodal boundary (red dashed line) in the ternary composition triangle (see Fig.~\ref{figplandis1D}.a and b). We then superimpose on the ternary plot the trajectories (solid black lines) of interfacial compositions obtained from the 1D phase-field simulations of planar-front dissolution. 
Dealloying is predicted to occur when the trajectory crosses the spinodal boundary, as illustrated in Fig.~\ref{figplandis1D}.a. In contrast, planar dissolution is stable when the trajectory does not reach the unstable domain, as displayed in Fig.~\ref{figplandis1D}.b. \nedit{Based on the combination of 1D phase-field simulations and the linear stability analysis for spinodal decompositions, this analysis} yields predictions in remarkable agreement with the 2D phase-field simulations shown in Fig.~\ref{figplandis} \nedit{from which the red boundary is reported in Fig.~\ref{figplandis1D}}.

\subsection{\editll{Spinodal decomposition versus planar dissolution on experimental time scales with finite solid-state diffusivity}}
\label{secSpinodalSolidD}

\nedit{
In the previous section, we demonstrated that the initial composition of the base alloy ($c_{Ta0}^s$) and the melt ($c_{Ti\infty}^l$) favor either a planar dissolution regime or the development of nanoporous structures triggered by spinodal decomposition. However, the analysis proposed above to predict the occurrence of these regimes still requires running long 1D phase-field simulations} \editll{under the assumption of zero solid-state diffusivity ($D_s=0$).
As already discussed in Section~\ref{dis_phaseEq}, if the solid-state diffusivity is finite, the interfacial concentrations will eventually relax to their phase equilibrium values on a long time scale, thereby affecting when spinodal decomposition occurs. This time scale is theoretically estimated in the next Section \ref{solid_diff} to correspond to a dealloying depth $x_i \gg 2pwD_l/D_s$. Very long 1D phase-field simulations with different $D_s/D_l$ ratios (see Fig.~\ref{figc3Dsphase}), which can reach such depth for $D_s/D_l$ as low $10^{-3}$, confirm that relaxation to interfacial equilibrium indeed occurs. Since it is computationally too costly to carry out 1D phase-field simulations on the time scale required to reach equilibrium for smaller $D_s/D_l$ in the experimentally relevant range $10^{-5}-10^{-4}$, which corresponds to substitutional solid-state diffusion, we can nonetheless predict the occurrence of spinodal decomposition by  assuming that interfacial concentrations are in phase equilibrium.}
\nedit{For this, we obtain numerically the interfacial compositions ($c_2^s$, $c_3^s$, $c_2^l$, and $c_3^l$) at the dealloying front by combining the results of the 1D dissolution model Eqs.~(\ref{c3eqkinc1})-(\ref{c3eqkinc2}) with the phase equilibrium conditions~(\ref{eqc3PhDEq1}-\ref{eqc3PhDEq3}). To evaluate the driving force for spinodal decomposition, we need to estimate interfacial concentrations ($\bar c_1$, $\bar c_2$, $\bar c_3$) at the interface where the spinodal decomposition occurs. A simple estimate from averaging the interfacial concentrations obtained from the 1D dissolution model ($\bar{c}_2 = \frac{c_2^s + c_2^l}{2}$ and $\bar{c}_3 = \frac{c_3^s + c_3^l}{2}$) yields poor predictions because it does not incorporate the non-linearity of the composition profiles at the interface.

A more reliable method consists in relaxing the composition profiles using the phase-field model as a free energy minimizer as in section~\ref{dis_phaseEq}: starting from initial compositions $c_2^s$, $c_3^s$, $c_2^l$ and $c_3^l$ on both sides of the solid/liquid interface, a short phase-field simulation is ran with $D_s = D_l$ to allow for the fast relaxation of the composition profiles to the chemical equilibrium (obtained when $\partial_t{c}_i=0$ and $\partial_t{\phi}=0$). From these equilibrium profiles, the interfacial compositions ($\bar c_1$, $\bar c_2$) are taken at $\phi=0.5$. %

}

\nedit{
Following this method, we can evaluate the driving force of spinodal decomposition for different combinations of initial compositions of the base alloy and melt. The results are reported in Fig.~\ref{figplandis1D}c with blue ``$+$" and ``$\times$" symbols denoting the occurrence of spinodal decomposition and planar dissolution regimes. The transition between these regimes is shown with a dash green line.
This boundary is slightly shifted compared to the prediction obtained from phase-field simulations (red line in Fig.~\ref{figplandis1D}c).
We expect that, for the composition domain on the left of the red line, spinodal decomposition occurs during the first stage of the dissolution. For compositions that fall between the red and green lines, spinodal decomposition should occur later in time, when the Ta peak slowly becomes wider and reaches chemical equilibrium. 
We believe that the green dash line reveals the experimentally relevant boundary distinguishing the planar dissolution and the spinodal decomposition regimes in the limit of small but finite solid-state diffusion. 
We note also that this boundary can be seen as an upper bound along the $c^l_{Ti\infty}$ axis because a larger solid diffusivity would reduce the height of the Ta peak at phase equilibrium (see Fig.~\ref{figc3Dsphase}) and left shift this boundary by an amount that depends on the ratio $D_s/D_l$. For the estimated experimental solid-state diffusivity $D_s/D_l \sim 10^{-5}$, the shift is expected to be small but finite. 
}

\nedit{
The analysis work presented in this section demonstrates the relevance of combining the ternary dissolution model (Section \ref{dis_phaseEq}) with the linear stability analysis for spinodal decomposition (Section \ref{initial_destabilization}) to rationalize the development of interconnected microstructures as a function of dealloying parameters such as the content of the precursor and the melt, the thermodynamics parameters and the temperature.
}

\section{Solid state diffusivity}
\label{solid_diff}

\editll{
In the previous Section~\ref{secSpinodalSolidD}, we used the fact that interfacial concentrations are expected to relax to equilibrium on a sufficiently long time scale in a situation where $D_s/D_l$ is small but finite.
 In the LMD context, $D_s/D_l$ is typically in the range $10^{-5}-10^{-4}$ for alloys with substitutional solid-state diffusion. While one would naively expect such a small ratio to have a negligible effect on dissolution kinetics, it actually has a strong effect on interfacial concentrations by enabling relaxation to local chemical equilibrium.
 In general, relaxation to local equilibrium should occur when the characteristic time for the interface to move a distance of one interface thickness $w$, 
 $\sim w/v$ where $v=\mathrm{d}x_i/\mathrm{d} t$ is the dissolution velocity, is longer than the characteristic time $\sim w^2/D_s$ for solid-state diffusion to occur on the scale $w$. Using the fact that $x_i=\sqrt{4pD_l t}$, we obtain that $v=2 p D_l/x_i$, and hence the condition for local equilibrium
 $w/v \gg w^2/D_s$ becomes $x_i\gg 2wpD_l /D_s$. 
Using for example a recent experiment where a Ta$_{15}$Ti$_{85}$ alloy was dealloyed by a pure Cu melt, \cite{lai2022topological}, the dealloying depth was approximately $270~\mu$m in $10~s$ of dealloying time. Using $w=1$ nm and the experimentally measured Peclet number $p=0.26$, the estimated cross-over dealloying depth $2wpD_l /D_s$ to reach local equilibrium is in the range $5-50~\mu$m for $D_s/D_l$ in the range $10^{-5}-10^{-4}$, and hence significantly shorter than the total $270~\mu$m dealloying depth. We would therefore expect interfacial concentrations to relax to equilibrium during the dealloying process.}

\nedit{In this section, we use 1D phase-field simulations with finite solid-state diffusivity to demonstrate that relaxation indeed occurs for sufficiently large dealloying depth. We also use 2D and 3D phase-field simulations to explore the role of solid-state diffusion on interfacial pattern formation. Those simulations are also relevant for our understanding of solid-state dealloying where the precursor alloy is placed in contact with a solid metal at moderate temperature. Solid-state dealloying experiments  show novel dealloyed structures that are qualitatively different from the one obtained by LMD \cite{wada2016evolution, mccue2017alloy, zhao2019bi}. The main difference of solid-state dealloying compared to LMD is that the diffusivity in both phases are comparable. Varying the diffusivity in the solid compared to the liquid can therefore potentially shed light on pattern formation during solid-state dealloying where the diffusivity contrast between both phases is small.}

\subsection{\nedit{Effect of solid-state diffusion on 1D dissolution kinetics}}
\label{1Dsoldiff}
\begin{figure}[htbp] 
	\centering
	\includegraphics[scale=0.5]{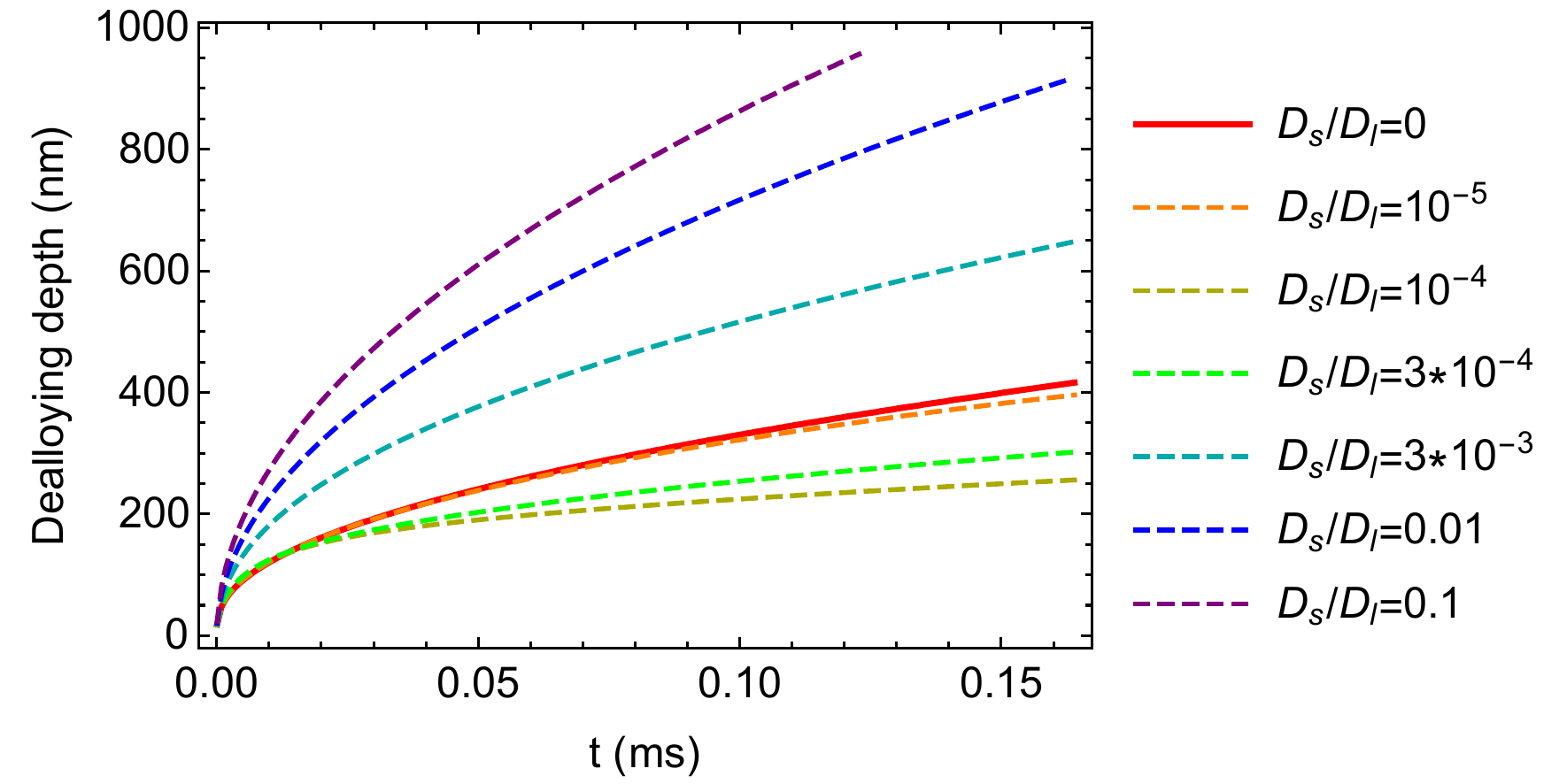}
	\caption[Effect on the dealloying kinetics of solid-state diffusivity.]{Dealloying depth versus time with different solid-state diffusivity.}
	\label{figc3Dskinecta}
\end{figure}
\begin{figure}[htbp] 
	\centering
	\includegraphics[scale=0.6]{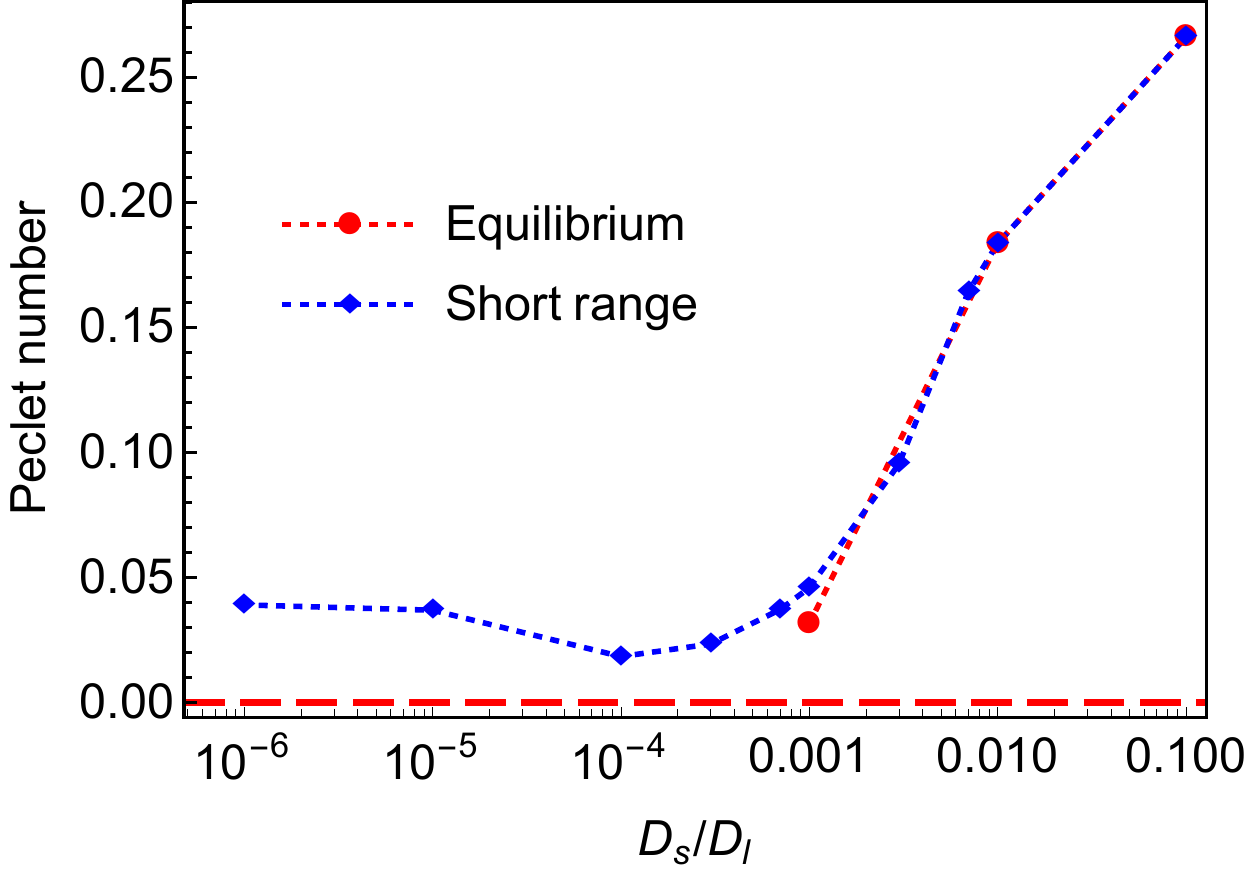}
	\caption[Effect on the dealloying kinetics of solid-state diffusivity.]{The Peclet number as function of the diffusivity ratio $D_s/D_l$. \nedit{The blue data points represent the Peclet number obtained from the short time simulations (see Fig.~\ref{figc3Dskinecta}). The red data points are obtained from the simulations that reach the phase equilibrium (see Fig.~\ref{figc3Dsphase}).} The \nedit{red} dashed line represents the Peclet number obtained \nedit{from the theoretical prediction for $D_s=0$ (see Section~\ref{dis_phaseEq}).}}
	\label{figc3Dskinectb}
\end{figure}
We first investigate the effect of solid-state diffusion on 1D phase-field simulations. To achieve fast dealloying kinetics, we consider a Ta$_{2}$Ti$_{98}$ precursor dealloyed in a pure Cu melt with various diffusivity ratios between solid and liquid phases. The results are shown in Fig.~\ref{figc3Dskinecta}. For all diffusivities, the dealloying front follows a square root diffusion law $x_i=\sqrt{4 p D_l  t}$.
Fig.~\ref{figc3Dskinectb} displays the evolution of the Peclet number obtained from Fig.~\ref{figc3Dskinecta} as a function of solid-state diffusivity. As expected, the dealloying kinetics is much faster when $D_s$ and $D_l$ are comparable, since Ta and Ti can diffuse away in both solid and liquid phases, such that no Ta peak can impede the dissolution. Interestingly, the \nedit{Peclet} number does not depend monotonically on the solid-state diffusivity: when $D_s/D_l\sim 10^{-4}$, the interface moves even slower than $D_s=0$. 

\begin{figure}[htbp] 
	\centering
	\includegraphics[scale=0.53]{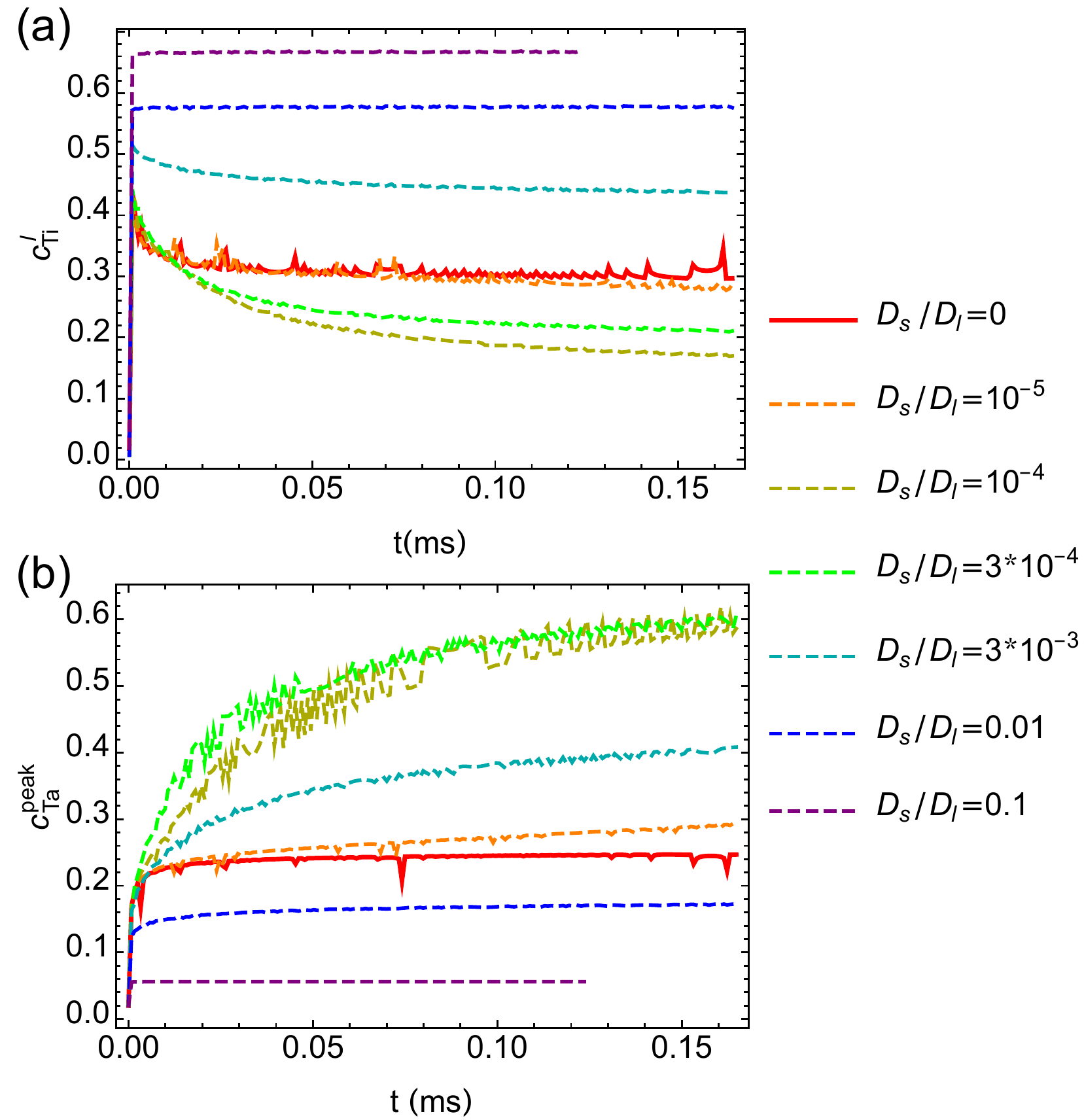}
	\caption[Concentration profiles during the dealloying with solid-state diffusion.]{The effect of solid-state diffusion demonstrated by the liquid concentration of Ti at the interface (a) and the concentration of Ta peak at the interface (b) versus time. }
	\label{figc3Dscon}
\end{figure}

This effect can be explained by looking in details at the interfacial concentrations of Ti and Ta in the liquid and the solid that control the flux of Ti in the liquid and the dealloying kinetics as pointed out in section \ref{1D}. 
Fig.~\ref{figc3Dscon} shows the evolution of these interfacial compositions for different solid diffusivities. For small values of $D_s$, the height of the Ta peak \nedit{($c_{Ta}^{peak}$)} increases to a value larger than the one obtained with zero solid-state diffusion. For larger values of $D_s$ however, the value of the Ta peak is reduced as compared to the $D_s=0$ case.

This nonlinear variation is due to the competition of two effects. First, a finite solid-diffusivity allows the spreading of the Ta peak in the solid phase, which reduces the influence of the concentration gradient terms on the height of the Ta peak that can reach a higher value. Consequently, the equilibrium concentration of Ti on the liquid side is reduced, according to the chemical equilibrium at the interface (see Fig.~\ref{figtCuTiTa}). If $D_s$ is further increased, it allows for a significant flux of Ta in the solid phase, which reduces the height of the peak, leading to larger Ti in the liquid and a faster kinetics.

\begin{figure}[htbp] 
	\begin{center}
		\includegraphics[scale=0.54]{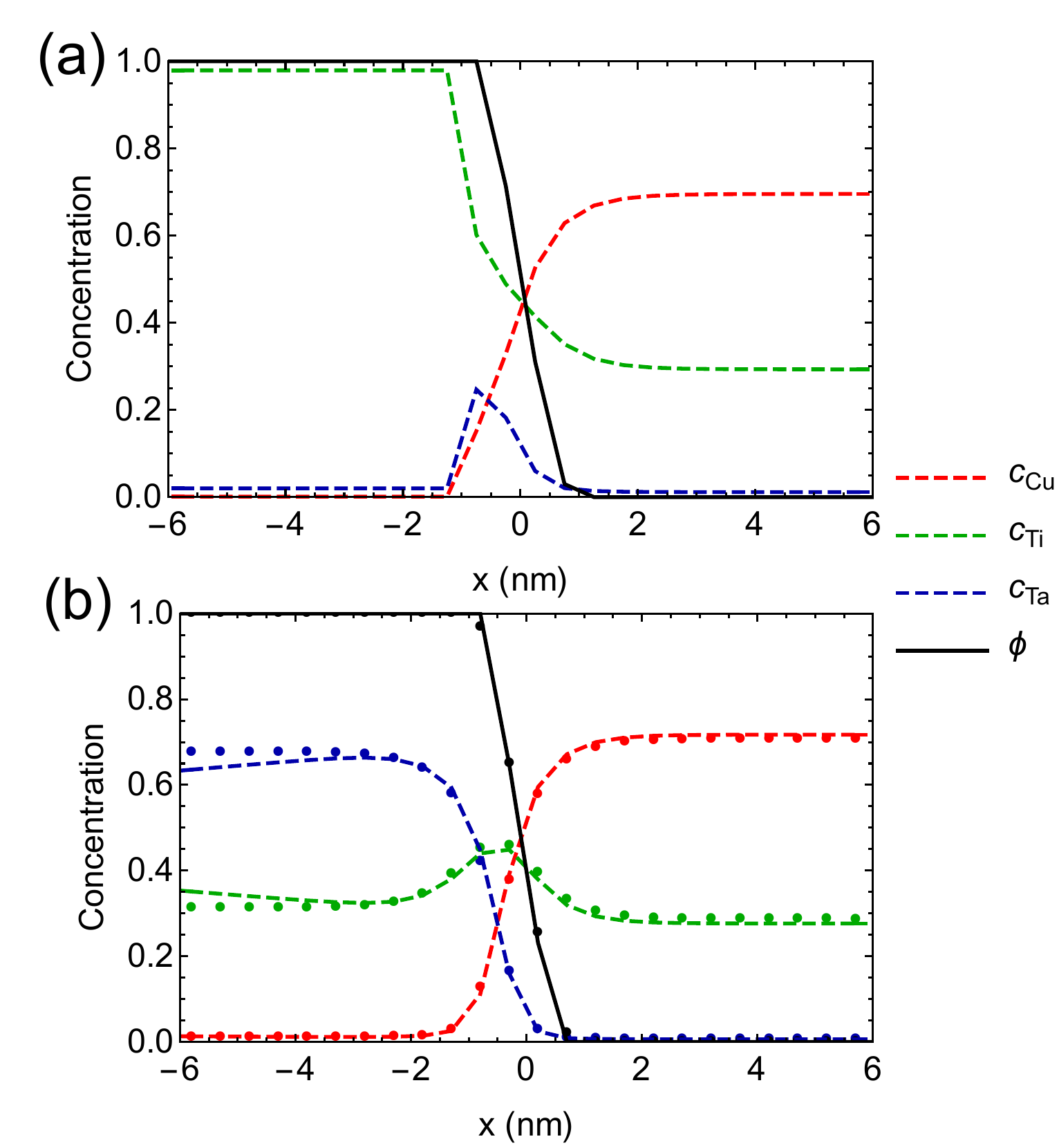}
		\caption[Interfacial concentration profiles.]{\nedit{Interfacial concentration profiles for a Ta$_{2}$Ti$_{98}$ precursor dealloyed in the pure Cu melt obtained for two solid-state diffusivities: (a) $D_s/D_l=0$. (b)$D_s/D_l=10^{-3}$. The dots in (b) represent the concentration profiles obtained from the phase equilibrium conditions (see text in Section~\ref{dis_phaseEq}).}}
		\label{figintconDsEq}
	\end{center}
\end{figure}

Our previous phase-field study \cite{geslin2015topology} has shown that the concentration profiles obtained from phase-field simulations vary significantly from the prediction of the phase diagram. As explained above, this discrepancy is attributed to the concentration gradient terms in the total free energy. For finite solid-state diffusivity, the width of the Ta peak increases, which reduces the role of these gradient terms, allowing the interfacial concentrations to reach a chemical equilibrium. 
\nedit{Interfacial concentration profiles obtained for $D_s/D_l=0$ and $D_s/D_l=10^{-3}$ are shown in Fig.~\ref{figintconDsEq}. For $D_s/D_l=0$, the width of the Ta peak is comparable to the interface thickness while for $D_s/D_l=10^{-3}$, it becomes much larger (Fig.~\ref{figintconDsEq} only shows part of the peak), which significantly reduces the effect of the concentration gradient terms.}
Therefore, we expect the interfacial concentrations to approach an equilibrium predicted by the phase diagram when the solid-state diffusivity is increased. 
\nedit{In addition, using the phase-field method with as a free-energy minimizer (with $D_s = D_l$) and following the same steps as for Fig.~\ref{figintDis} yields equilibrium concentration profiles shown with dots in Fig.~\ref{figintconDsEq}b. 
As expected, allowing for a small but finite solid-state diffusivity allows the relaxation of the interfacial concentration profiles towards chemical equilibrium. As discussed above, this relaxation can be achieved on time scales where the dealloyed depth satisfies $x_i/w \gg 2pD_l /D_s$.}

\begin{figure}[htbp] 
	\begin{center}
		\includegraphics[scale=0.5]{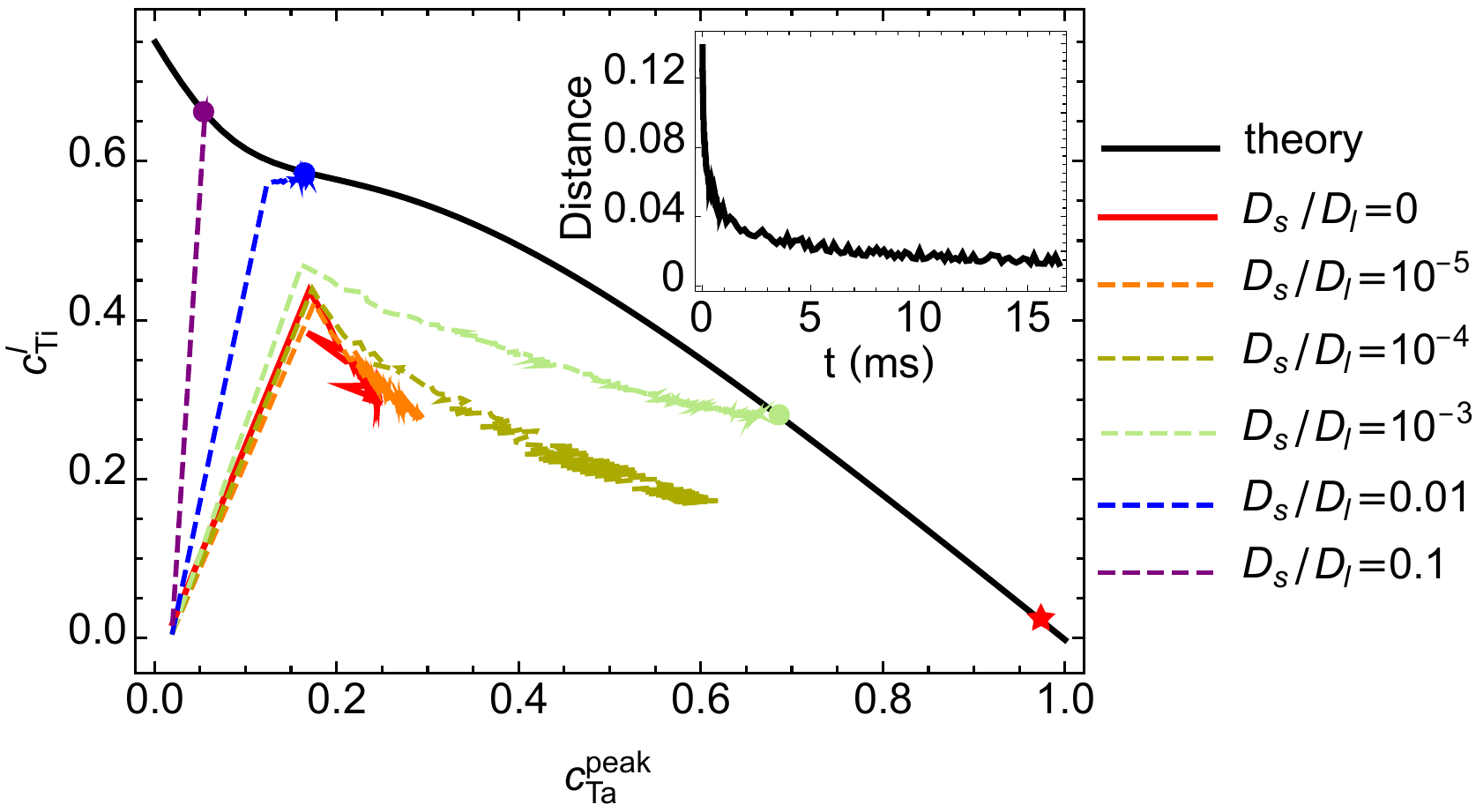}
		\caption[Finite solid-state diffusivity promotes the equilibrium concentrations reaching the phase diagram.]{Interfacial concentrations obtained from 1D phase-field simulations with various solid-state diffusivities. The black solid line is the equilibrium concentrations extracted from the phase diagram. The green, blue, and purple dots are the intersection of the theoretical phase diagram and the interfacial concentration profile with $D_s/D_l=10^{-3}$, $D_s/D_l=0.01$, and $D_s/D_l=0.1$, respectively. The red star represents the theoretical prediction of the interfacial concentrations with $D_s/D_l=0$. The insert shows the time-evolution of the distance - computed in the $c_\mathrm{Ti}^l$-$c_\mathrm{Ta}^{\text{peak}}$ diagram - between the analytical prediction from the phase-diagram and the simulated interfacial concentrations plot versus time for $D_s/D_l=10^{-3}$.}
		\label{figc3Dsphase}
	\end{center}
\end{figure}

From sections \ref{secpfmodel} and \ref{dealloying_kinetics_ternary}, we know that multiple interfacial equilibria are possible which are represented by the tie-lines of the ternary phase-diagram (Fig.~\ref{figtCuTiTa}).
These equilibria can be represented by the thick black line in the $c^\mathrm{peak}_\mathrm{Ta}$ - $c^l_\mathrm{Ti}$ diagram of Fig.~\ref{figc3Dsphase}. The colored lines represent the evolution of the interfacial concentrations obtained from phase-field simulations with different solid diffusivities. 
In addition, the calculations of section \ref{dis_phaseEq} can be used to predict the interfacial concentrations obtained in the limit of vanishing solid-state diffusivity: this prediction is obtained by combining the phase equilibrium conditions Eqs.~(\ref{eqc3PhDEq3}-\ref{eqc3PhDEq1}) with the constraint of the concentration profiles Eqs.~(\ref{c3eqkinc1}-\ref{c3eqkinc2}) and is represented as a red star symbol on Fig.~\ref{figc3Dsphase}. 

From Fig.~\ref{figc3Dsphase}, we find that the interfacial concentrations vary significantly with time.
As expected, when the solid diffusivity is large, the interface concentrations converge towards an equilibrium on the black line and reach a steady-state dissolution regime. This convergence is reached quickly for large solid-state diffusivity but requires longer time for smaller $D_s/D_l$ ratios, such that this steady-state equilibrium remains out of reach of our phase-field simulation for $D_s/D_l<10^{-3}$.
For the specific case of $D_s/D_l=10^{-3}$, we show in the inset of Fig.~\ref{figc3Dsphase} the time evolution of the concentration gap between the simulated interfacial concentrations and the converged value, revealing the slow convergence of the interfacial concentrations towards the phase-diagram prediction.
\nedit{The variation of the interfacial concentrations, especially the value of the Ta peak $c_{Ta}^{peak}$, reduces the Peclet number during this convergence for lower solid-state diffusivity (see red data points for $D_s/D_l=10^{-3}$ in Fig.~\ref{figc3Dskinectb}). For smaller $D_s/D_l$ ratios and longer times, the Peclet number will eventually reach the limit obtained for $D_s=0$ from the theoretical prediction and shown with a red dashed line in Fig.~\ref{figc3Dskinectb}. 
}

Interestingly, \nedit{the data points (purple, blue, green and red) in Fig.~\ref{figc3Dsphase} representing the concentrations reached for the steady-state dissolution regime indicate that} the final equilibrium concentrations (and the corresponding tie-line) vary significantly with the solid-state diffusivity. This reveals the significance of this parameter that may in turn influence the spinodal decomposition process and the resulting morphologies.

\subsection{\nedit{Effect of solid-state diffusion on 2D and 3D dealloyed morphologies}}
\begin{figure}[htbp]
	\begin{center}
		\includegraphics[scale=0.22]{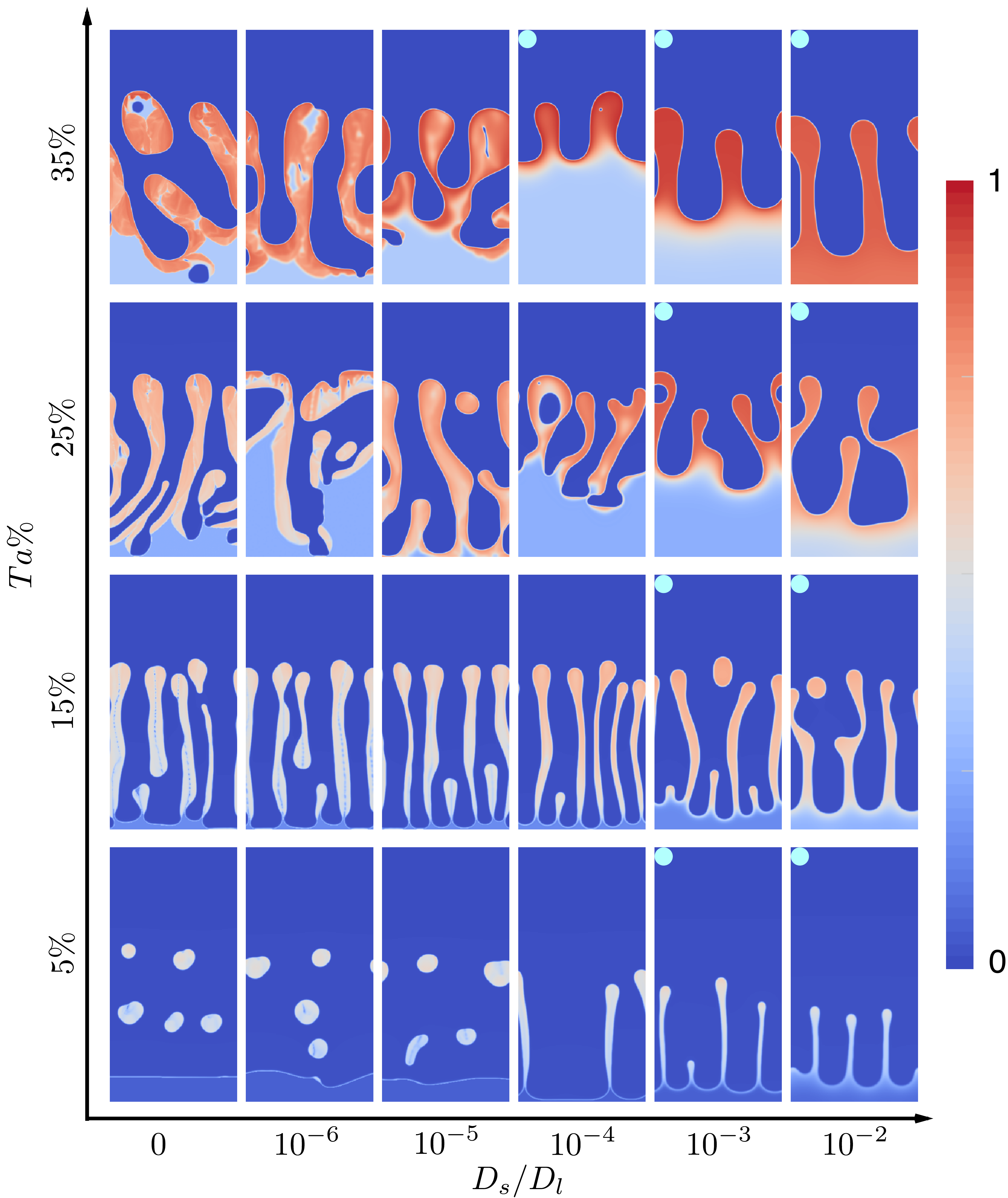}
		\caption[2D phase-field simulations for various Ta content in the base alloy and the solid-state diffusivity.]{2D phase-field simulations obtained for various Ta content in the precursor alloy and solid-state diffusivities. The colormap represents the Ta concentration field. The light blue circles on the upper left corner of some snapshots indicate that these simulations started with an initially perturbed interface to promote the microstructure evolution. The domain size for all the simulations is 256nm$\times$128nm.}
		\label{figc3DsTaDs}
	\end{center}
\end{figure}

In section \ref{1Dsoldiff}, we showed that the solid-state diffusivity has a significant effect on the interfacial compositions and allows the diffusion of Ta in the solid phase. In 2D and 3D phase-field simulations, we expect that these effects will also modify the spinodal decomposition process and the further morphology development, thereby changing the morphology of the dealloyed structure.

In this section, we use 2D phase-field simulations to show how the solid-state diffusion affects the morphologies of the dealloyed microstructures. 
As shown in Fig.~\ref{figc3DsTaDs}, the finite solid-state diffusivity has three effects on the morphological evolution. 
First, large solid-state diffusivities ($D_s/D_l > 0.001 $) inhibit the interfacial spinodal decomposition, thereby promoting a planar dissolution regime. To force the development of a dealloyed microstructure, the initial condition of the simulations is taken from an intermediate configuration obtained for $D_s=0$. The resulting microstructures are marked with a light blue circle on the upper left corner in Fig.~\ref{figc3DsTaDs}. 
Second, we note that a finite solid-state diffusivity promotes more connected dealloyed structures for all Ta compositions. This effect is shown in the 5\% Ta simulations, where blobs appear for low solid-state diffusivity, whereas lamellae form for $D_s/D_l>10^{-4}$.  
Finally, we find that the finite solid-state diffusivity stabilizes the diffusion-coupled growth of lamellar structures \cite{geslin2015topology}, thereby favoring the formation of aligned structures over high-genus topologically connected structures.
\begin{figure}[htbp]
	\begin{center}
		\includegraphics[scale=0.22]{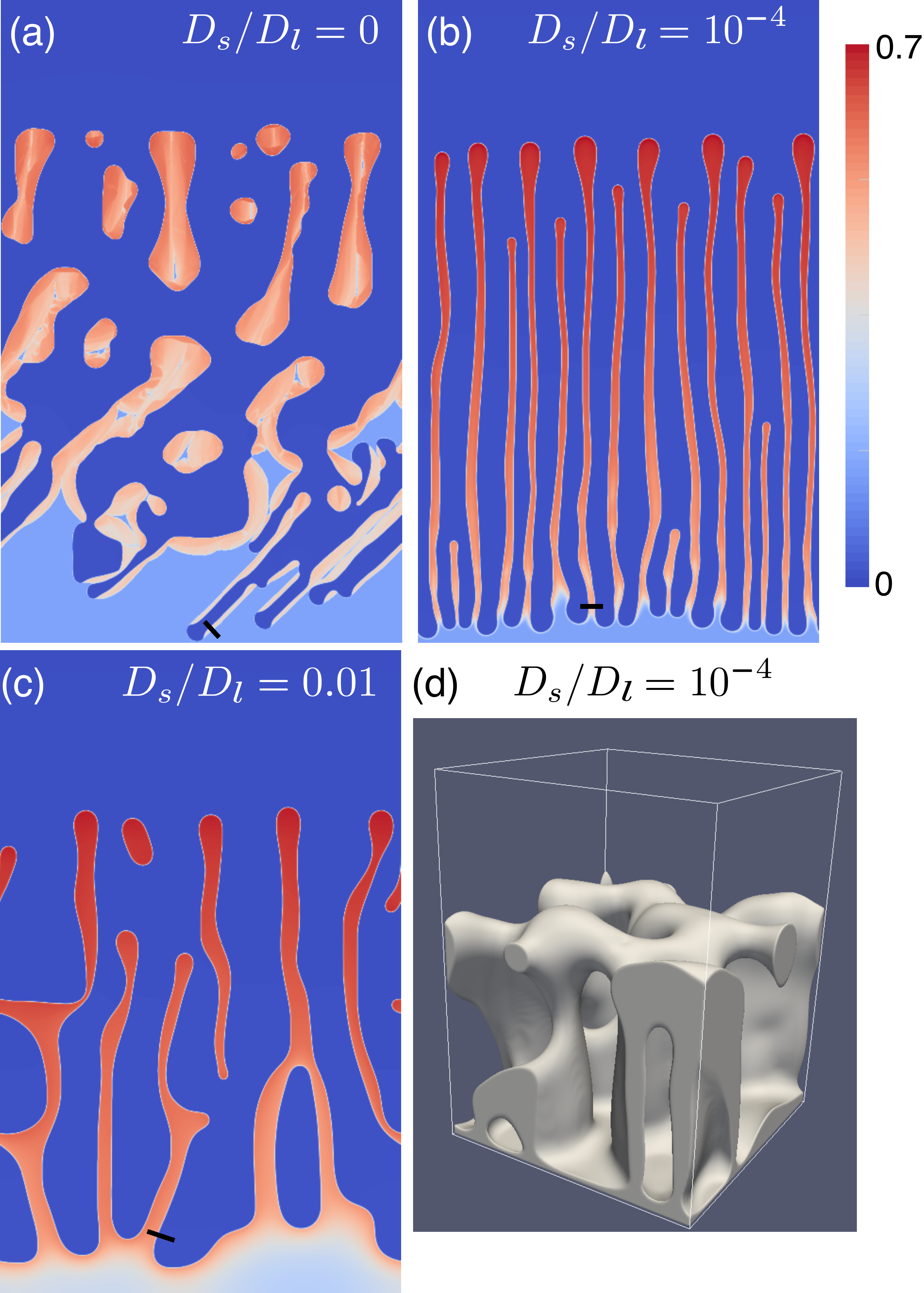}
		\caption[phase-field simulations with variant solid-state diffusivities.]{phase-field simulations of Ta$_{15}$Ti$_{85}$ alloys dealloyed in the pure Cu melt with various solid-state diffusivities indicated in the plots. The black lines indicate the sampling positions for Fig.~\ref{figc3TerPltCon}. The domain size for all 2D simulations is 1024 nm$\times$640 nm and for 3D simulation is 128 nm$\times$96 nm$\times$96 nm. }
		\label{figc3DsLong}
	\end{center}
\end{figure}

A more significant comparison is also shown in Fig.~\ref{figc3DsLong}, where the size of the simulation domain is much larger. 
For a finite solid-state diffusivity $D_s/D_l=0.01$, the dealloyed structure first forms aligned ligaments, eventually merging when the velocity decreases. We observed that the merged lamellae break later due to the dissolution of the solid branches in the liquid. 
For the solid-state diffusivity $D_s/D_l=10^{-4}$,  the dealloyed structure forms aligned ligaments, and the shorter ligaments will be dissolved later as the spacing of ligaments is increasing. 
We also performed phase-field simulations to check if these findings holds in 3D. Fig.~\ref{figc3DsLong}d shows that the dealloyed structure forms parallel walls at the dealloying front, matching the morphology obtained in 2D. Due to coarsening, the top layer of the dealloyed structure becomes eventually connected. Interestingly, this type of elongated yet connected microstructure was observed in experiments of solid-state dealloying where the diffusivity in both phases are comparable \cite{wada2016evolution}.

\begin{figure}[htbp]
	\begin{center}
		\includegraphics[scale=0.5]{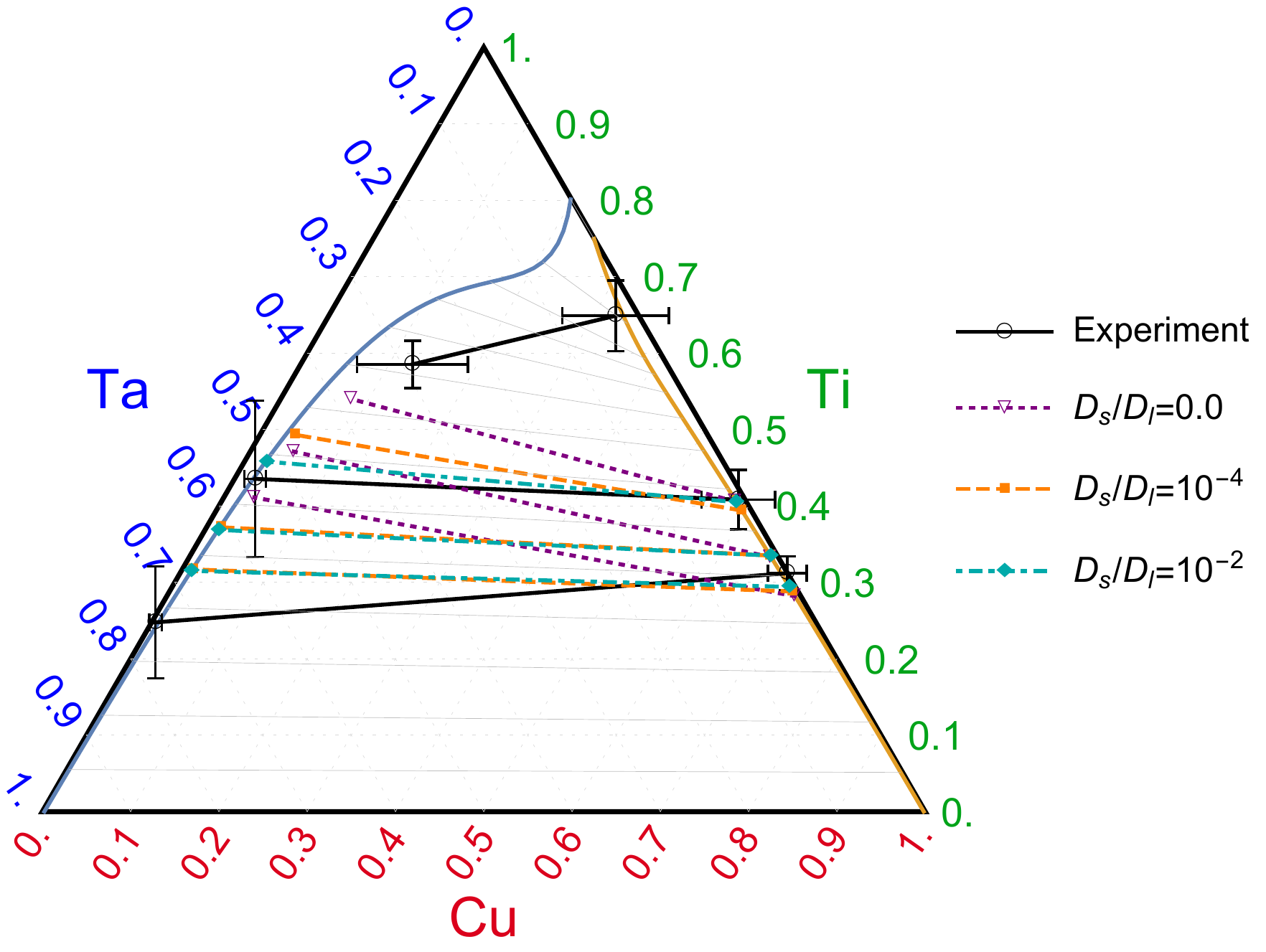} 
		\caption[phase-field simulations with varying solid-state diffusivity.]{Ternary phase diagram with the equilibrium interfacial concentrations extracted from the phase field simulations (Solid lines with different solid-state diffusivities) and experiment (Dashed line with Ta$_{15}$Ti$_{85}$ dealloyed in the pure Cu melt). }%
		\label{figc3TerPltCon}
	\end{center}
\end{figure}
Another effect of the solid-state diffusivity observed in the simulations is that the interfacial concentrations relax to a local chemical equilibrium on long time-scales.
To examine the influence of this effect in 2D simulations, we perform a quantitative analysis of the larger size 2D phase-field simulations shown in Fig.~\ref{figc3DsLong}. 
We first realized that the averaged liquid concentrations of Ti at the dealloying front do not vary significantly, which indicates that the dealloying kinetics remains similar when the solid-state diffusivity varies from 0 to $D_s/D_l=0.01$. 
We measure the concentrations at the solid-liquid interface close to the dealloying front, at the position indicated by a black marker in Fig.~\ref{figc3DsLong}. 
The results are reported on the ternary phase diagram of Fig.~\ref{figc3TerPltCon} where two sets of interfacial concentrations obtained at the top and center of the dealloyed region are reported. 
For the different solid diffusivities, the concentrations in the liquid remain similar (between $0.3$ and $0.4$) but the solid concentrations vary significantly.
For vanishing solid-state diffusivity, the equilibrium concentrations do not follow a possible equilibrium indicated by a tie-line of the phase diagram. 
When the solid-state diffusivity is increased, the equilibrium concentration of Ta in the solid increases to reach the local chemical equilibrium (brown line on Fig.~\ref{figc3TerPltCon}). This convergence towards a chemical equilibrium is not instantaneous and follows a transient regime as shown in the previous section. \nedit{Based on the calculation at the beginning of this section,} the dealloying depth necessary to reach this local chemical equilibrium is given by $x_i/w=2pD_l/D_s$, which gives $x_i=3500$ nm for $D_s/D_l=10^{-4}$,  and $x_i=35$ nm for $D_s/D_l=0.01$. As a comparison, the total dealloying depth in the phase-field simulations presented on Fig.~\ref{figc3DsLong} is $900$ nm and this estimate is therefore consistent with our numerical results. This estimate also reveals that this transient off-equilibrium concentrations could have an influence on the early stage of LMD experiments, for which $D_s/D_l=10^{-4}$ is a realistic ratio.

\subsection{Discussion on the discrepancy between phase-field simulations and experiments}
\label{seccpExp}

In the ternary phase diagram of Fig.~\ref{figc3TerPltCon}, we also added experimental measurements of concentrations obtained by postmortem chemical analysis of a Ta$_{15}$Ti$_{85}$ alloy dealloyed for $10$ s in a pure Cu melt \cite{lai2022topological}. 
Each experimental tie line links two points that correspond to concentration measurements of the three elements in the Ti-rich phase (corresponding to the liquid phase during dealloying) and the Ta-rich phase (corresponding to the solid ligaments).%
Concentration measurements are spatially averaged in each phase over a line parallel to the dealloying front at five different distances from the dealloying front. 
The horizontal and vertical error bars indicate the standard deviation obtained from multiple measurements of the concentration of Cu and Ti respectively.

\begin{figure}[htbp]
	\begin{center}
		\includegraphics[scale=0.5]{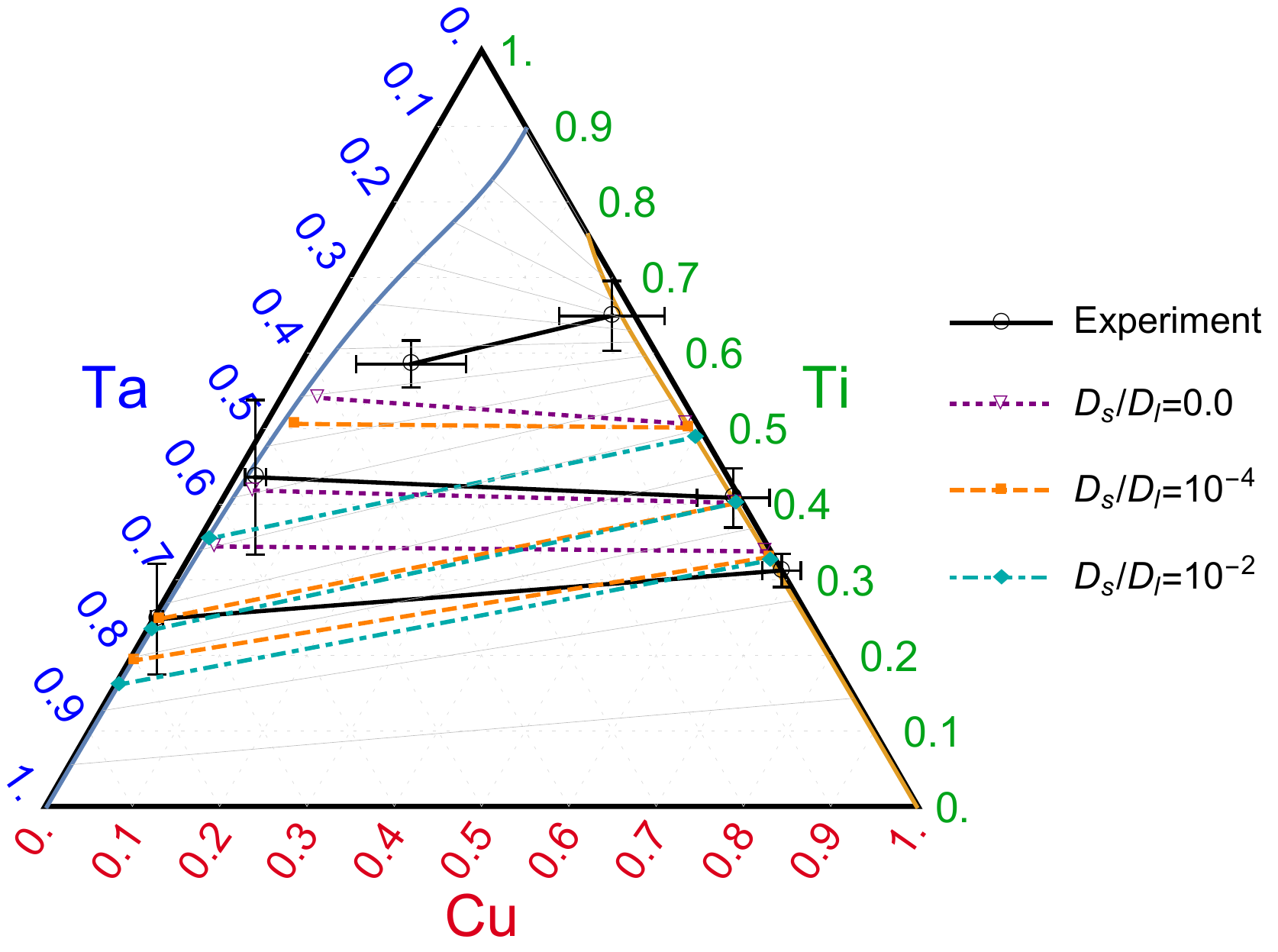} 
		\caption[phase-field simulations with varying solid-state diffusivity.]{Ternary phase diagram improved with mixing enthalpies and the equilibrium interfacial concentrations extracted from the phase field simulations for various solid-state diffusivities and experiments. }%
		\label{figc3TerPltConnPD}
	\end{center}
\end{figure}

The Ti concentration in the Ti rich phase decreases from about 0.7 close to the dealloying front to 0.3 close to the edge of the dealloyed layer.
This experimental result differs significantly from the phase diagram where the Ti concentration remains significantly smaller.
The discrepancies between experimental and numerical results may come from the simplified thermodynamic model employed in our simulations. As our phase diagram is generated from a set of simplified parameters, it may not model quantitatively the experimental system.

To improve this point, we can employ a richer thermodynamic model by replacing the original mixing enthalpy ($ \sum_{i<j}^{i,j\leq3} \Omega_{ij} c_i c_j$ in Eq.~\ref{eq:free_energy_ch}) by 
\begin{equation}
 \sum_{i<j}^{i,j\leq3} c_i c_j  \big[\Omega_{ij}^s\phi +\Omega_{ij}^l(1-\phi )+(c_i-c_j) ( {}^1 L_{ij}^s\phi+{}^1 L_{ij}^l(1-\phi))\big]
\end{equation}
The parameters obtained from the thermodynamic assessments of the real Cu-Ti and Ti-Ta system are listed in Table~\ref{tnPD} \cite{dinsdale1991sgte, ansara1998mg, arroyave2003thermodynamic}. The mixing enthalpy of the Cu-Ta system is adapted from the previous phase diagram to maintain the same solubility at the temperature 1513K.

\begin{table}[]
	\centering
\begin{tabular}{c|ccc}
\hline
$T$ (K)					& 	&		1513		&		\\ \hline	 \hline
		           			 & Cu-Ti           & Ti-Ta           &Cu-Ta           \\ \hline
$\Omega_{ij}^s$ (eV/nm$^3$)       & 3.512              & 12.44               & 75.62              \\ \hline
$\Omega_{ij}^l$ (eV/nm$^3$)       & -8.036               & 1.036              & 65.12              \\ \hline \hline 
${}^1L_{ij}^s$ (eV/nm$^3$)       & 0               & 2.591               & 0              \\ \hline
${}^1L_{ij}^l$ (eV/nm$^3$)       & 0               & 7.255               & 0              \\ \hline  
\end{tabular}
	\caption[Mixing enthalpy for the improved phase diagram.]{Parameters of the mixing enthalpy and simulation temperature for the improved phase diagram \cite{dinsdale1991sgte, ansara1998mg, arroyave2003thermodynamic}.}
	\label{tnPD}
\end{table}
The improved phase diagram with mixing enthalpies obtained from \cite{dinsdale1991sgte, ansara1998mg, arroyave2003thermodynamic} is shown in Fig.~\ref{figc3TerPltConnPD}. The tie-lines are different from the previous phase diagram which slightly improve the comparison with experimental results. We attribute the remaining discrepancy to (i) the significant measurement errors as shown with the error bars and (ii) the fact that the thermodynamic model remains incomplete. The equilibrium can be influenced by several material parameters such as ternary interaction (not considered here) that may change significantly the tie-lines of the phase-diagram.

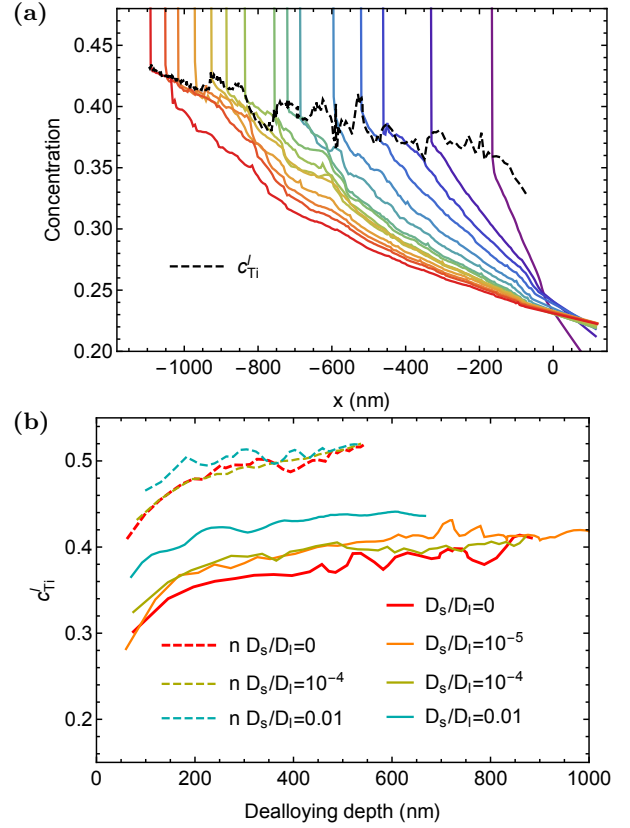
\begin{figure}[htbp] 
	\centering
	\includegraphics[scale=0.54]{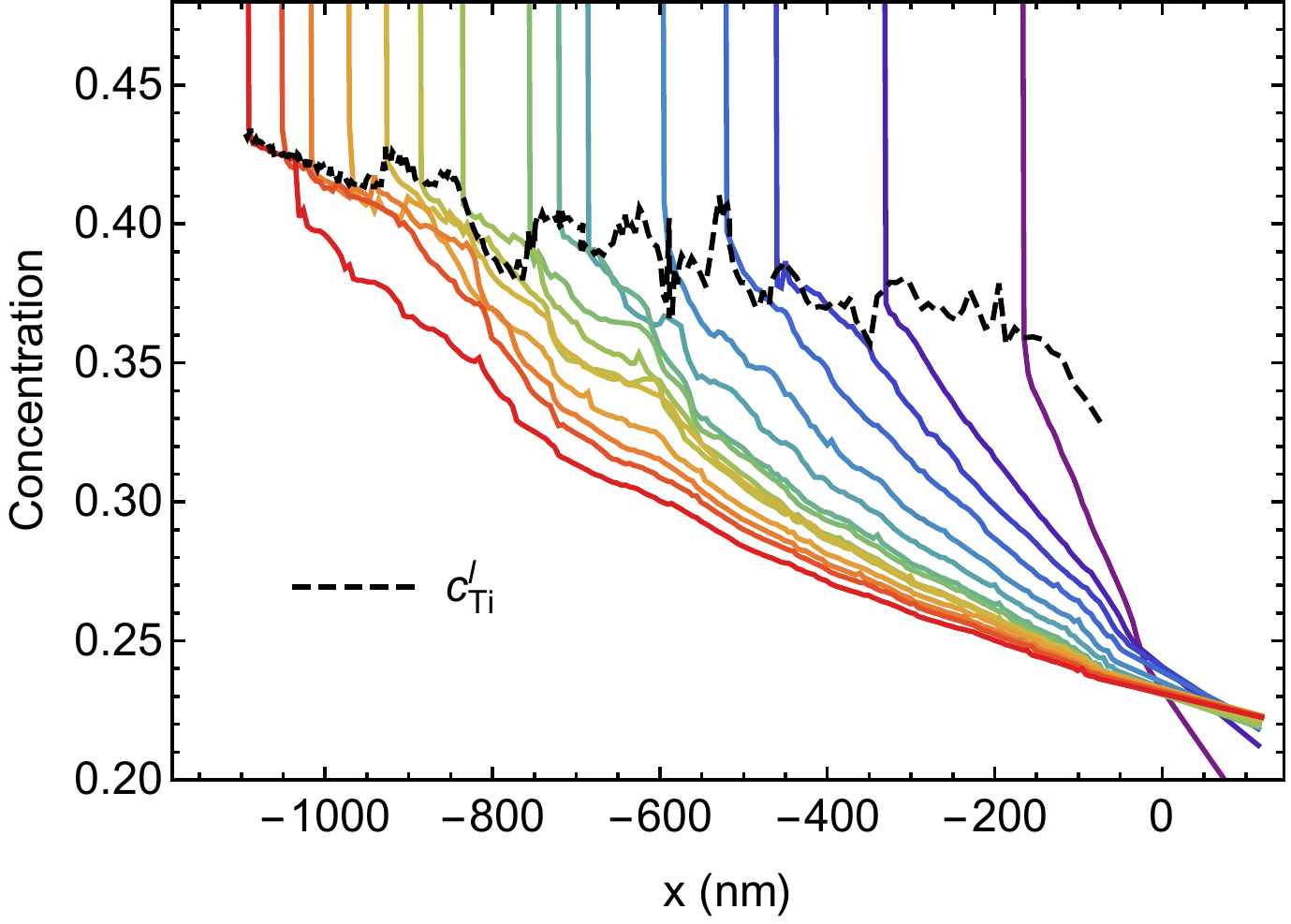}
	\includegraphics[scale=0.54]{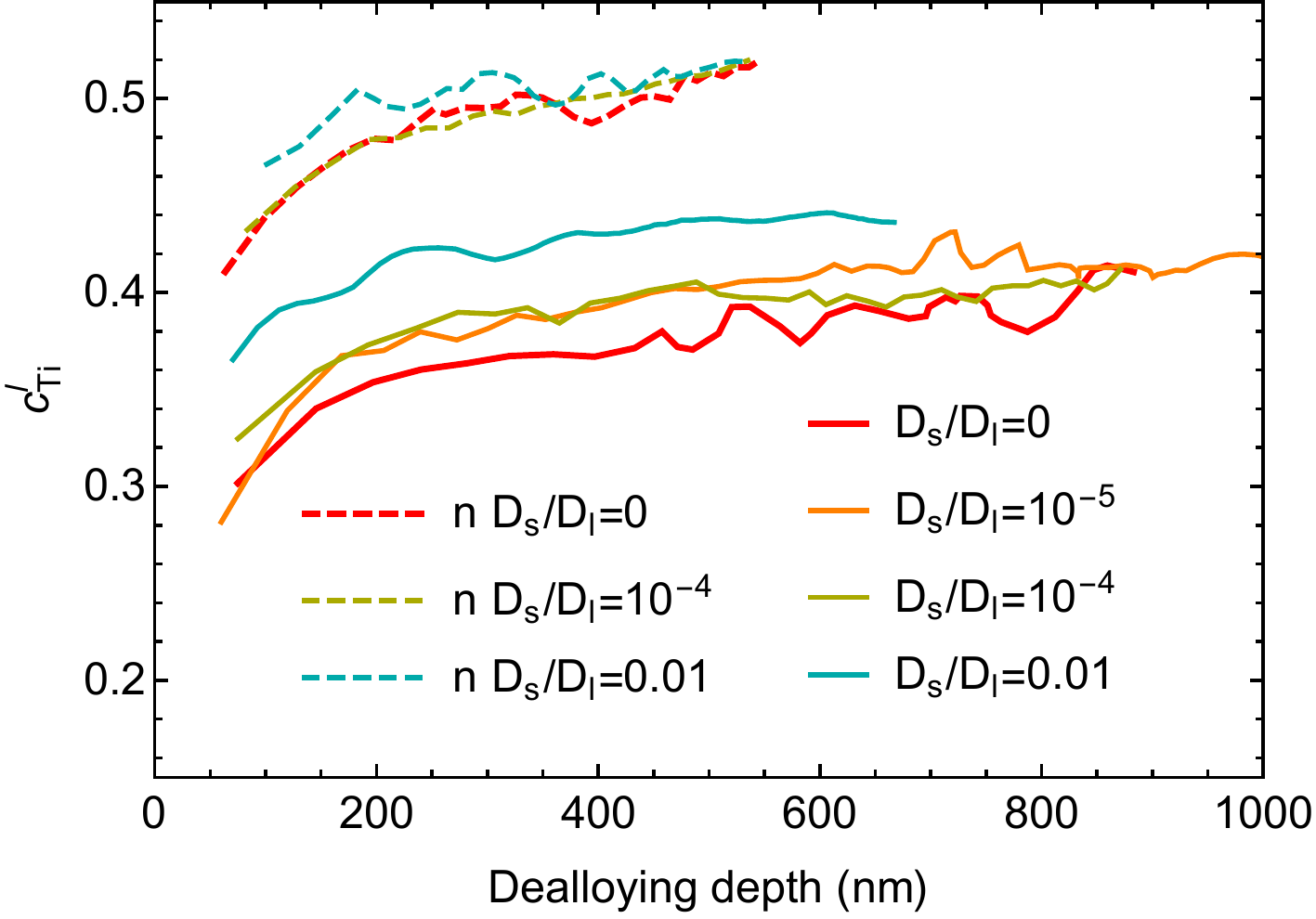}
	\begin{picture}(1,0)(0,0)
   		\put(-230.,305.) {\text{{\textbf{(a)}}}}
		\put(-230.,150.) {\text{{\textbf{(b)}}}}
	\end{picture}
	\caption[Evolution of the Ti concentration in the liquid at the dealloying front during the 2D dealloying with various solid-state diffusivities.]{(a) Ti concentration profiles extracted from 2D phase-field simulations of Ta$_{15}$Ti$_{85}$ alloys dealloyed in the pure Cu melt for several time frames. The colors varying from purple to red indicate the concentration profiles extracted from early to later time frames. The negative coordinate indicates the penetration of liquid channels (See section~\ref{1D}). The dashed line is the variation of the Ti concentration in the liquid at the dealloying front $c_{Ti}^l$ during dealloying.  (b) 2D Phase-field simulations of Ta$_{15}$Ti$_{85}$ alloys dealloyed in the pure Cu melt quantifying the increasing trend of the Ti concentration in the liquid at the dealloying front. The dashed lines are results from the improved phase-diagram.}
	\label{figTiLxi}
\end{figure}

We also note that the liquid Ti concentration is higher for the same solid Ta concentration compared to the simple phase diagram of Fig.~\ref{figtCuTiTa}. Phase-field simulations with this improved thermodynamic model indicate that the liquid Ti concentration at the solid-liquid interface $c_{Ti}^l$ approaches the experimental value. 
{To show quantitatively the difference, we first obtain the Ti concentration profiles from different time frames of simulations and then extract the Ti concentration at the dealloying front versus the corresponding dealloying depth (Fig.~\ref{figTiLxi}a).} As shown in Fig.~\ref{figTiLxi}b, the solid-state diffusivity has a small effect on the selection of $c_{Ti}^l$, but with the improved phase diagram (dash lines on Fig.~\ref{figTiLxi}b), $c_{Ti}^l \simeq 0.5$, which is closer to the experimental value of $0.7$. 

 {
In addition to the effect of the improved phase diagram, Fig.~\ref{figTiLxi} also reveals that the Ti concentration in the liquid at the dealloying front increases during dealloying, while interfacial concentrations were assumed to be constant during dealloying in a previous study \cite{geslin2015topology, mccue2016kinetics}.
We conjecture that the interfacial concentrations potentially have a relationship with the dealloying kinetics.
The dealloying kinetics is quantified by the velocity of the dealloying front $v=2pD_l/x_i$, where $x_i$ is the dealloying depth and $p$ is Peclet number from the diffusion law $x_{i}=\sqrt{4 p D_{l} t}$.
In Fig.~\ref{figcTiLFit}a, the continuous line represents the evolution of Ti concentration in the liquid at the dealloying front against the dealloying velocity obtained for dealloying in pure Cu melt. 
While fluctuating, $c^l_{Ti}$ decreases proportionally to $\ln(v)$, which can be fitted to
\begin{equation} 
	c_{Ti}^l=-k\ln (v/v_0), 
\label{eqFitcTiL}
\end{equation}
with the fitting parameters $k=0.0476$ and $v_0=8.03\times 10^9$ $\mathrm{nm/s}$ (dash line on Fig.~\ref{figcTiLFit}a). In addition, dots shown in Fig.~\ref{figcTiLFit}a report data obtained for different initial Ti content in the melt. These data points also follow approximatively Eq.~(\ref{eqFitcTiL}), which shows that this relation also holds for different initial bath compositions.

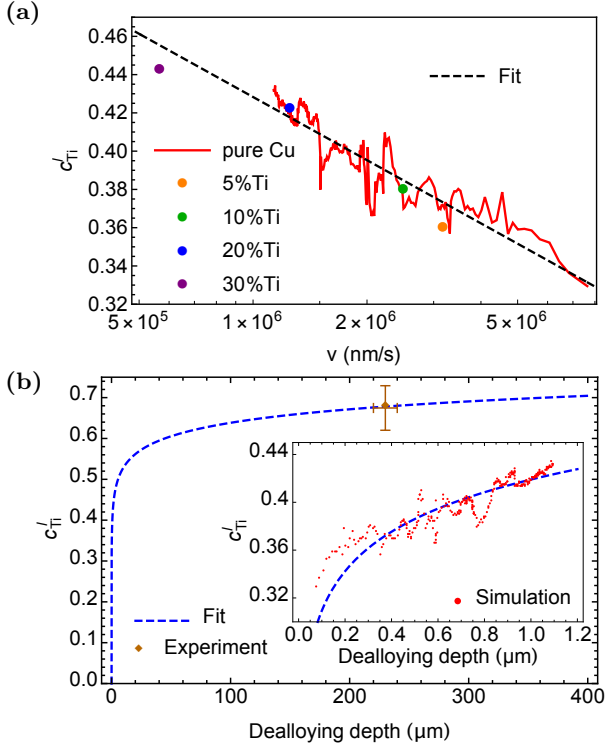
\begin{figure}[htbp] 
	\centering
	\includegraphics[scale=0.55]{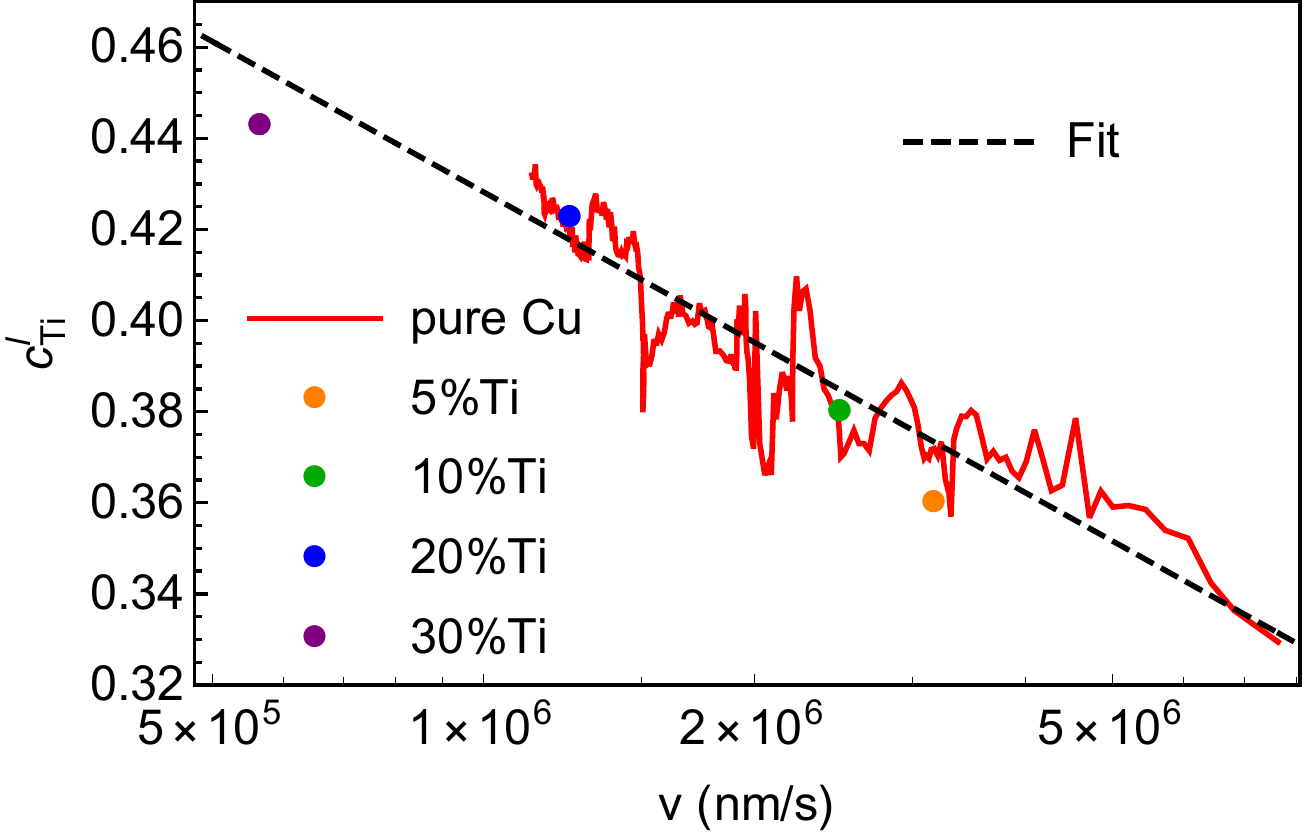}
	\includegraphics[scale=0.55]{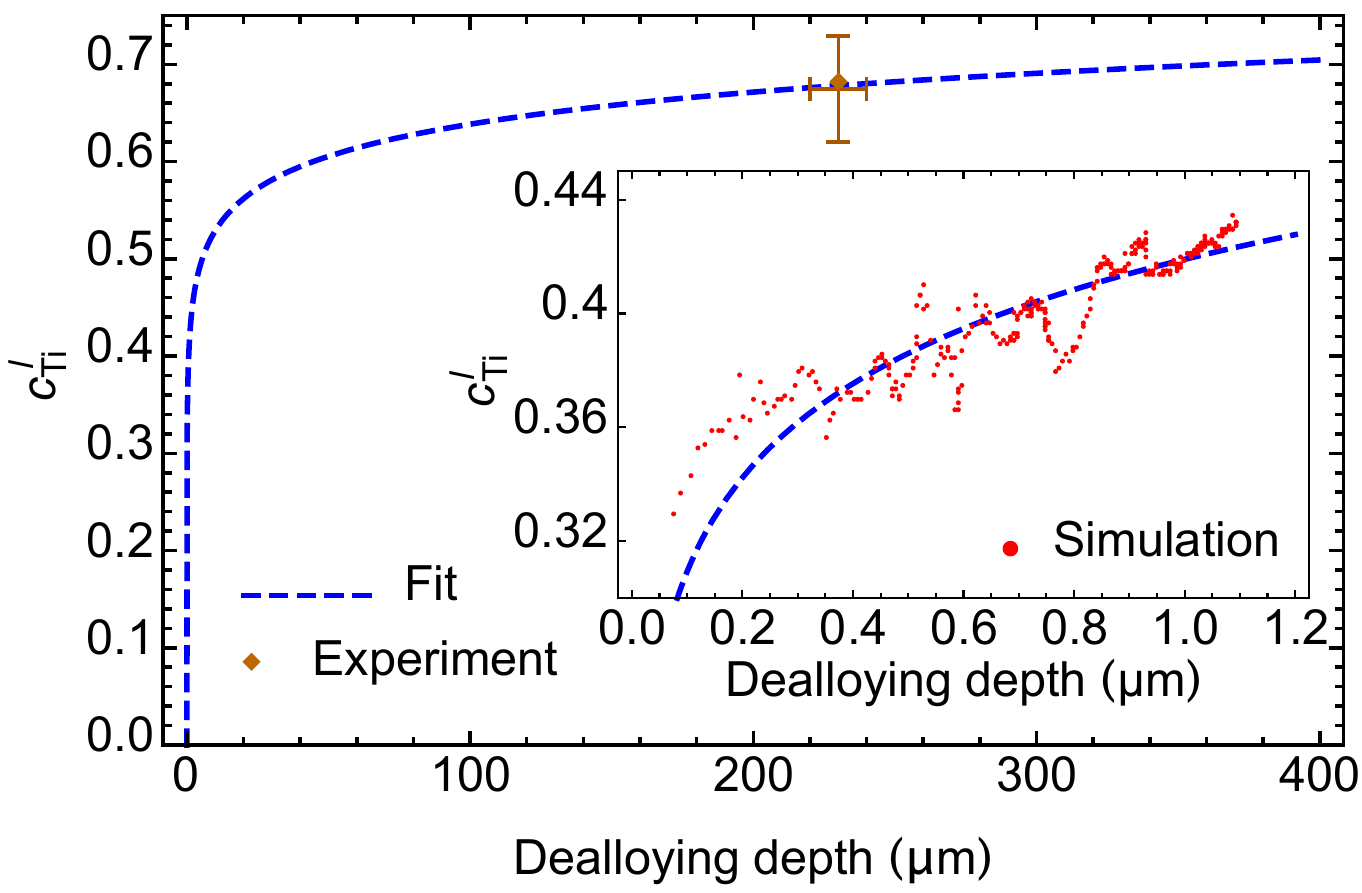}
	\begin{picture}(1,0)(0,0)
   		\put(-230.,275.) {\text{{\textbf{(a)}}}}
		\put(-230.,135.) {\text{{\textbf{(b)}}}}
	\end{picture}
	\caption[Evolution of the Ti concentration in the liquid at the dealloying front with log-linear fit.]{(a) Log-linear plot of the Ti concentration in the liquid at the dealloying front $c_{Ti}^l$ versus the dealloying velocity. Four dots are $c_{Ti}^l$ obtained from the last frame of phase-field simulations with various CuTi melts, where the dealloying depth is around $400$ nm. The red line is reported from Fig.~\ref{figTiLxi}a. The black dashed line is the best fit to the red line. (b) Fitting result of $c_{Ti}^l$ versus the dealloying depth extended to the experimental length scale with the comparison of the experiment \cite{lai2022topological}. The insert is the comparison between the fit and the simulation within the simulation length scale.}
	\label{figcTiLFit}
\end{figure}
Combining Eq.~\ref{eqFitcTiL} with the relation $x_iv=2pD_l$, we can rewrite the relation between $c_{Ti}^l$ and dealloying depth $x_i$ as
\begin{equation}
 	c_{Ti}^l=k \ln (x_i/x_0), 
\end{equation}
where the parameters $x_0=2pD_l/v_0=0.15$ nm and $p=0.0868$ are for dealloying of a Ta$_{15}$Ti$_{85}$ alloy by the pure Cu melt.
As shown in the insert in Fig.~\ref{figcTiLFit}b, at the length scale of simulations, the fit matches well with the simulation. 
While we extrapolate the fit to the experimental length scale, $c_{Ti}^l$ varies within the range of 0.1 for the change of the dealloying depth from 100 $\mu$m to 400 $\mu$m (Fig.~\ref{figcTiLFit}b). 
This extrapolation satisfies the observation of experiments \cite{mccue2016kinetics} which treats $c_{Ti}^l$ as constant within the range of the measurement error during dealloying.
Especially, as shown in Fig.~\ref{figcTiLFit}b, the predicted value of $c_{Ti}^l$ is in good agreement with the experimentally observed value.

A limitation of the fit is that $c_{Ti}^l$ will increase beyond the solubility limit of Ti in the liquid for very large dealloying depth beyond the experimental range. This limitation stems from the fact that the relation between $c_{Ti}^l$ and the dealloying depth is extrapolated from phase-field simulations
over a limited range of dealloying depths where $c_{Ti}^l$ only varies in the range $0.3$ to $0.45$. Therefore, we cannot expect the fit to be accurate 
for arbitrarily large depth. Despite this limitation, our extrapolation scheme successfully predicts the observed value of $c_{Ti}^l$, allowing us to bridge at least empirically phase-field simulations and experimental length and time scales. 
For larger dealloying depth than those probed experimentally, we expect that $c_{Ti}^l$ should slowly approach a plateau corresponding to the solubility limit of Ti in the liquid. 
}

\section{Conclusion}
In summary, we have used a combination of theoretical analysis and phase-field simulations to clarify several aspects of the liquid metal dealloying process. This study goes beyond our previous work on the topic \cite{geslin2015topology, mccue2017alloy} by (i) presenting a ternary diffusion model accounting for the diffusion of both Ta and Ti in the melt, (ii) further developing the linear stability analysis of the interfacial spinodal decomposition \nedit{\cite{morral1971spinodal, de1972analysis} mentioned in Ref.~\cite{geslin2015topology} and using it to predict the formation of connected morphologies as a function of the dealloying parameters} and (iii) investigating the role of solid state diffusivity on the kinetics and morphology of the dealloyed microstructure.

We first proposed a theoretical analysis for the 1D dealloying kinetics and the time evolution of the concentration profiles of the different elements.
This analysis reveals that the dealloying kinetics includes two regimes. 
At first, the dissolution kinetics slows down due to the build-up of the Ta peak at the solid-liquid interface. 
After the Ta peak stabilizes, the dissolution reaches a stationary regime with a small but steady Ta leak in the melt.
This dissolution kinetics follows the same $x_i \sim t^{1/2}$ diffusion kinetics as a binary dissolution assumed before \cite{mccue2016kinetics}.
We have shown that the predictions of our 1D dissolution model match well the results obtained from numerical simulations. 
\nedit{Furthermore, combining this 1D dissolution model with the phase equilibrium conditions enables us to predict the planar dissolution kinetics and interfacial concentrations in the limit of very small but finite solid diffusivity, which is relevant experimentally.}

In other situations than the ideal 1D case, Ta and Cu diffuse laterally along the solid-liquid interface, promoting a spinodal decomposition. A linear stability analysis detailed in section \ref{stability_analysis} allows to derive an analytical expression for the growth rate and fastest-growing wave vector during the spinodal decomposition. We showed that the wave-length predicted theoretically is of the same order of magnitude as the microstructure obtained from phase-field simulations, the discrepancy being attributed to the effect of non-linearities.
The analysis of the phase-field simulations indicates that the selection of the initial spacing is determined by the interplay between the development of the fastest growing wave-length and the slow dissolution kinetics of Ta-rich regions.
Furthermore, we apply the criterion of spinodal decomposition to investigate the planar dissolution obtained when Ti is added into the melt. If the driving force for spinodal decomposition remains negative, the dealloying interface remains planar while a positive driving force leads to spinodal decomposition.  
The 2D phase-field simulation results are found to be in good agreement with the prediction of the driving force of spinodal decomposition when the concentrations are taken from the 1D phase-field simulations.
\nedit{We go further by combining the 1D dissolution model proposed in Section~\ref{dis_phaseEq} with the criterion of spinodal decomposition to theoretically predict the occurrence of dealloying as a function of the composition of the base alloy and the melt. This analysis provides a prediction for the boundary between connected morphologies and planar-dissolution regime within the limit of very small but finite solid-state diffusivity.}

While we generally assume the solid-state diffusivity negligible in these phase-field simulations, the dealloying kinetics and morphologies are shown to be affected by this parameter, even though the solid-state diffusivity is four to five orders of magnitude smaller than the liquid-state diffusivity.
In 1D simulations, the solid-state diffusion enables the solid-liquid interface to relax to the local chemical equilibrium, thereby influencing the concentrations on the liquid side of the interface and in turn the dissolution kinetics. 
\nedit{This effect strongly influences the dealloying for the large solid-state diffusivity (e.g., $D_s/D_l\sim 10^{-2}$). For experimentally relevant values of the solid-state diffusivity ($ 10^{-4} D_l \sim 10^{-5} D_l $), the interfacial concentrations are shown to converge towards a chemical equilibrium. Interestingly, we showed that this chemical equilibrium depends on the specific value of the solid-state diffusivity.}
In 2D simulations, a finite solid-state diffusivity is found to promote the formation of lamellar structures, thereby favoring the formation of aligned microstructure over high-genus topologically connected structures. 
Despite the work presented in this paper, a discrepancy persists between experiments and phase-field results, in particular concerning the equilibrium concentration of Ti in the liquid ($c_{Ti}^l$)  that remains high in experiments ($\sim 0.7$) compared to numerical results ($\sim 0.4$).
This discrepancy can be explained by the limitation of the phase-field models. First, it can be attributed to the lack of accuracy of the simplified thermodynamic model employed in this work. In the last section, we employ a richer thermodynamic model that demonstrates a slight improvement of the numerical/experiment comparison.
{However, this thermodynamics model may not be precise enough. In particular, ternary interaction terms proportional to $c_1c_2c_3$ are neglected and may be important to take into account to achieve a quantitative comparison with experimental results. 
In addition, other parameters, such as the values of the coefficients $\sigma_i$ of the composition gradient terms are not easily determined and can influence significantly the results.}
{Second, $c_{Ti}^l$ may be higher in experiments than in phase-field simulations due to the fact that even long simulations only access dealloying depths on the $\mu$m scale that are one to two orders of magnitude smaller than those typically studied experimentally. This possibility is suggested by the finding that $c_{Ti}^l$ slowly increases with the dealloying depth in phase-field simulations. By fitting this behavior against a logarithmic law, we were able to extrapolate $c_{Ti}^l$ to experimentally relevant depths and found that this prediction agrees well with the measured value. \nedit{The logarithmic behavior is however only phenomenological and }the agreement with experimental observation is therefore only suggestive. Further work is needed to understand the physical mechanism of this slow logarithmic increase of $c_{Ti}^l$ to determine if it remains valid over the entire range of dealloying depth that spans both phase-field simulations and experiments.}
This work paves the way to several prospects towards the predictive modeling of the liquid metal dealloying process. 
Firstly, we have shown that combining the 1D ternary diffusion model with the linear stability analysis for spinodal decomposition could be used to predict the occurrence of the initial destabilization and therefore the development of a connected microstructure as function of the composition of the melt and the precursor alloy. This line of work could be applied to other systems in order to predict which combination of elements in the precursor and the melt can be used to obtain connected microstructures. 
Secondly, section \ref{solid_diff} shows that solid diffusion has to be incorporated in phase-field modeling in order to capture the appropriate chemical equilibrium at the solid-liquid interface and to yield quantitative results. As discussed in section \ref{solid_diff}, incorporating diffusion in both phases can also bring new insights to the morphologies evidenced in solid-state dealloying \cite{wada2016evolution,mccue2017alloy}. 
Finally, the important role of the Ta diffusion in the liquid phase evidenced in section \ref{1D} can also bring new insights into the coarsening mechanism of the connected microstructure. Indeed, most previous studies assumed that coarsening occurs by surface diffusion \cite{wada2013three, kim2015optimizing, wada2011dealloying, geslin2019phase}, while Ta diffusion in the liquid phase could contribute significantly to the coarsening mechanism \cite{lai2022topological} and better explain experimental observations.

\begin{acknowledgements}
This research was supported by Grant No. DE-FG02-07ER46400 from the U.S. Department of Energy, Office of Basic Energy Sciences.  \end{acknowledgements}

\small
\bibliographystyle{apsrev4-1}
\bibliography{dealloying.bib}

\end{document}